\documentclass[fleqn,usenatbib]{mnras}

\usepackage{newtxtext,newtxmath}
\usepackage{multirow}
\usepackage{hyperref}
\usepackage[T1]{fontenc}

\usepackage{graphicx}	% Including figure files
\usepackage{subfig}
\usepackage{amsmath}	% Advanced maths commands

\usepackage{amssymb}	% Extra maths symbols

\title{A Chemical Map of the Outbursting V883 Ori system: Vertical and Radial Structures}

\author[D. A. Ru\'iz-Rodr\'iguez et al.]{
D. A. Ru\'iz-Rodr\'iguez,$^{1}$\thanks{E-mail: druiz@nrao.edu; Jansky Fellow of the National Radio Astronomy Observatory.}
J. P. Williams,$^{2}$
J. H. Kastner,$^{3}$
L. Cieza,$^{4}$
M. Leemker,$^{5}$
and D. A. Principe.$^{6}$
\\
$^{1}$National Radio Astronomy Observatory, 520 Edgemont Road, Charlottesville, VA 22903-2475, USA\\
$^{2}$Institute for Astronomy, University of Hawaii, Honolulu, HI 96822, USA\\
$^{3}$ Chester F. Carlson Center for Imaging Science, School of Physics $\&$ Astronomy, and Laboratory for Multiwavelength Astrophysics,\\ Rochester Institute of Technology, 54 Lomb Memorial Drive, Rochester, NY 14623, USA\\
$^{4}$Facultad de Ingenier\'ia y Ciencias, N\'ucleo de Astronom\'ia, Universidad Diego Portales, Av. Ejercito 441. Santiago, Chile\\
$^{5}$Leiden Observatory, Leiden University, P.O. box 9513, 2300 RA Leiden, The Netherlands\\
$^{6}$Massachusetts Institute of Technology, Kavli Institute for Astrophysics and Space Research, Cambridge, MA, 02138, USA
}

\date{Accepted XXX. Received YYY; in original form ZZZ}

\pubyear{2022}

\begin{document}
\label{firstpage}
\pagerange{\pageref{firstpage}--\pageref{lastpage}}
\maketitle

\begin{abstract}

We present the first results of a pilot program to conduct an Atacama Large Millimeter/sub-millimeterArray (ALMA) Band 6 (211 -- 275 GHz) spectral line study of young stellar objects (YSO) that are undergoing rapid accretion episodes, i.e. FU Ori objects (FUors). Here, we report on molecular emission line observations of the FUor system, V883 Ori. In order to image the FUor object with a full coverage from $\sim$0.5$^{''}$ to the map size of $\sim$30$^{''}$, i.e. from disc to outflow scales, we combine the ALMA main array (the 12-m array) with the Atacama Compact Array (7-m array) and the total power (TP) array. We detect HCN, HCO$^{+}$, CH$_{3}$OH, SO, DCN, and H$_{2}$CO emission with most of these lines displaying complex kinematics. From PV diagrams, the detected molecules HCN, HCO$^{+}$, CH$_{3}$OH, DCN, SO, and H$_{2}$CO probe a Keplerian rotating disc in a direction perpendicular to the large-scale outflow detected previously with the $^{12}$CO and $^{13}$CO lines. Additionally, HCN and HCO$^{+}$ reveal kinematic signatures of infall motion. The north outflow is seen in HCO$^{+}$, H$_{2}$CO, and SO emission. Interestingly, HCO$^{+}$ emission reveals a pronounced inner depression or ``hole'' with a size comparable to the radial extension estimated for the CH$_{3}$OH and 230 GHz continuum. The inner depression in the integrated HCO$^{+}$ intensity distribution of V883 Ori is most likely the result of optical depth effects, wherein the optically thick nature of the HCO$^{+}$ and continuum emission towards the innermost parts of V883 Ori can result in a continuum subtraction artifact in the final HCO$^{+}$ flux level.

\end{abstract}


%%%%%%%%%%%%%%%%%%%%%%%%%%%%%%%%%%%%%%%%%%%%%%%%%%

%%%%%%%%%%%%%%%%% BODY OF PAPER %%%%%%%%%%%%%%%%%%

\section{Introduction} \label{sec:intro}

The chemistry governing planet formation is set early during the formation and evolution of discs surrounding young stellar objects (YSOs). Highly energetic events, such as outflows and outbursts, beginning at the earliest stellar formation phases likely impact and mold the final outcome of disc evolution, ultimately determining the architectures of exoplanet systems \citep{Hubbard2017, Tychoniec2020}. It is believed that outflows and their effects on discs and planet building are intimately connected to the accretion process \citep[][and refs. therein]{Frank2014} which, in turn, is punctuated by occasional sharp increases in accretion rate \citep[][and refs. therein]{Audard2014}. Thus, it is not surprising that YSOs that are undergoing rapid accretion episodes, such as FU Orionis star-disc systems (FUors), also show high levels of outflow activity \citep{Hartmann1996, Ruiz2017a, Ruiz2017b}.

These classical FUors are known to present accretion rates between 10$^{-6}$ and 10$^{-4}$ $M_{\odot}$ yr$^{-1}$, reaching bolometric luminosities of 100-400 $L_{\odot}$, during every outburst episode, usually a few years or decades of duration \citep[e.g.][]{Audard2014}. In effect, this very brief, but powerful, interaction between the central source and its surroundings alters the physical structure and chemical abundances of the young disc itself \citep[e.g.][]{Cieza2016, vantHoff2018, Lee2019, Leemker2020}. For example, \citet{Cieza2016} used 1.3 mm continuum observations  to claim the first-time detection of a water snow-line in a circumstellar disc via Atacama Large Millimeter/sub-millimeter Array (ALMA) observations of an FUor object, V883 Ori, with the enlarged radius of the snow-line ($\sim$ 40 au) being a direct result of the heating of the disc during the outburst. Similarly, it has been suggested that the presence of different molecules (e.g. HCO$^{+}$, CH$_{3}$OH) in the gas phase as a function of radius can be useful in the location of the snow-line, which in turn are likely being affected by the ongoing outburst episode \citep{vantHoff2018, Lee2019, Leemker2020}.

Since only a handful of FUors have been identified, meaning that most YSOs are currently in a low luminosity phase, it is difficult to estimate a time scale since the last outburst occurred and/or predict an oncoming outburst for any given system \citep{Molyarova2018, Principe2018}. Without a clear estimate of an evolutionary time scale and periodicity of these outbursting phenomena, the impact of the FUor-outburst on the disc and envelope chemical structure remains unknown, preventing us to set the initial conditions for planet formation. Fortunately, besides shaping the disc physical structure \citep[e.g.][]{Cieza2016}, outbursts must leave footprints on the disc and envelope chemical composition \citep{Rab2017}. The intricate connection between high energy emission generated during massive outbursts (e.g. X-rays, UV radiation) and the properties of burst events (e.g. their duration, repetition time, and peak flux) results in an enhanced molecular ionization, which leads to rich chemistry that can be probed by different tracers. For instance, \citet{Jorgensen2013} observed the low-mass proto-star IRAS 15398-3359 on 0.5$^{''}$ (75 au diameter) scales with ALMA at 340 GHz and detected a ring-like structure with a radius of about 1-1.5$^{''}$  (150-200 au) from H$^{13}$CO$^{+}$ emission. These authors proposed that inner-disc water vapor, produced by an accretion burst within the last 100-1000 yr, may have destroyed any present HCO$^{+}$. Also, \citet{Cleeves2017} reported variation in the optically thin H$^{13}$CO$^{+}$ emission in the IM Lup disc, which can potentially be explained via X-ray driven chemistry enhancing the HCO$^{+}$ abundance in the upper layers of the disc atmosphere during large or prolonged flaring events. Thus, ongoing and past outbursts likely play a significant role in the chemical and physical evolution of YSOs, whose impact can be detected from different molecular tracers at different vertical layers and radial scales \citep[e.g.][]{Ruiz2020}. Nevertheless, the impact of high-energy-induced molecular ionization and the potential importance of energetic outbursts during FUor and FUor-like events at the earliest evolutionary stages remains unexplored.

  In this paper, we report on ALMA Band-6 observations of V883 Ori, searching for chemical signatures observed in density tracers (HCN, HCO$^{+}$, DCN), and shock tracers (CH$_{3}$OH, SiO, SO), which can be used to identify and distinguish regions of shock-excited gas, and thermal desorption of these species by high-energy  (UV and X-ray)  radiation. V883 Ori is a $\sim$1.3 $M_{\odot}$ object located in the Orion molecular cloud at a distance of 417 pc \citep{Menten2007}. Initially, this object was found to have a detectable water snow-line at a radius of $\sim$ 45 au from the central source, considering a sublimation temperature of water of $>$ 100 K at the mid-plane \citep{Cieza2016}. However, more recent HCO$^{+}$ and CH$_{3}$OH observations have suggested that the snowline may be much further out \citep{vantHoff2018, Leemker2020} if the location of the snow line is taken as the radius where half of the water is in gas-phase and the other half is frozen out onto the grains. To date, several molecular species have been detected in V883 Ori with impressive morphological structures, however, these studies are focused only on the disc structures \citep[e.g.][]{vantHoff2018, Lee2019}, independent of cavities and outflows traced by $^{12}$CO and $^{13}$CO emission lines \citep{Ruiz2017b}. Characterizing these systems in their entirety offers valuable insights of density and thermal structures related to other sources of energy such as X-rays and UV radiation.

Here, we present the emission detections of HCN, HCO$^{+}$, SO, DCN, and H$_{2}$CO in the V883 Ori environment, together with other well-utilized molecular tracers ($^{12}$CO, $^{13}$CO, CH$_{3}$OH, H$^{13}$CO$^{+}$) already presented in previous studies \citep[e.g.][]{Ruiz2017b, Lee2019}. These emission lines are collectively used as a probe of the circumstellar disc and outflows as well as their ``transition" layers. Thus, these chemical tracers offer a unique view of the vertical and radial structure of the disc-outflow system at the scale of $\sim$ 100 - 2000 au, whose significance in understanding the chemical properties in FUor objects is relevant to the final outcome, i.e. planetary systems. The aim of this work is to investigate whether or not chemistry alone can explain the radial emission profiles of these molecules, or if other high-energy radiation sources are required to produce the observed emission features.

%whose interpretation is key  to the understanding of the final outcome, planetary systems.
This  paper is  organized as follows. In Section \ref{sec:observations}, we  describe the observing setup and the data reduction process. In Section \ref{Sec:Results}, we present the results of the analysis, while in Section \ref{Sec:Discussion} we discuss our findings. To finalize, we state our conclusions in Section \ref{Sec:Conclusions}.

\section{ALMA Observations and Data Analysis} \label{sec:observations}

\begin{table*}
%\contcaption{Deconvolved radial extensions fr gom ALMA observations.}
\caption{Observational details.}
\begin{tabular}{cccccccc}
\hline Array & Observation  & Bandpass & Flux & Phase & Time On & Median PWV  \\
& Date & Calibrator & Calibrator  &Calibrator& Source [mm:ss] & [mm]\\
\hline
&&\textbf{ALMA Cycle 5}&Project ID: &2017.1.00015.S &PI: J. Williams&\\
\hline

12-m $\times$ 44 & 21-Jan-2018   &J0607-0834 & J0607-0834 & J0541-0541 & 17:21 & 0.98 \\
12-m $\times$ 43 & 12-Jul-2018   & J0542-0913 & J0522-3627 & J0542-0913 & 08:48 & 2.25 \\
7-m $\times$ 11 & 31-Oct-2017  & J0522-3627& J0607-0834 & J0607-0834 & 34:45  & 1.43 \\
7-m $\times$ 11 & 01-Nov-2017  & J0522-3627& J0607-0834 & J0607-0834 & 34:11 & 0.58 \\
TP & 03-May-2018   &J0542-0913&J0542-0913 &J0542-0913 & 38:01 & 0.63 \\
TP & 04-May-2018   &J0542-0913& J0542-0913 &J0542-0913 & 38:01 & 0.71 \\
TP & 06-May-2018  & J0542-0913 & J0542-0913 &J0542-0913 & 38:01 &  1.40 \\
TP & 15-May-2018  &J0607-0834 & J0607-0834 &J0607-0834 & 38:01 & 0.89 \\

\hline
&&\textbf{ALMA Cycle 6}& Project ID: &2018.1.01131.S & PI: D. Ruiz-Rodriguez &\\
\hline

12-m $\times$ 41 & 28-Apr-2019  & J0423-0120 & J0532-0307 & J0529-0519 & 05:03  & 1.0 \\
7-m $\times$ 11 & 06-Mar-2019 & J0423-0120& J0423-0120 & J0501-0159 & 05:02 & 2.7  \\
TP & 28-Apr-2019  & J0607-0834& J0607-0834 &J0607-0834 &  39:25 & 0.95 \\
TP & 02-May-2019  & J0607-0834& J0607-0834 &J0607-0834   & 39:25 & 1.1 \\

\hline
\end{tabular}
\label{Table:Observations}
%{FHWM value obtained from Gaussian fit. For details, see Sec}
%{Radial extension obtained after deconvolving FWHM from beamsize.}
\end{table*}

\subsection{ALMA Observations}
\label{Sec:Observations}

\begin{table*}
%\contcaption{Deconvolved radial extensions fr gom ALMA observations.}
\caption{Spectral Windows}
\begin{tabular}{lcccccccc}
\hline Cycle& Line & Rest Freq. & \multicolumn{2}{c}{12-m} &  \multicolumn{2}{c}{7-m} & \multicolumn{2}{c}{TP}  \\
\hline
%&&&\bf{ALMA}& \bf{Cycle 5}&&\\
&&&Bandwidth&Channel&Bandwidth&Channel&Bandwidth&Channel\\
&&[GHz]&[MHz]&[$\#$]&[MHz]&[$\#$]&[MHz]&[$\#$]\\

\hline
\multicolumn{1}{|l|}{\multirow{8}{*}{\rotatebox{90}{\textbf{Cycle 5}}}}& SiO v=0 5-4 & 217.105 &58.5 & 480& 62.5&512& 62.5&512\\
\multicolumn{1}{|l|}{} & DCN v=0 J=3-2& 217.238 & 58.5 & 480 & 62.5&512&62.5&512 \\
\multicolumn{1}{|l|}{} & H$_{2}$CO 3(2,1)-2(2,0) & 218.760 & 58.5 & 480 & 62.5&512& 62.5&512 \\
\multicolumn{1}{|l|}{} & C$^{18}$O  v=0 2-1$^a$  & 219.560 & 58.5 &960 & 62.5&1024 & 62.5&1024   \\
\multicolumn{1}{|l|}{} & SO 3$\Sigma$ v=0 6(5)-5(4)& 219.949 & 58.5 &960 & 62.5&1024& 62.5&1024\\
\multicolumn{1}{|l|}{} & $^{13}$CO  v=0 2-1$^a$  & 220.398 & 58.5 &960 & 62.5&1024 & 62.5&1024 \\
\multicolumn{1}{|l|}{} & $^{12}$CO  v=0 2-1$^a$ & 230.55   & 117.2 & 960 & 62.5&1024 & 125 & 1024 \\
\multicolumn{1}{|l|}{} & $^{13}$CS v=0 5-4 & 231.221 & 117.2 & 960 & 125 & 1024 & 125 & 1024 \\
\hline

 \multicolumn{1}{|l|}{\multirow{3}{*}{\rotatebox{90}{\textbf{Cycle 6}}}} & CH$_{3}$OH v t=0 4(3,1)-4(2,2) & 251.867 & 187.5 & 480  & 2000 & 512 & 2000& 512\\
\multicolumn{1}{|l|}{} & HCN v=0 J=3-2 & 265.886 & 937.5 & 1920 & 1000& 2048 & 1000& 2048 \\
\multicolumn{1}{|l|}{}  & HCO$^{+}$ v=0 3-2& 267.558 & 937.5 & 1920 &  1000& 2048 & 1000& 2048\\
\hline
\end{tabular}
\\
\footnotesize{$^a$ CO isotopologues will be presented in Ruiz-Rodriguez et al. In prep. }\\
\label{Table:Width}

\end{table*}

We obtained ALMA Cycle 5 and Cycle 6 Band 6 observations toward V883 Ori with a center position of $\alpha$ (J2000) = 18$^{\rm h}$ 14$^{\rm m}$ 10.47$^{\rm s}$ ; $\delta$ (J2000) = -32$^{\rm o}$47$^{\rm '}$34.50$^{\rm ''}$. In order to image the FUor object with a full coverage from 0.5$^{''}$ to the map size of 30$^{''}$, we combine the ALMA main array (12-m array) with the Atacama Compact Array (7 m array) and the total power (TP) array (12-m diameter, single-dish). By combining the 12-m, 7-m, and TP data, the observations are sensitive to emission, from disc to outflow scales. Table \ref{Table:Observations} summarizes the observational logs for this project.

In Cycle 5, V883 Ori was observed (project ID: 2017.1.00015.S PI: J. Williams) using the compact (baselines $\sim$ 55 to 91 m) and the extended configurations, C43-2 (15 - 315 m) and C43-5 (15 - 1400 m), on 2017 October 31, 2018 Jan 21, and July 12, respectively (Table \ref{Table:Observations}). The FoV of 30$^{''}$ $\times$ 30$^{''}$ was covered by a mosaic of 27 pointing (12-m array). The shorter u$-$v distance range was fulfilled by observations using a 7 pointing mosaic (7-m array) with a maximum resolvable size of 30$^{''}$, and the TP observations covered the diffuse emission over the entire mapped region. The spectral windows targeted $^{12}$CO(J = 2-1), $^{13}$CO(J = 2-1), C$^{18}$O (J = 2-1), SO (J = 6(5)-5(4)), DCN (J = 3-2), SiO (J= 5-4), $^{13}$CS (J = 5-4) and continuum, at rest frame frequencies of 230.588, 220.398, 219.560, 219.949, 217.238, 217.104, 231.220, and 232.50 GHz, respectively (Table \ref{Table:Width}). The correlator for these spectral windows were set to have bandwidths between 58.5 MHz and 2000 MHz. More details can be found in table \ref{Table:Width}.

In cycle 6, the 12-m array observations were obtained on 2019 April 28, the 7-m observations on 2019 March 06, and the TP observations on 2019 April 28 and May 02 (see Table \ref{Table:Observations}), and as part of the program 2018.1.01131.S (PI: D. Ruiz-Rodriguez). The array configuration was C43-4 with the shortest and longest baselines of 15 m and 784 m, respectively. The correlator was set up with three spectral windows in dual polarization mode, centered at 265.886 GHz (HCN J = 3-2 line), 267.557 GHz (HCO$^{+}$ J = 3-2 line), and 251.866 GHz (CH$_{3}$OH). The bandwidths used in the spectral windows are summarized in Table \ref{Table:Width} for each configuration.

\begin{table*}
\caption{Integrated line intensities}
\begin{tabular}{lccccccc}
\hline Line & Rest Freq. & Peak Flux & Integrated$^a$ & Beam & P.A. & rms\\
& [GHz] & [Jy beam$^{-1}$ km s$^{-1}$] & [Jy km s$^{-1}$] & [" $\times$ "] & [deg.] & [$\times$ 10$^{-2}$ Jy/beam] & \\
\hline
&&&\textbf{ALMA Cycle 5}&&\\
\hline
SiO v=0 5-4 & 217.105 & Not detected & -- & 1.5 $\times$ 1.1 & -77.0 &1.5\\
DCN v=0 J=3-2& 217.238 & 0.80 $\pm$ 0.02 & 1.12 $\pm$ 0.04 & 1.5 $\times$ 1.1 & -77.0&2.4\\
H$_{2}$CO 3(2,1)-2(2,0) & 218.760 & 0.95 $\pm$ 0.02 & 1.20  $\pm$ 0.04 & 1.5 $\times$ 1.1 & -77.1&5.5 \\
SO 3$\Sigma$ v=0 6(5)-5(4)& 219.949 & 0.80  $\pm$ 0.04 & 1.10   $\pm$ 0.10& 1.5 $\times$ 1.1 & -77.8 &1.6\\
$^{13}$CS v=0 5-4 & 231.221 & Not detected & -- & 1.5 $\times$ 1.1  &-77.3 &2.6 \\

\hline
&&&\bf{ALMA Cycle 6}&\\
\hline

CH$_{3}$OH v t=0 4(3,1)-4(2,2) & 251.867 & 0.47 $\pm$ 0.01 & 0.71  $\pm$ 0.02& 0.5 $\times$ 0.4 & -83.1 &4.5 \\
HCN v=0 J=3-2 & 265.886 & 0.97  $\pm$ 0.03 & 4.60   $\pm$ 0.16& 0.5 $\times$ 0.4 & -84.1 &1.1 \\
HCO$^{+}$ v=0 3-2& 267.558 & 0.31 $\pm$ 0.03 & 6.50  $\pm$ 0.50 & 0.5 $\times$ 0.4 &-86.2 &7.0\\
\hline
\end{tabular}
\\
\footnotesize{$^a$ The uncertainty includes a 10$\%$ systematic flux calibration error. }\\
\label{Table:Imaging}

\end{table*}

\begin{figure*}
\centering   
     \includegraphics[width=0.3\textwidth]{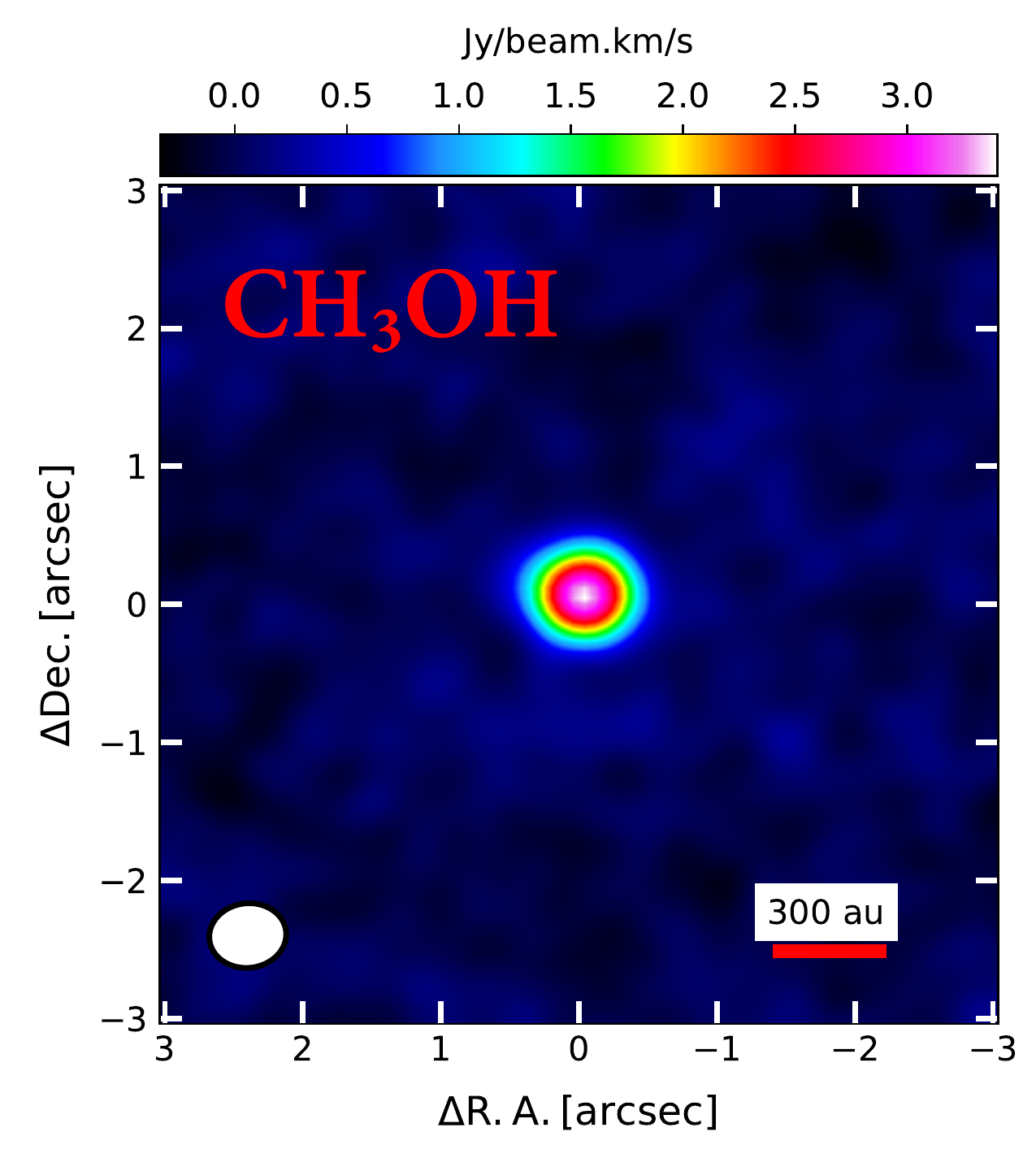}
    \includegraphics[width=0.3\textwidth]{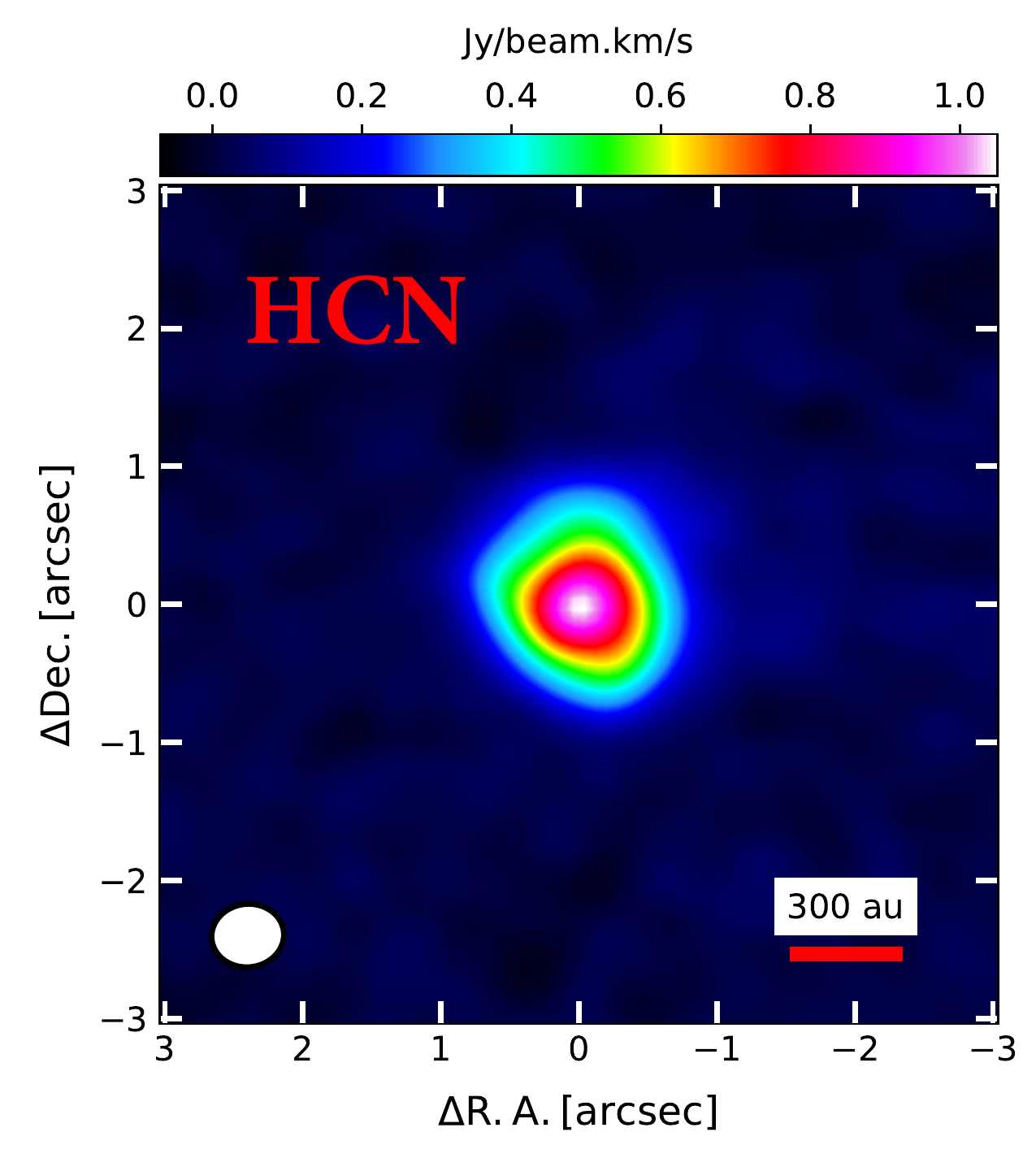}
    \includegraphics[width=0.3\textwidth]{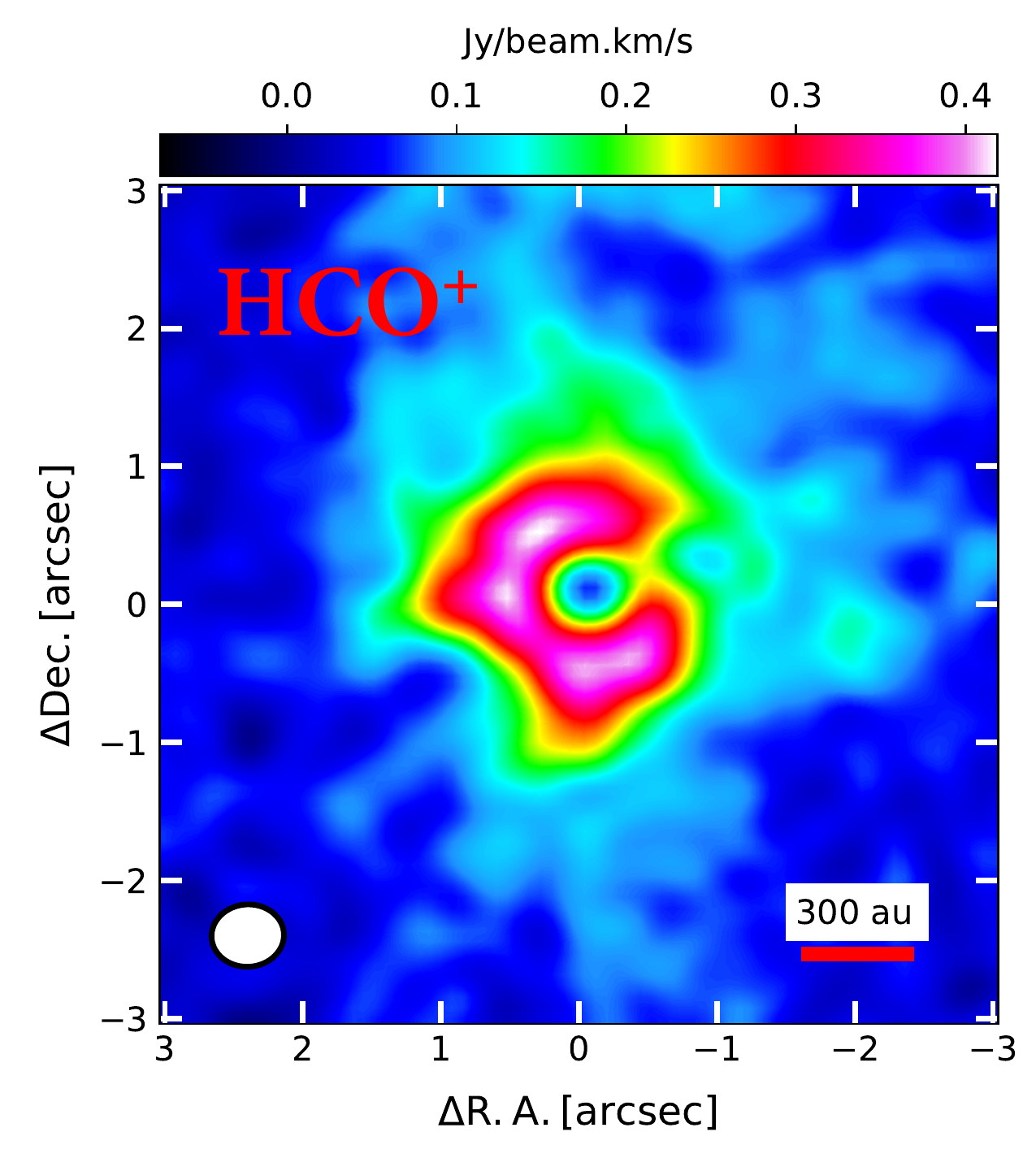}
%\begin{subfigure}{.5\textwidth}
    \centering
    \includegraphics[width=0.3\textwidth]{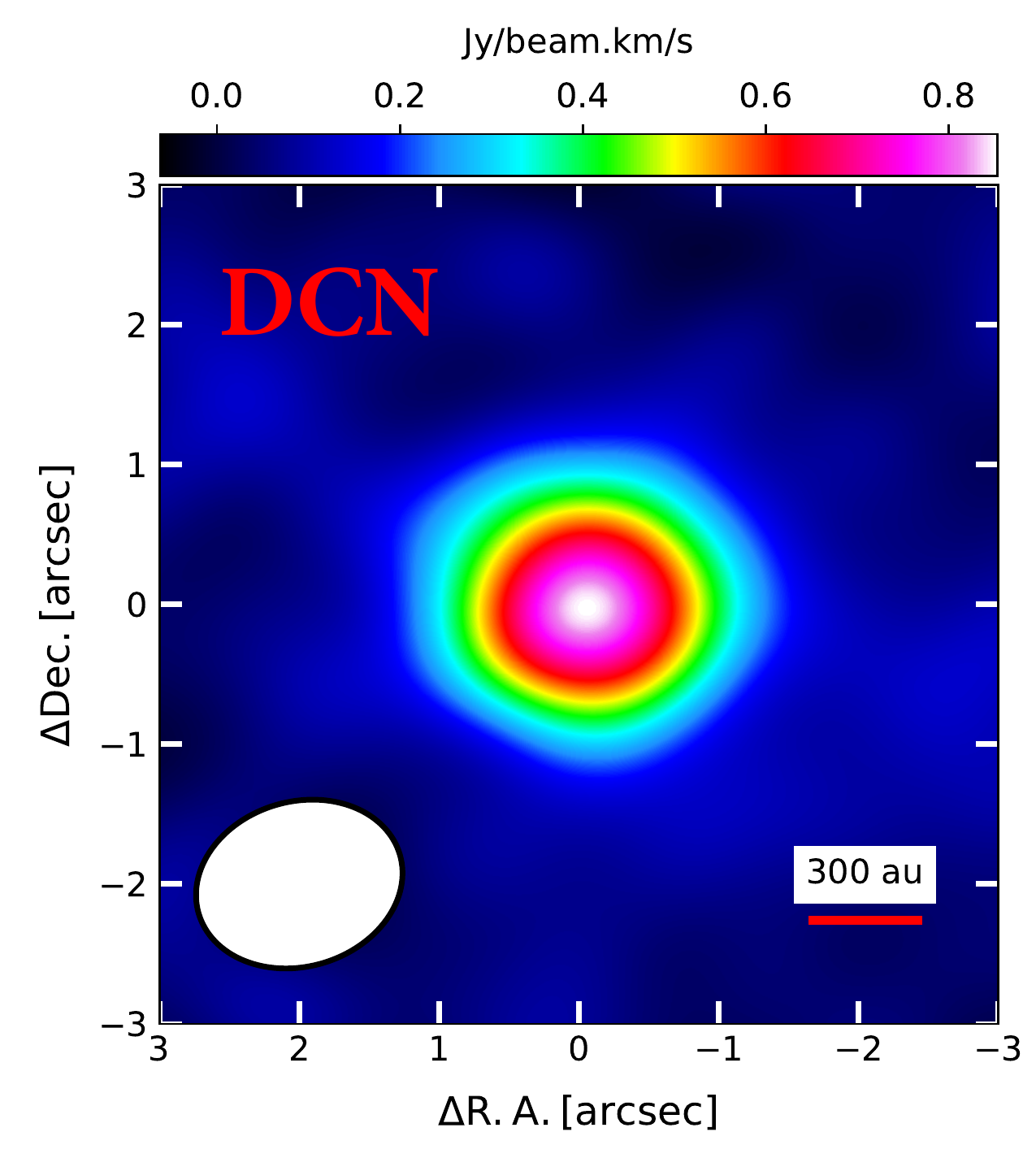}
    \includegraphics[width=0.3\textwidth]{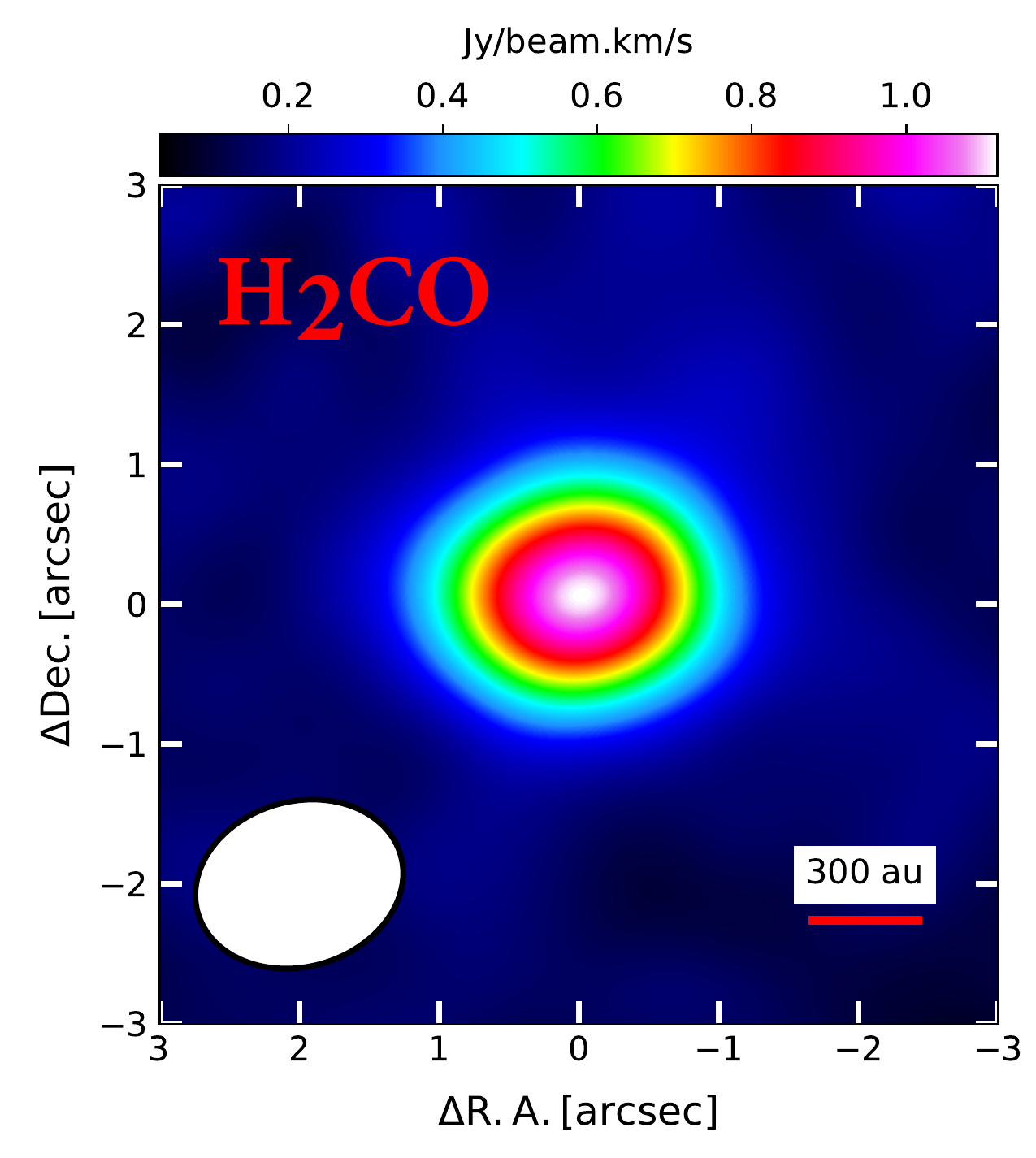}
    \includegraphics[width=0.3\textwidth]{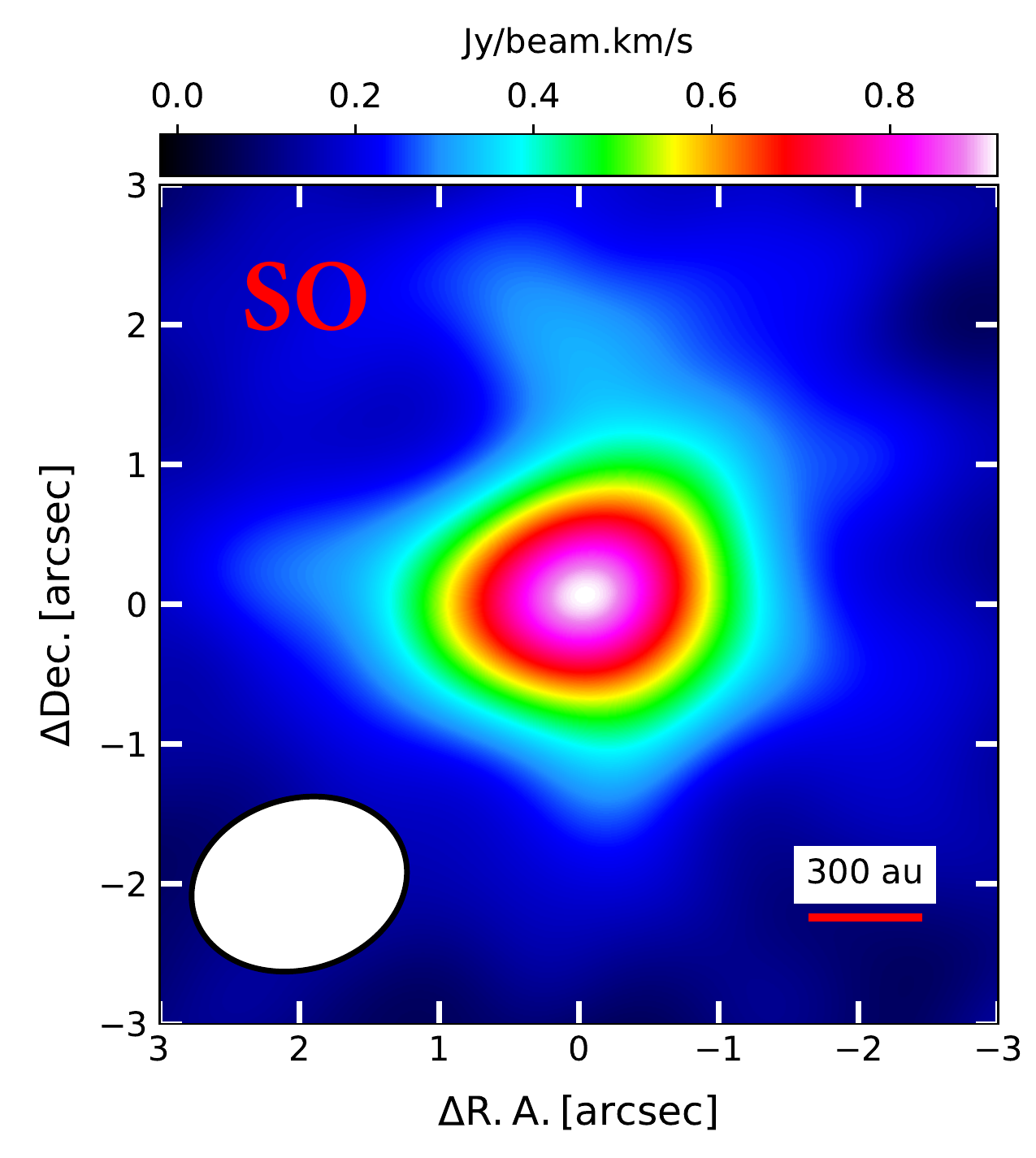}
%\end{subfigure}
\caption{Moment 0 maps of molecular emission toward V883 Ori. Molecular lines are the CH$_{3}$OH (top-left panel), HCN (top-middle panel), and HCO$^{+}$ (top-right panel), DCN (bottom-left panel), H$_{2}$CO (bottom-middle panel), and SO (bottom-right panel). The line integrated emission of the CH$_{3}$OH, HCN, HCO$^{+}$, DCN, H$_{2}$CO, and SO lines was performed in the range from -5.0 to 10 km s$^{-1}$, -0.2 to 8.8 km s$^{-1}$, 2.2 to 6.8 km s$^{-1}$, 2.2 to 6.52 km s$^{-1}$, 2.2 to 6.52 km s$^{-1}$, and 2.56 to 6.88 km s$^{-1}$, respectively. In each panel, the corresponding beam size is shown at the bottom left. North is up, east is left}
\label{Fig:mom0}
\end{figure*}

\subsection{Data Analysis}\label{Sec:Analysis}

Data were calibrated using the standard calibration script in the CASA software package (Version 4.7.2- Cycle 5, and 5.4.0 - Cycle 6). In order to concatenate all data sets, we initially converted the TP map into visibilities by using the Total Power to Visibilities (TP2VIS) package that runs on the CASA platform \citep[and references therein]{Koda2019}. We adopt the root-mean-square (rms) noise in the TP map\footnote{A more detailed description of the reduction process can be found in Appendix \ref{App:C}.} as the optimal weight of the TP visibilities\footnote{https://github.com/tp2vis/distribute}. For line imaging, we initially subtracted the continuum emission by fitting a first-order polynomial to the continuum in the uv-plane. After subtracting continuum emission determined at the emission-free channels, we then combined the 12-m, 7-m, and TP data in u-v space using the CASA task \textit{concat} and deconvolved them jointly. We determined weights of the visibility data using the task \textit{statwt}. Unfortunately, the  sensitivity of the extended configuration C43-5 from Cycle 5 observations is not optimal for the combination. We attempted to combine the 12-m, 7-m, and TP data by concatenating the visibility data together using different weights, however, the resulting rms becomes higher when including the C43-5 array.  We opted to not include these observations in the final products. Nevertheless, for completeness, we list in Table \ref{Table:Observations} the cycle 5 C43-5 array observations.

After concatenating the visibility data, the reduction process was done using the TCLEAN task with a multi-scaled deconvolver of 0, 20, 60 for cycle 5 and scales of 0, 6, and 20 for Cycle 6, which applies an iterative procedure with a decreasing threshold parameter to automatically mask regions during the cleaning process \citep[e.g.][]{Kepley2020}. To achieve a good balance between sensitivity and angular resolution the Briggs weighting parameter R was set to 0.5 for the spectral line image cubes. For Cycle 5 data, the spectral data cubes were constructed on a 512 $\times$ 512 pixel grid with 0.05$^{"}$ pixel size, and with 0.18 km s$^{-1}$ velocity resolution. For Cycle 6 data, the spectral data cubes were constructed on a 512  $\times$ 512 pixel grid with 0.08$^{"}$ pixel size, and with 0.6 km s$^{-1}$ velocity resolution. The parameters of the final images for the two sets of observations are shown in Table \ref{Table:Imaging}.

\begin{figure*}
\centering

    \centering
    \includegraphics[width=0.302\textwidth]{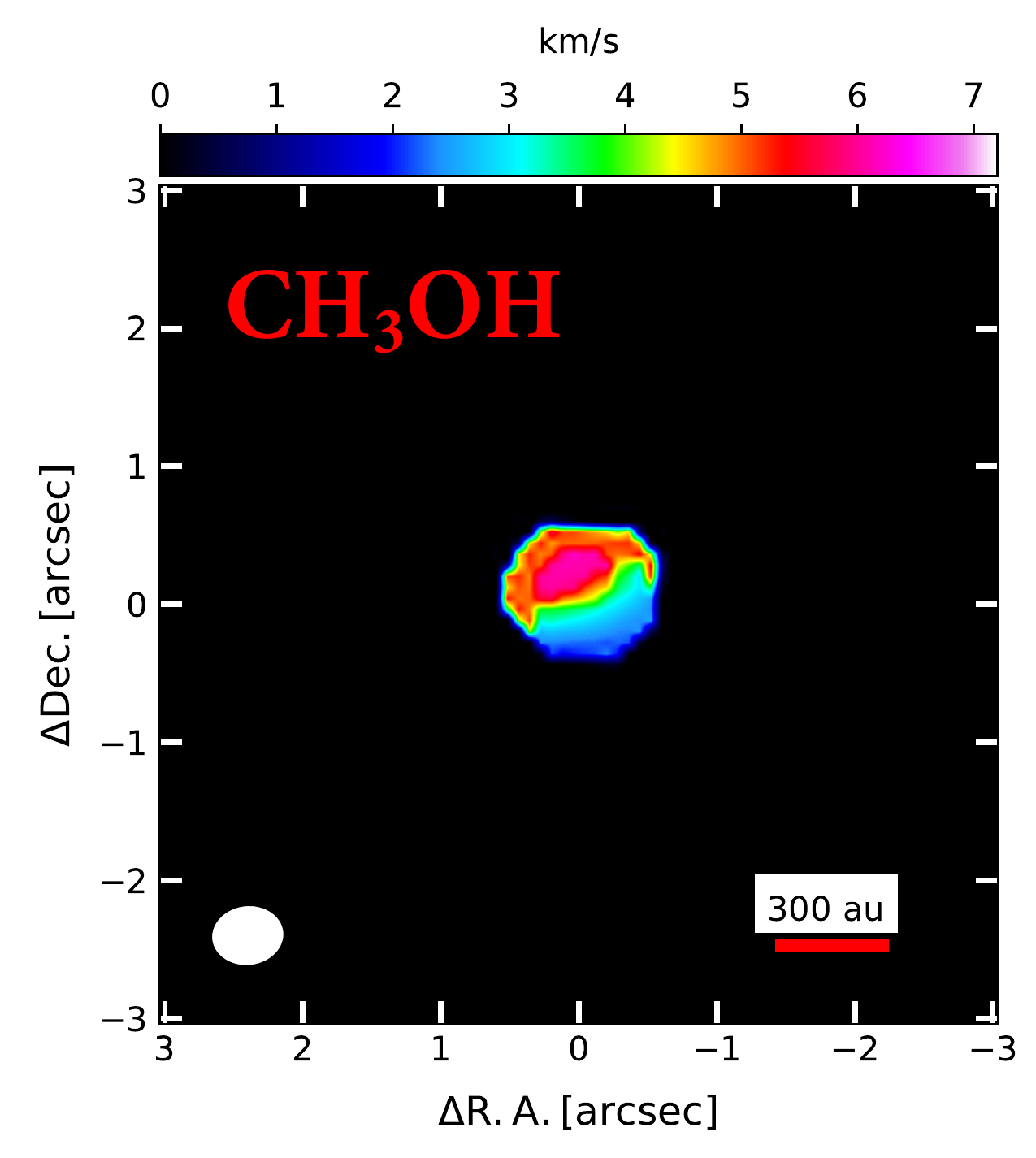}
    \includegraphics[width=0.305\textwidth]{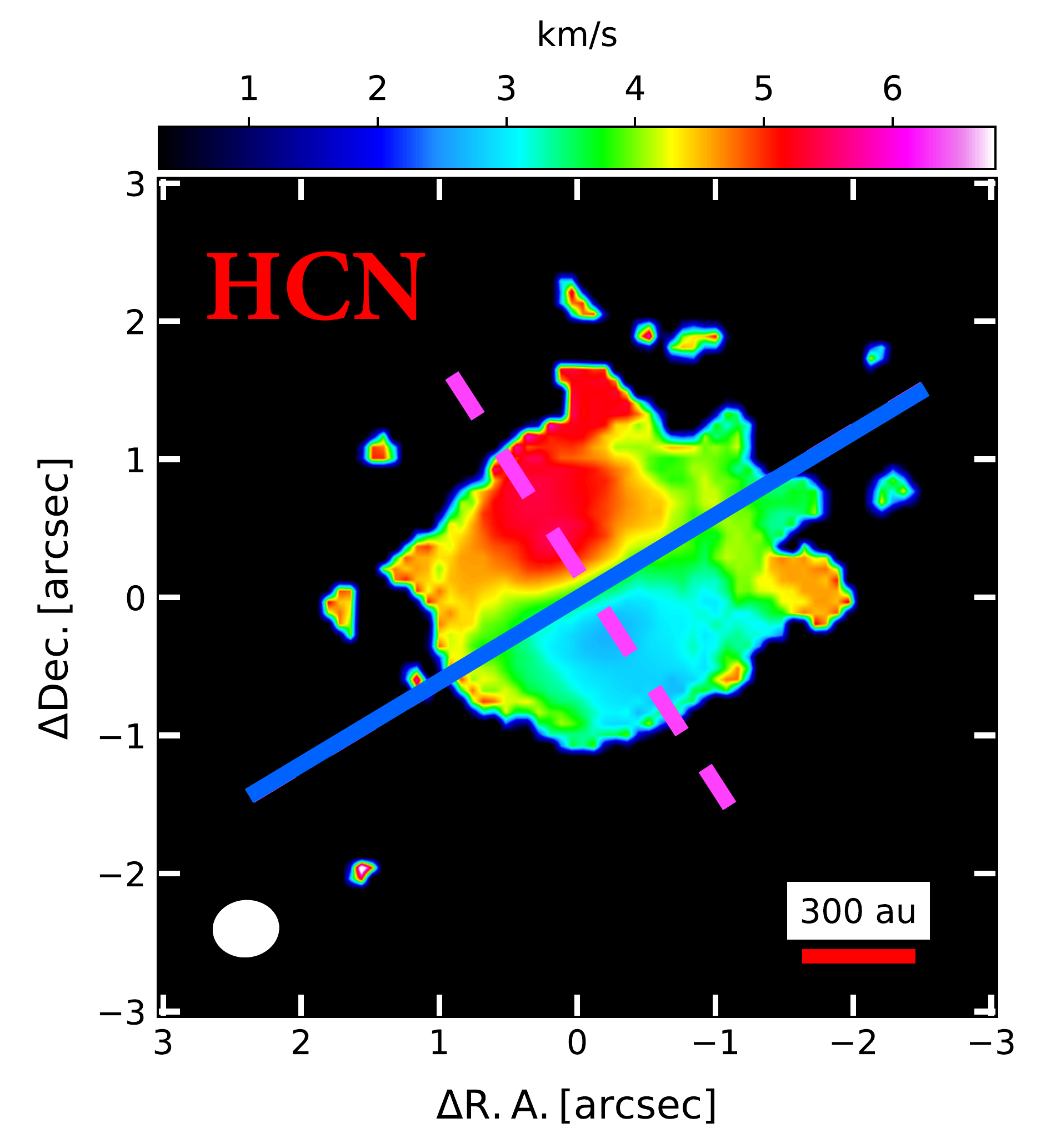}
     \includegraphics[width=0.295\textwidth]{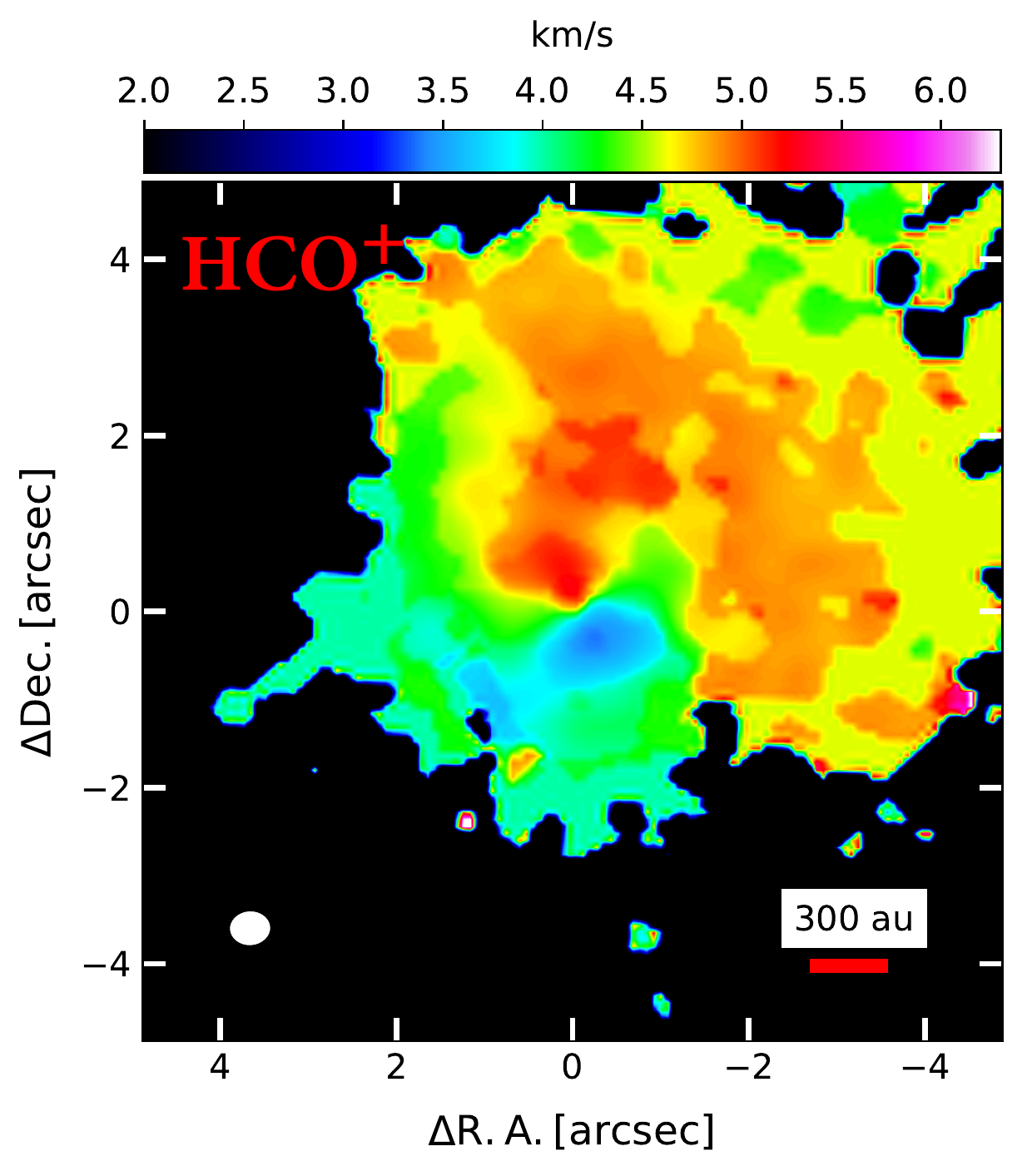}
    \includegraphics[width=0.3\textwidth]{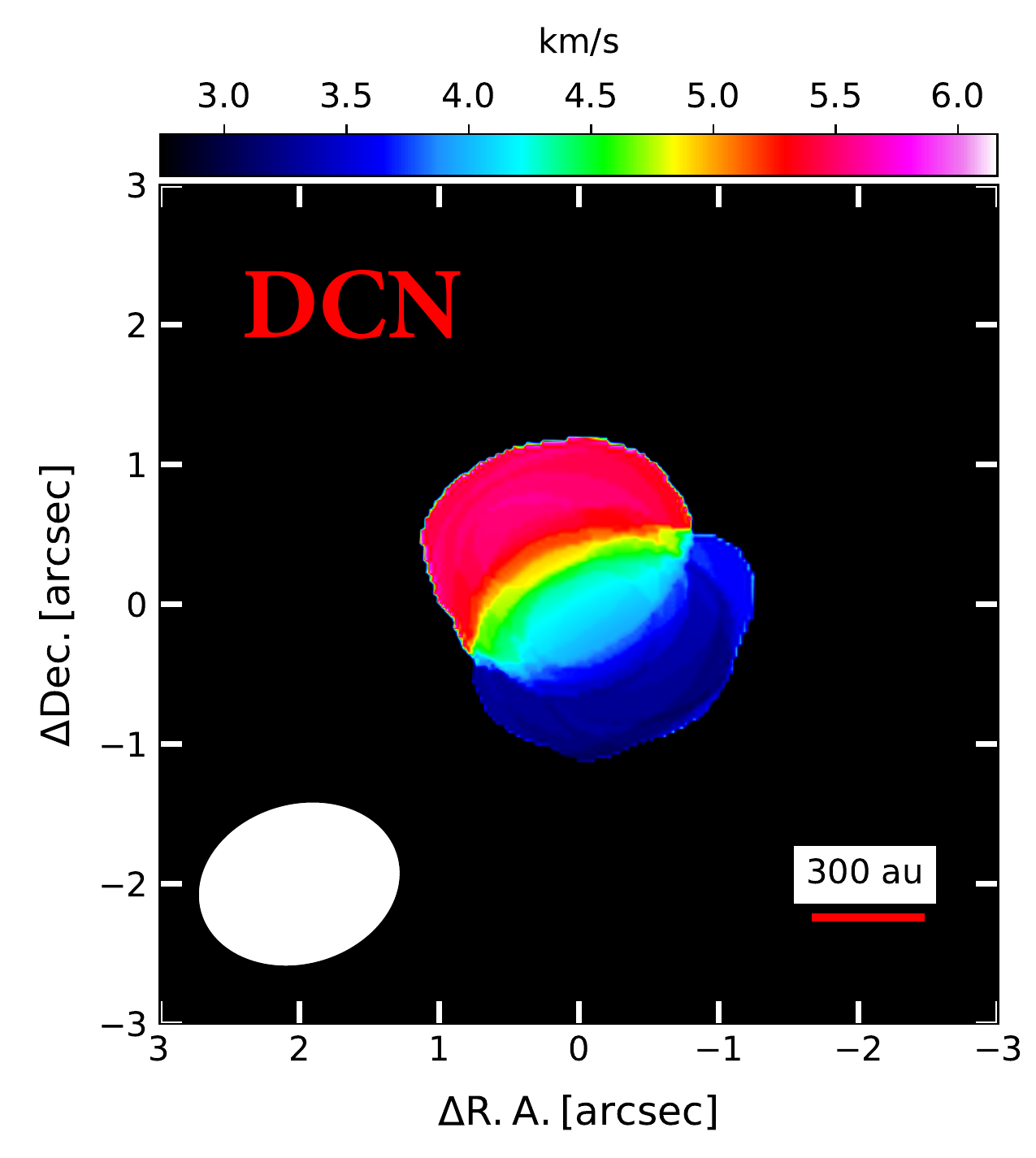}
    \includegraphics[width=0.3\textwidth]{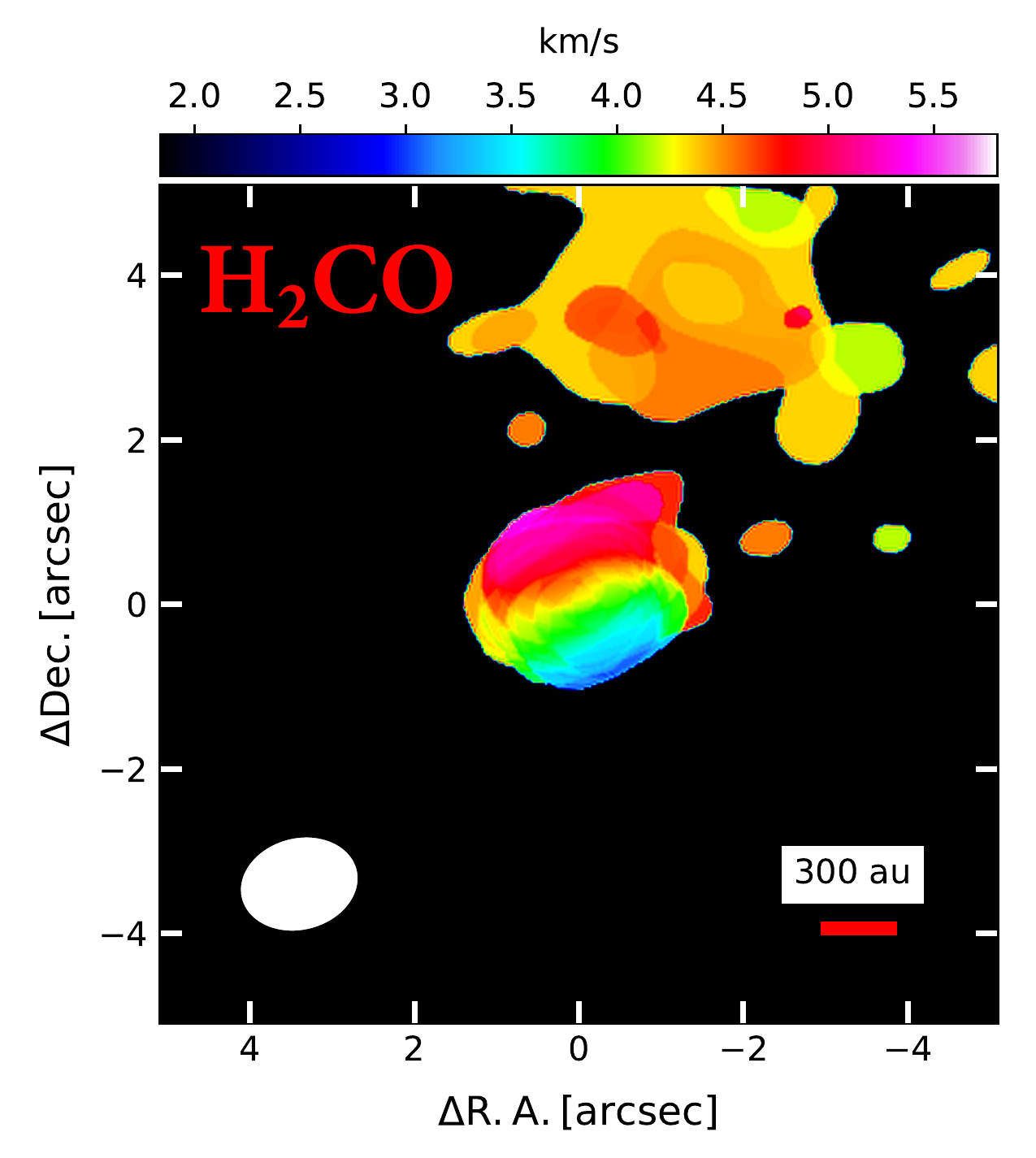}
     \includegraphics[width=0.3\textwidth]{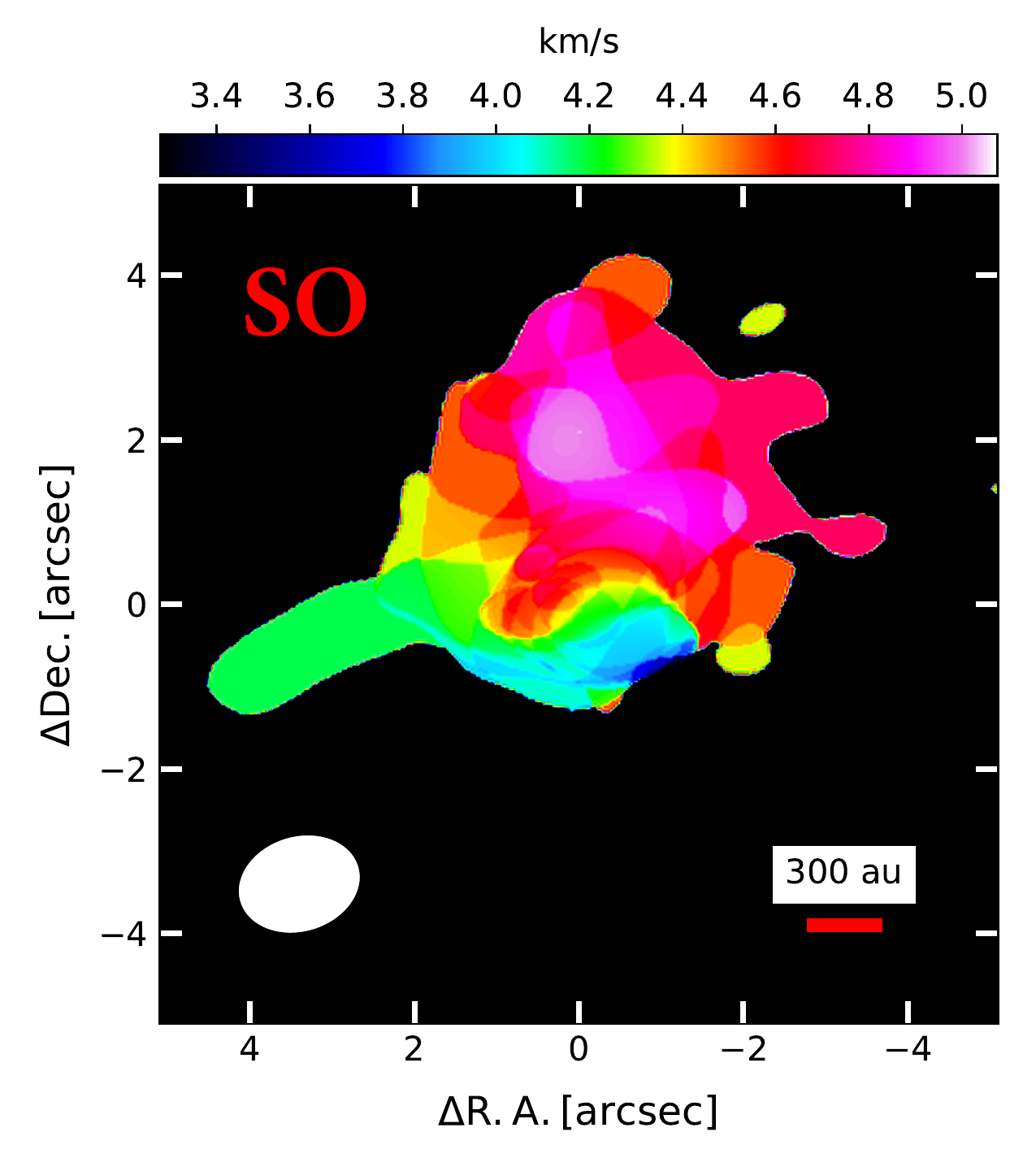}

\caption{V883 Ori intensity-weighted mean velocity maps of  CH$_{3}$OH (top-left panel), HCN (top-middle panel),  HCO$^{+}$ (top-right panel), DCN (bottom-left panel), H$_{2}$CO (bottom-middle panel), and  SO (bottom-right panel) lines. These moment 1 maps were created by clipping the data at 3$\sigma$ in each channel. The beam size is shown at the bottom left. These velocity maps were integrated as in Figure \ref{Fig:mom0}. The dashed magenta and solid blue lines at the top-middle panel (HCN emission) indicate the regions where the PV diagrams are extracted, i.e. P.A. of $\sim$ 30 $\rm ^{o}$ and $\sim$ 120 $\rm ^{o}$, respectively. }
\label{Fig:mom1}
\end{figure*}

\section{Results} \label{Sec:Results}

\subsection{Line detections and emission morphologies}
\label{Sec:RP}

\begin{table*}
\caption{Deconvolved radial extensions from ALMA detected lines in V883 Ori.}
\begin{tabular}{lccccccccc}
\hline Line & $\rm FWHM_{RP}$ & Beamwidth & Radial Extension\\
&  [au] & [au] & [au] & \\
\hline
& & ALMA cycle 6&\\
\hline
CH$_{3}$OH$^a$ &  607 & 560 & 131  $\pm$ 10\\
HCN$^a$ &  831 & 560 & 315  $\pm$ 20 \\
HCO$^{+}$$^a$ &  3238 & 560 & 1610  $\pm$ 100\\

\hline
& & ALMA cycle 5&\\
\hline
DCN &  880 & 558 & 340 $\pm$ 15 \\
H$_{2}$CO$^b$ & 846 & 560 & 324 $\pm$ 20 \\
SO & 2570& 580 &  1250 $\pm$ 100\\

\hline
\end{tabular}\\
\footnotesize{$^a$ Radial extension estimated from the degraded ALMA images, see Sec. \ref{Sec:RP}.}\\
\footnotesize{$^b$ Radial extension of the bulk H$_{2}$CO emission at $>$7$\sigma$. }
\label{Table:RP}
\end{table*}

We detected nine out of eleven of our target emission lines. Table \ref{Table:Imaging} shows a summary of the line detections and non-detections presented in this paper\footnote{We will report CO and its isotopologues from Cycle 5 observations in a coming paper (Ruiz-Rodriguez et al. In prep.) with a more detailed kinematic analysis.}. Figure \ref{Fig:mom0} displays the velocity-integrated intensity (moment-0) maps for CH$_{3}$OH, HCN, HCO$^{+}$, DCN, H$_{2}$CO, and SO emission. These moment maps are integrated over the velocity range -5.0 to 10 km s$^{-1}$, -0.2 to 8.8 km s$^{-1}$, 2.2 to 6.8 km s$^{-1}$, 2.2 to 6.52  km s$^{-1}$, 2.2 to 6.52 km s$^{-1}$, and 2.56 to 6.88 km s$^{-1}$ for HCN, CH$_{3}$OH, HCO$^{+}$, DCN, H$_{2}$CO, and SO, respectively, using thresholds according to their rms values (see Table \ref{Table:Imaging}). The velocities are with respect to the local standard of rest (LSR). From our observations, the integrated CH$_{3}$OH, HCN, DCN, H$_{2}$CO, and SO emission are centrally peaked on the V883 Ori position at 0.47 Jy beam$^{-1}$ km s$^{-1}$, 0.97 Jy beam$^{-1}$ km s$^{-1}$, 1.00 Jy beam$^{-1}$ km s$^{-1}$, 1.1 Jy beam$^{-1}$ km s$^{-1}$, and 0.80 Jy beam$^{-1}$ km s$^{-1}$, respectively. In contrast, the integrated HCO$^{+}$ emission is off center, revealing a ring-like emission structure that is discussed in Section \ref{Sec:Discussion}.

\begin{figure*}
\centering
    \centering
    \includegraphics[width=0.38\textwidth, height=0.38\textwidth]{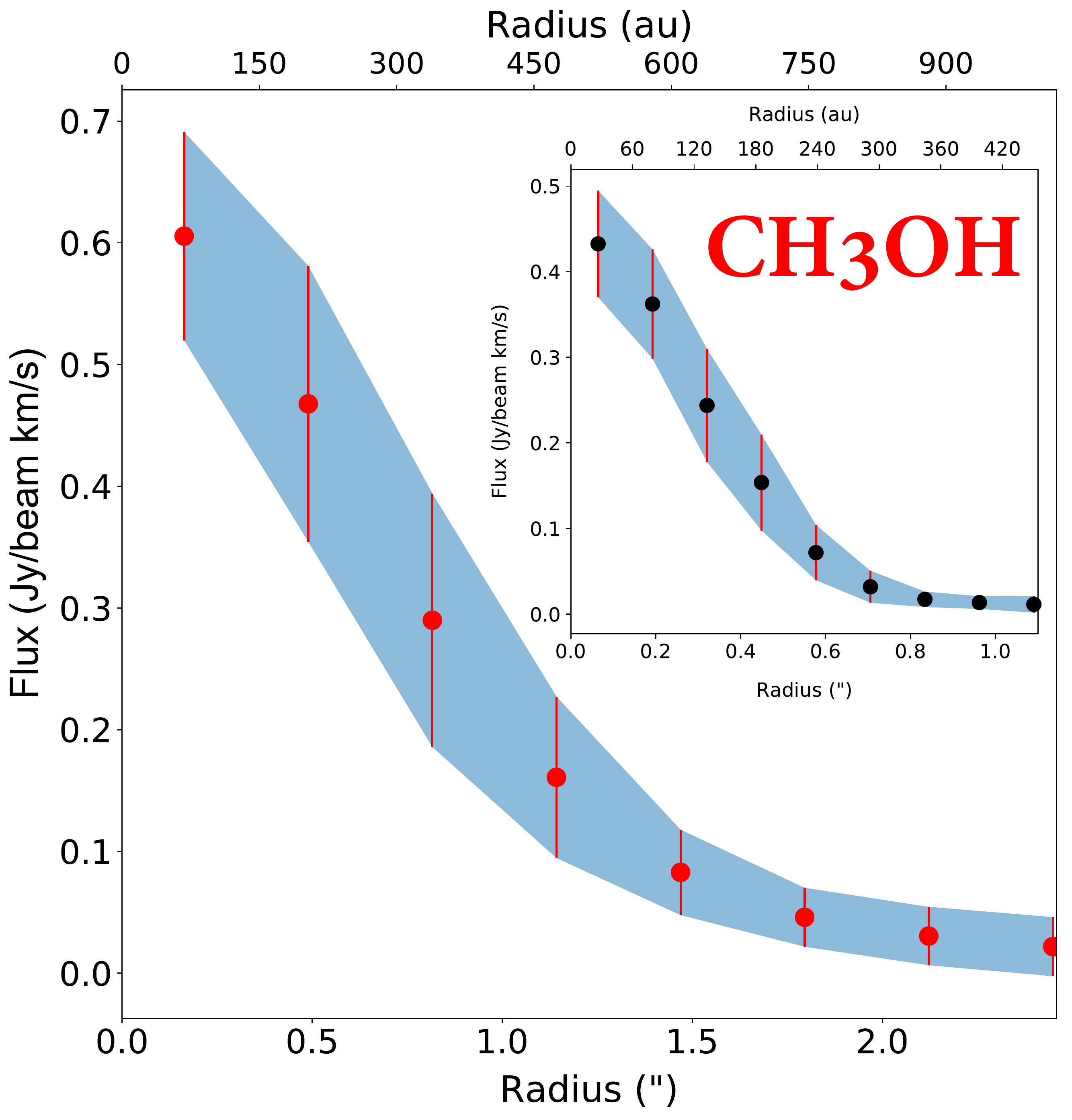}
     \includegraphics[width=0.38\textwidth, height=0.38\textwidth]{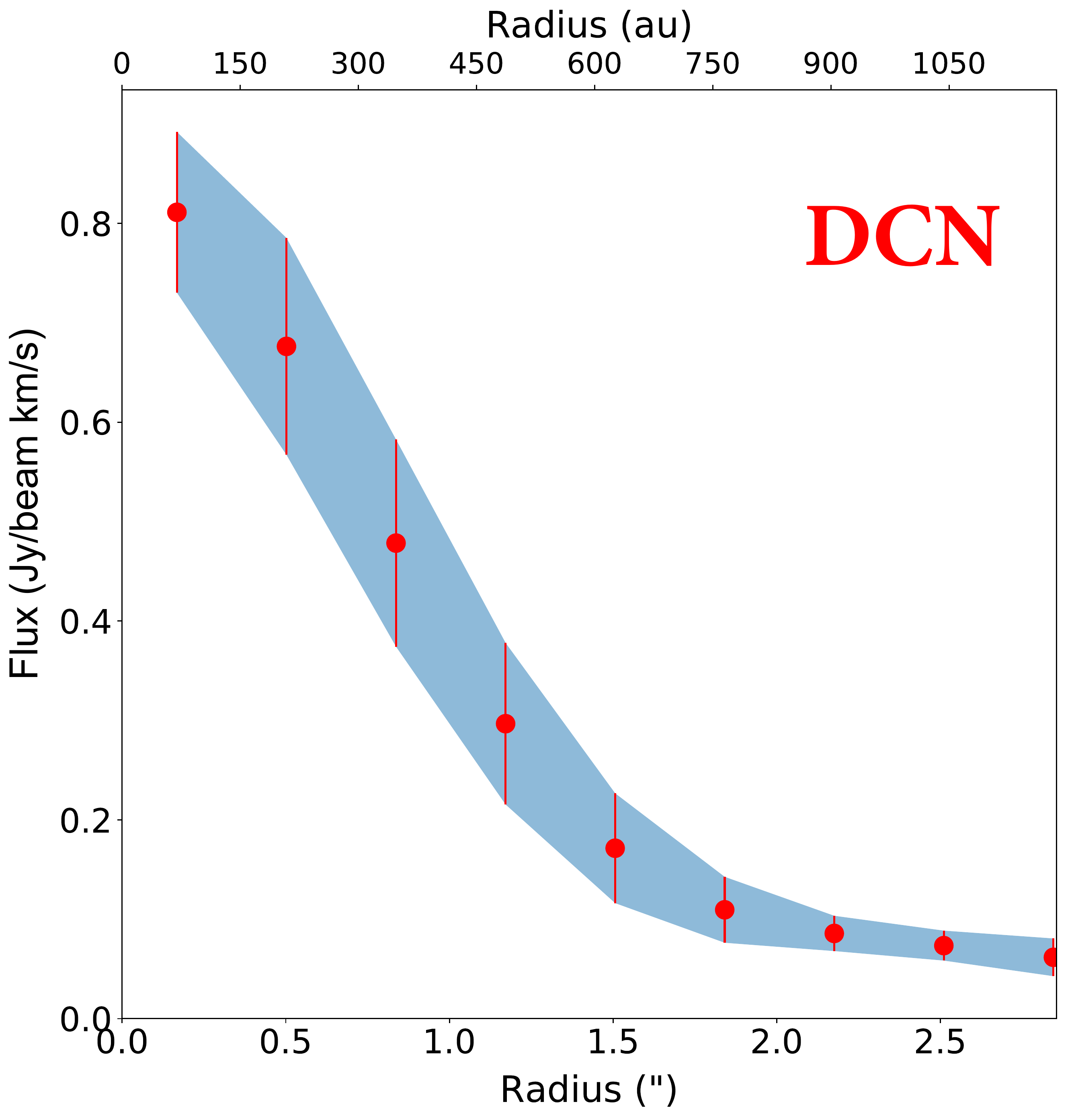}
     
     \includegraphics[width=0.38\textwidth, height=0.38\textwidth]{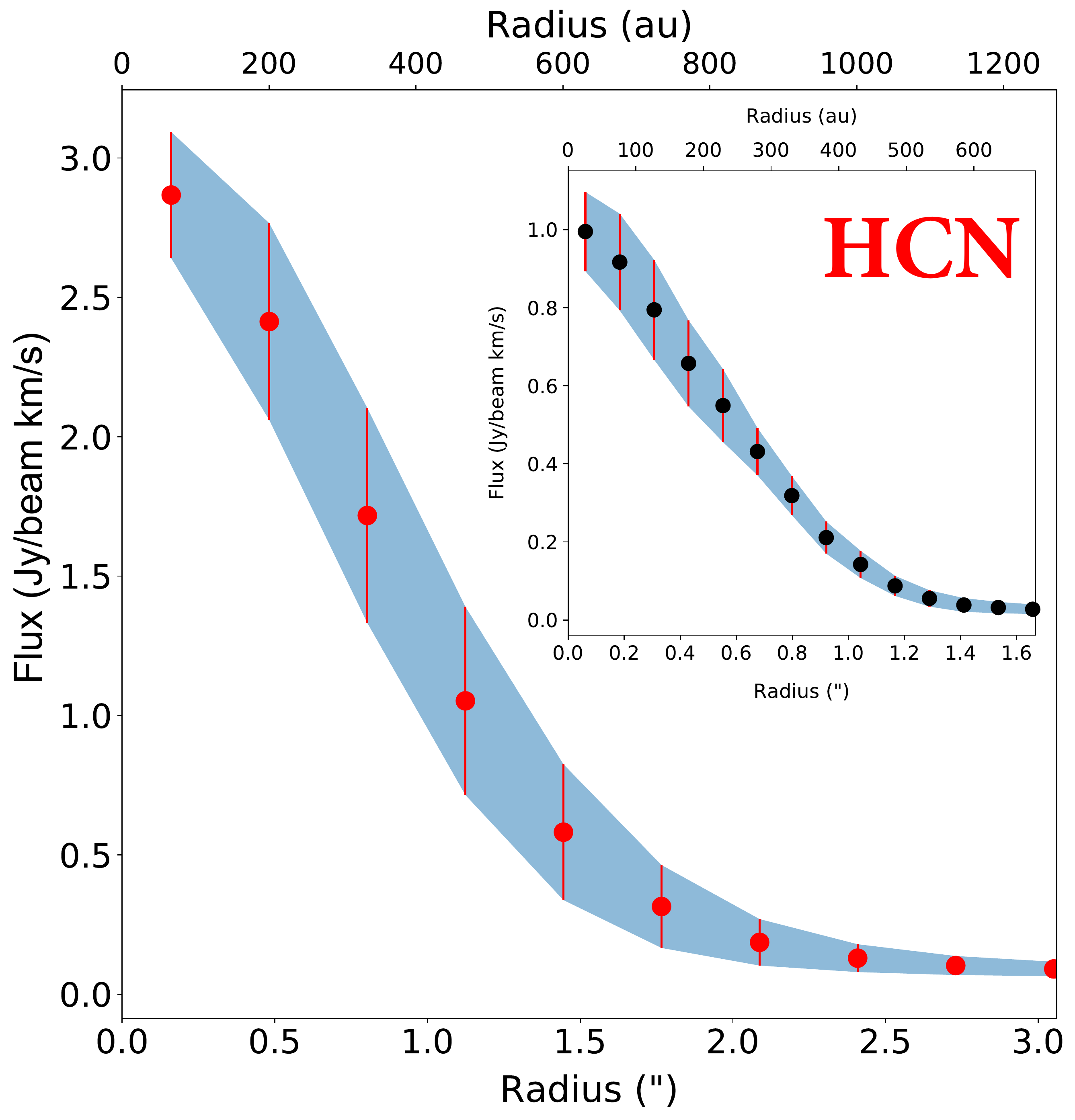}
      \includegraphics[width=0.38\textwidth, height=0.38\textwidth]{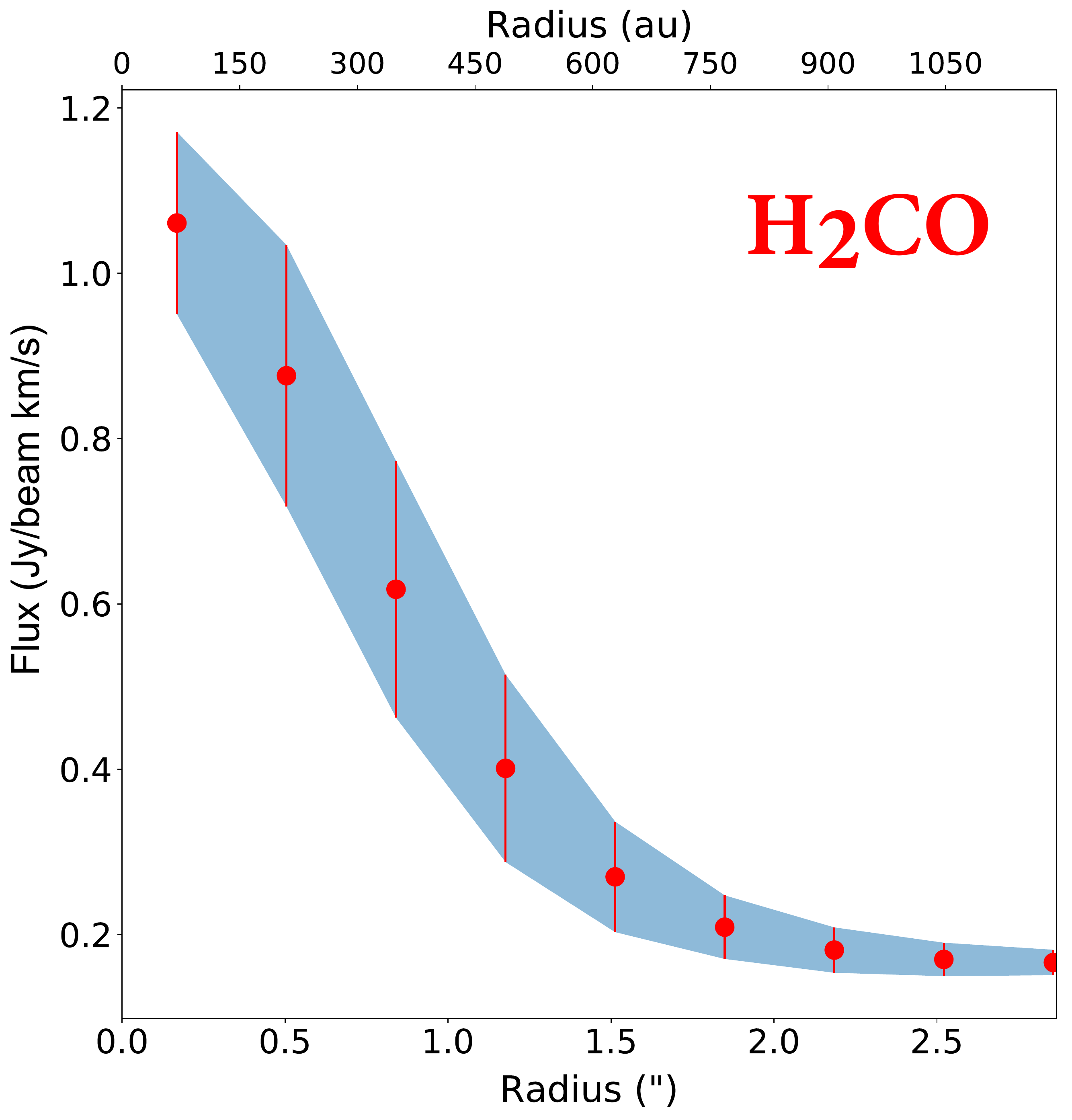}
      
    \includegraphics[width=0.38\textwidth, height=0.38\textwidth]{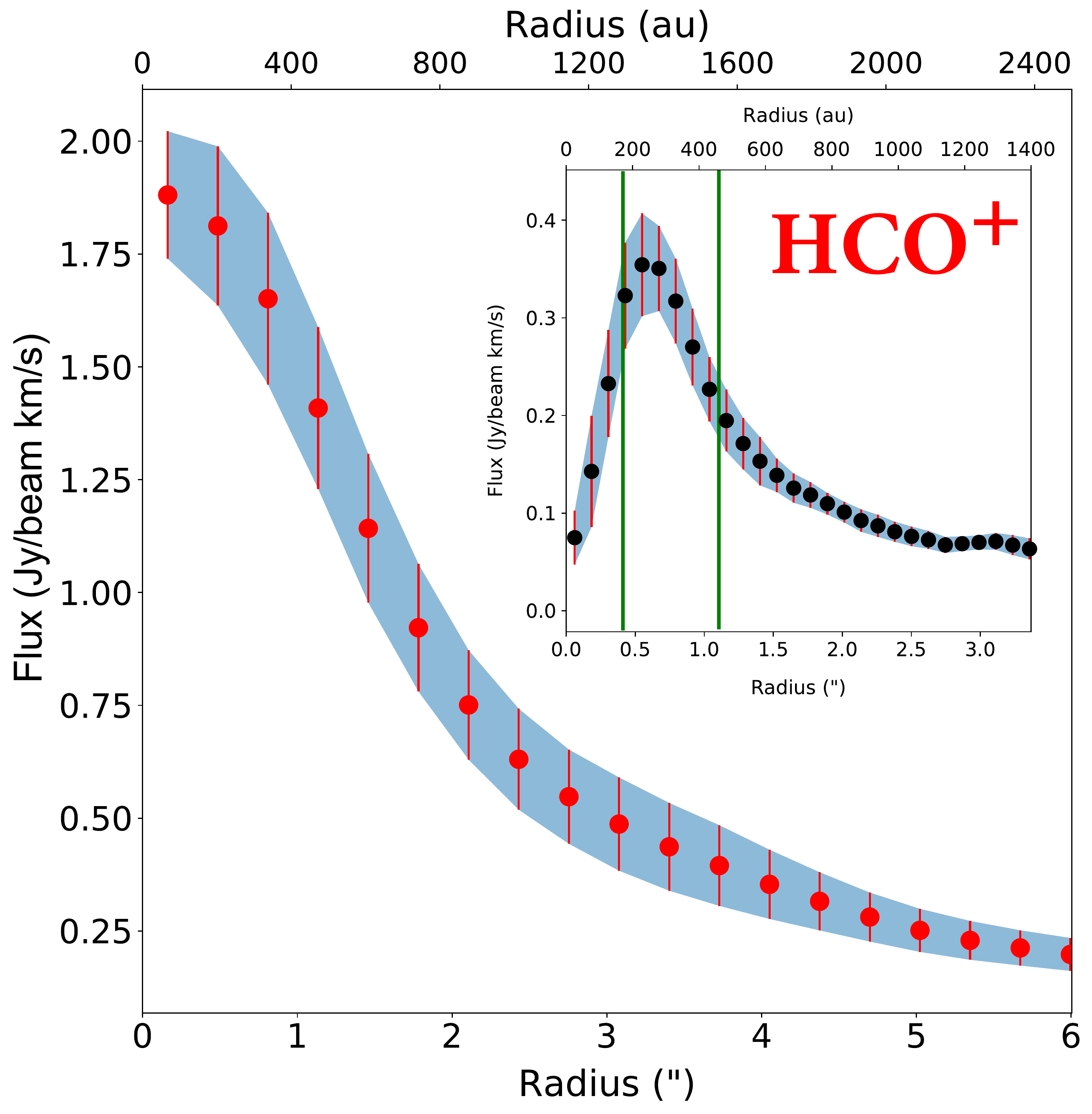} 
    \includegraphics[width=0.38\textwidth, height=0.38\textwidth]{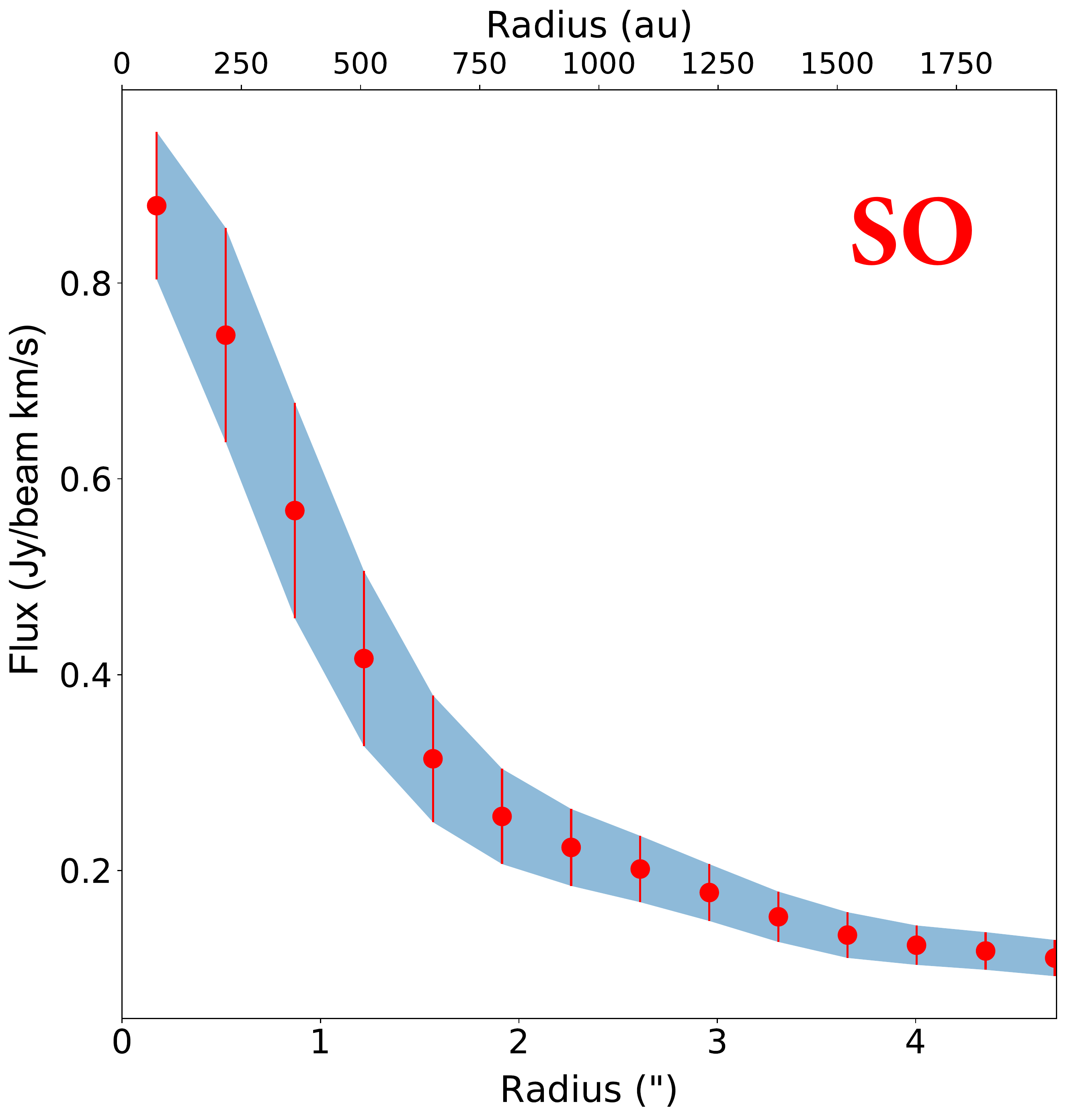}
    
\caption[short]{Azimuthally averaged and de-projected radial intensity profiles for ALMA cycle 6 (left panels), and cycle 5 (right panels) observations. The blue-shaded area represents the 1$\sigma_{\rm bin}$ error on the flux mean at a given radius, while the red error bars show the uncertainty on the flux mean values for each bin, both divided by the square root of beams. The CH$_{3}$OH (top-left panel), HCN (left-middle panel), HCO$^{+}$ (left-bottom panel) radial profiles displayed in the insets are extracted from the $\sim$0.5$^{"}$ resolution images, while radial parameters shown in Table \ref{Table:RP} are estimated from the degraded ALMA images of $\sim$1.5$^{"}$. Vertical solid green lines represent the intensity break and inner edge of the ring-like structure observed in HCO$^{+}$. More details in Sec. \ref{Sec:RP}. }
\label{Fig:RP}
\end{figure*}

The relative spatial distributions of the line emissions close to the central source can be parameterized by extracting azimuthally averaged radial profiles from all observed integrated intensities. These profiles are derived by computing mean values of pixels in radial bins of one-quarter of a Beam size (see Table \ref{Table:Imaging}) over the full azimuthal range of 2$\pi$. The extracted radial intensity profiles are azimuthally averaged after deprojecting for the V883 Ori disk inclination\footnote{All position angles are specified north through east.}  ($\it{i}$ $=$ 38$\rm ^{o}$ and P.A. $=$120 $\rm ^{o}$, see \citet{Cieza2016} and \citet{Ruiz2017b}, respectively). To bring both data sets to comparable resolutions, we convolved the Cycle 6 data with the Cycle 5 ALMA Clean synthetic beam of size $\sim$1.5$^{"}$ $\times$ 1.1$^{"}$. These data are called the degraded ALMA images. The extracted azimuthally averaged radial profiles are displayed in Fig. \ref{Fig:RP}. For our purposes, we used the full widths at half maximum ($\rm FWHM_{RP}$) of the profiles to derive deconvolved sizes. To that end, we have estimated $\rm FWHM_{RP}$ as the region enclosing $\sim$70$\%$ of the flux and considering only emission above 3$\sigma$, where $\sigma$ is the rms noise as determined from the nearby line-free channels. Thus, the radial extent of the structure dominating the emission can be estimated through deconvolution using the formula:

\begin{equation}
\rm{R_{ext}} = \frac{{\sqrt{FWHM_{RP}^{2} - FWHM_{Beam}^{2}}}}{2}
\label{Eq:FWHM}
\end{equation}

where the flux within $\rm FWHM_{RP}$ marks the radial extension, and $\rm FWHM_{Beam}$, the mean FWHM of the beam. The uncertainties for the radial flux profiles are estimated as the standard deviation of the pixel values in each bin ($\sigma_{\rm bin}$) at a given radius divided by the square root of the beam area. The resulting radial extensions are listed in Table \ref{Table:RP}.

As the ring-like structure disappears in the degraded HCO$^{+}$ emission image owing to the larger beam, we also used the sharp (resolved) break in the intensity profile detected in the original HCO$^{+}$ emission at $\sim$460 au and the inner edge of the HCO$^{+}$ ring-like structure at $\sim$170 au in order to estimate a deconvolved size of these features (see solid green lines in Fig. \ref{Fig:RP}). Based on eq. \ref{Eq:FWHM}, and with a half of a beam size of $\sim$102 au, we assume that $\rm FWHM_{RP}$ $=$ $\rm{R_{r}\sqrt{8ln2}}$, where $\rm R_{r}$ represents the position of these features in the HCO$^{+}$ intensity profile. We then estimate that the HCO$^{+}$ ring emission has an inner radius at $\sim$130 au and an outer radius at $\sim$450 au.

\

From these estimates, we note that the observed HCO$^{+}$ ring structure extends out to $\sim$450 au, well beyond the outer radius of dust emission \citep[$\sim$125 au; ][]{Cieza2016}. However, as the channel maps in Appendix \ref{Fig:HCOChannel} show, the HCO$^{+}$ emission appears more extended to the northern cavity \citep[see][]{Ruiz2017b}, tracing an outflow that reaches up to $\sim$ 1600 au with a narrow velocity range between $\sim$ 4.0 and 5.5 km s$^{-1}$ (see Fig. \ref{Fig:HCOChannel}). In contrast, HCN is detected mostly close to the central object within a radius of $\sim$ 320 au (see Table \ref{Table:RP} and Fig. \ref{Fig:HCNChannel}), and at what seems to be a transition layer between the disc and a starting point of the outflows (see Sec. \ref{Sec:Infall} ). The DCN emission line traces similar regions as the HCN emission (see Fig. \ref{Fig:DCNChannel}), rising from within a radius of $\sim$ 330 au (Table \ref{Table:RP}). Similarly to HCO$^{+}$, but to a smaller extent and slower, SO emission is detected around the source and traces an outflow with a very narrow velocity range (4-5 km s$^{-1}$; see Fig. \ref{Fig:SOChannel}), reaching distances of $\sim$ 1200 au. The H$_{2}$CO emission displays the most complex structure, with distinct core and extended components (Fig. \ref{Fig:H2COChannel}; see Sec \ref{Sec:mom1}); thus, probing material that belongs to the outflow-envelope structure with velocities close to the systemic velocity (i.e. 4.5 km s$^{-1}$) of $\sim$ 4-5 km s$^{-1}$ (see Fig. \ref{Fig:H2COChannel}) that may trace the inner part of the large envelope, where the rotation starts to dominate, see Fig. \ref{Fig:mom1}. However, the bulk of the H$_{2}$CO emission ($>$7$\sigma$) originates around the central protostar (Fig. \ref{Fig:mom0}) where the greatest velocity dispersion is found within a radius of  325 au.

Using the CASA task $\rm IMFIT$, we estimated the total integrated fluxes and peak intensities by fitting two-dimensional Gaussians to the moment zero maps. Integrated fluxes of 0.71 Jy km s$^{-1}$, 4.6 Jy km s$^{-1}$, 1.12 Jy km s$^{-1}$ 1.20 Jy km s$^{-1}$, and 1.10 Jy km s$^{-1}$ were estimated for CH$_{3}$OH, HCN, DCN, H$_{2}$CO, and SO within circular regions of radius 0.7$\rm ^{"}$, 1.3$\rm ^{"}$, 2.1$\rm ^{"}$, 2$\rm ^{"}$, and 2.8$\rm ^{"}$, respectively. For the HCO$^{+}$ emission ring, we used an annulus region centered at the V883 Ori position with r$_{1}$ = 0.2$\rm ^{"}$ and r$_{2}$ = 1.5$\rm ^{"}$, yielding a total flux of $\sim$ 6.5 Jy km s$^{-1}$. The derived values are summarized in Table \ref{Table:Imaging}.

\subsection{Velocity field}
\label{Sec:mom1}

Figure \ref{Fig:mom1} presents the intensity-weighted velocity (moment 1) maps for V883 Ori. The integration range for all detected lines are chosen to match the moment-0 maps (see Fig. \ref{Fig:mom0}) and display the kinematic structure of these molecular lines. When creating these maps, the data were clipped at 3$\sigma$ in each channel, except for H$_{2}$CO and SO, where we clipped at 3$\sigma$ and $\sim$7$\sigma$ in order to distinguish two distinct regions, as further presented below.

The moment 1 maps of HCN, HCO$^{+}$, CH$_{3}$OH, DCN, H$_{2}$CO, and SO in Fig. \ref{Fig:mom1} all reveal a velocity shift indicative of rotation about the central YSO. That is, each of these molecules displays a pattern of blue-shifted and red-shifted emission to the south-west and north-east, respectively, along P.A. $\sim$ 30$^{\rm o}$, which is approximately perpendicular to the outflow rotation axis (P.A.$\sim$ 120$^{\rm o}$), as determined from the CO emission \citep{Cieza2016, Ruiz2017b}. On the other hand, after clipping the data at 3$\sigma$, the red- and blue-shifted SO emission reveals outflow material at the north-west and south-east sides of the object tracing the bottom part of the cavities previously detected in CO emission \citep{Ruiz2017b}, 
while, when clipped at 7$\sigma$, the SO data show signs of rotation close to the central star. Similarly, the H$_{2}$CO moment-1 map reveals two distinct regions of the FUor system with a strong velocity gradient on small scales, indicating rotational motion along the southwest to northeast direction of the disc; whereas, at larger distances, the H$_{2}$CO line also traces material that appears to trace the stellar envelope, given its radial velocities (4.0 and 5.0 km s$^{-1}$; Fig. \ref{Fig:H2COChannel}). 

\subsubsection{Position - Velocity Diagrams}
\label{Sec:PV}

\begin{figure*}
\centering
%\begin{subfigure}{1.0\textwidth}
    \centering
     \includegraphics[width=0.33\textwidth]{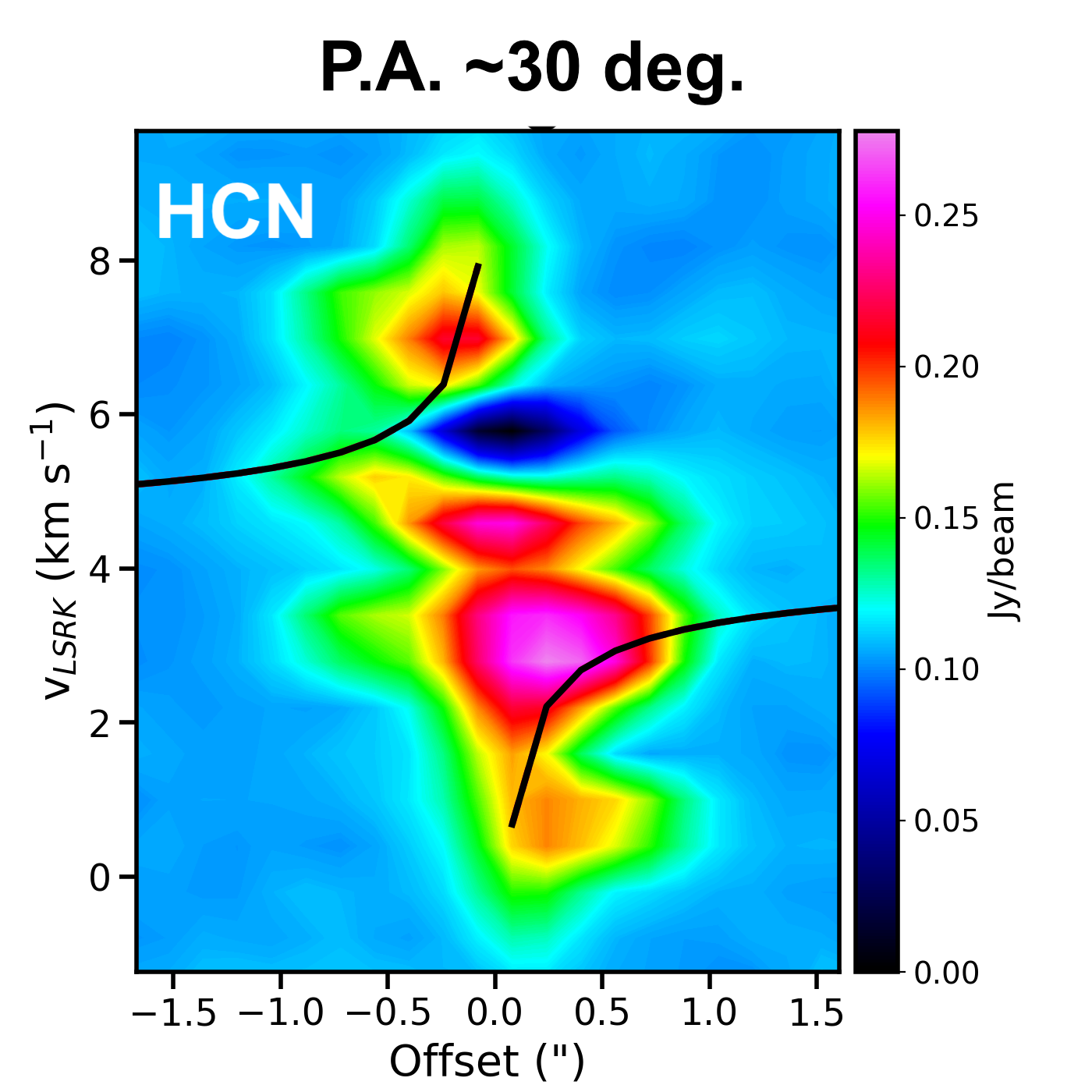}
      \includegraphics[width=0.33\textwidth]{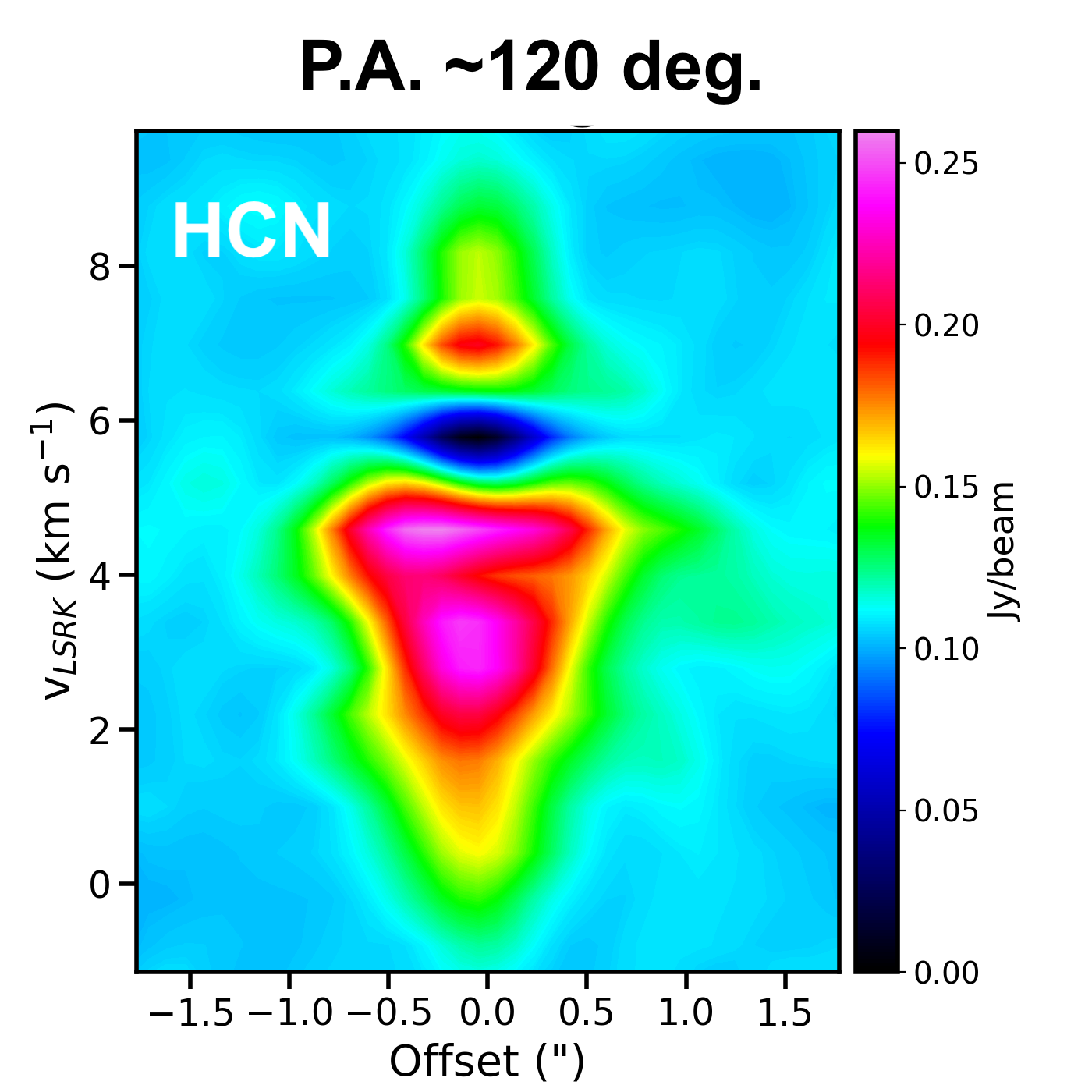}
      
    \includegraphics[width=0.33\textwidth]{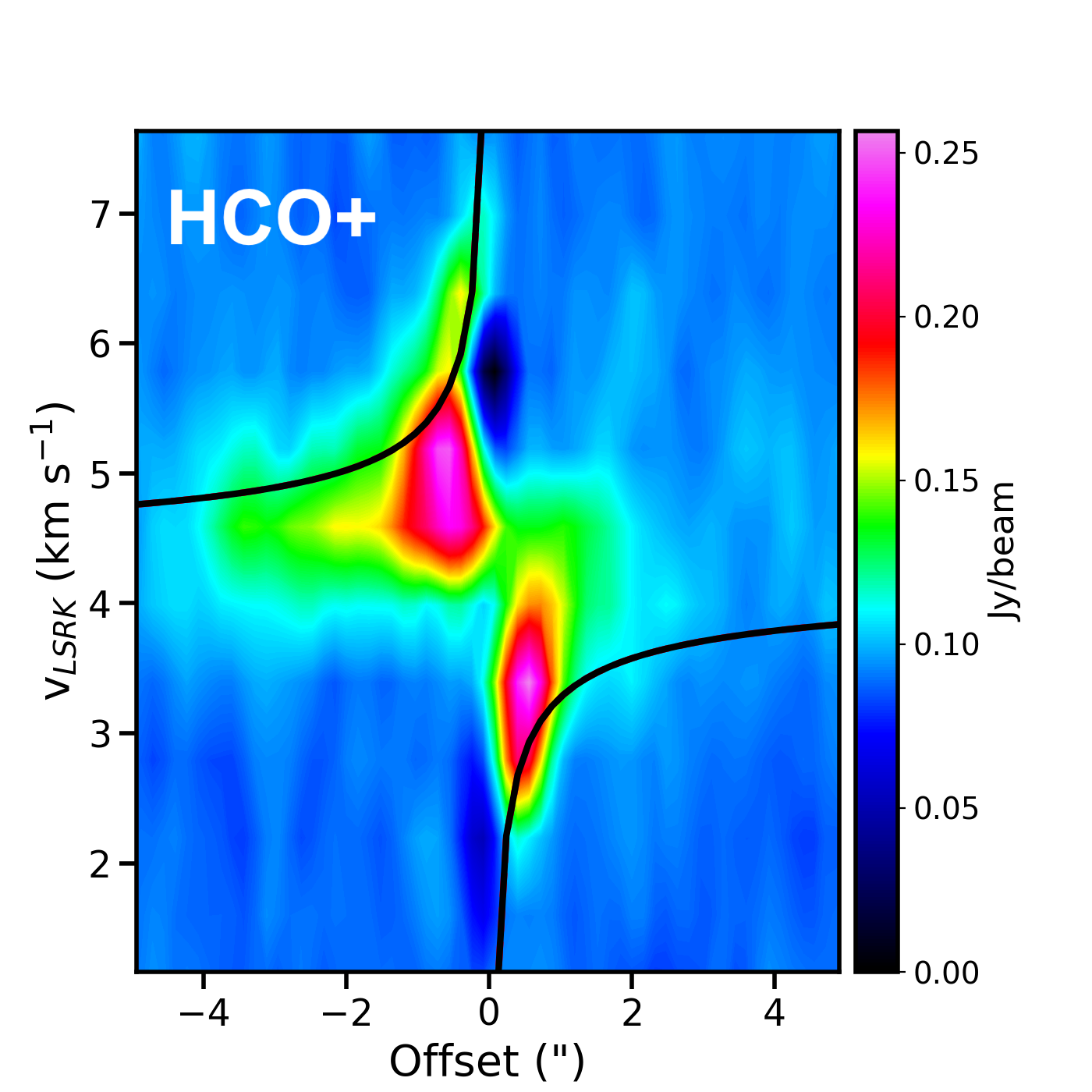}
     \includegraphics[width=0.33\textwidth]{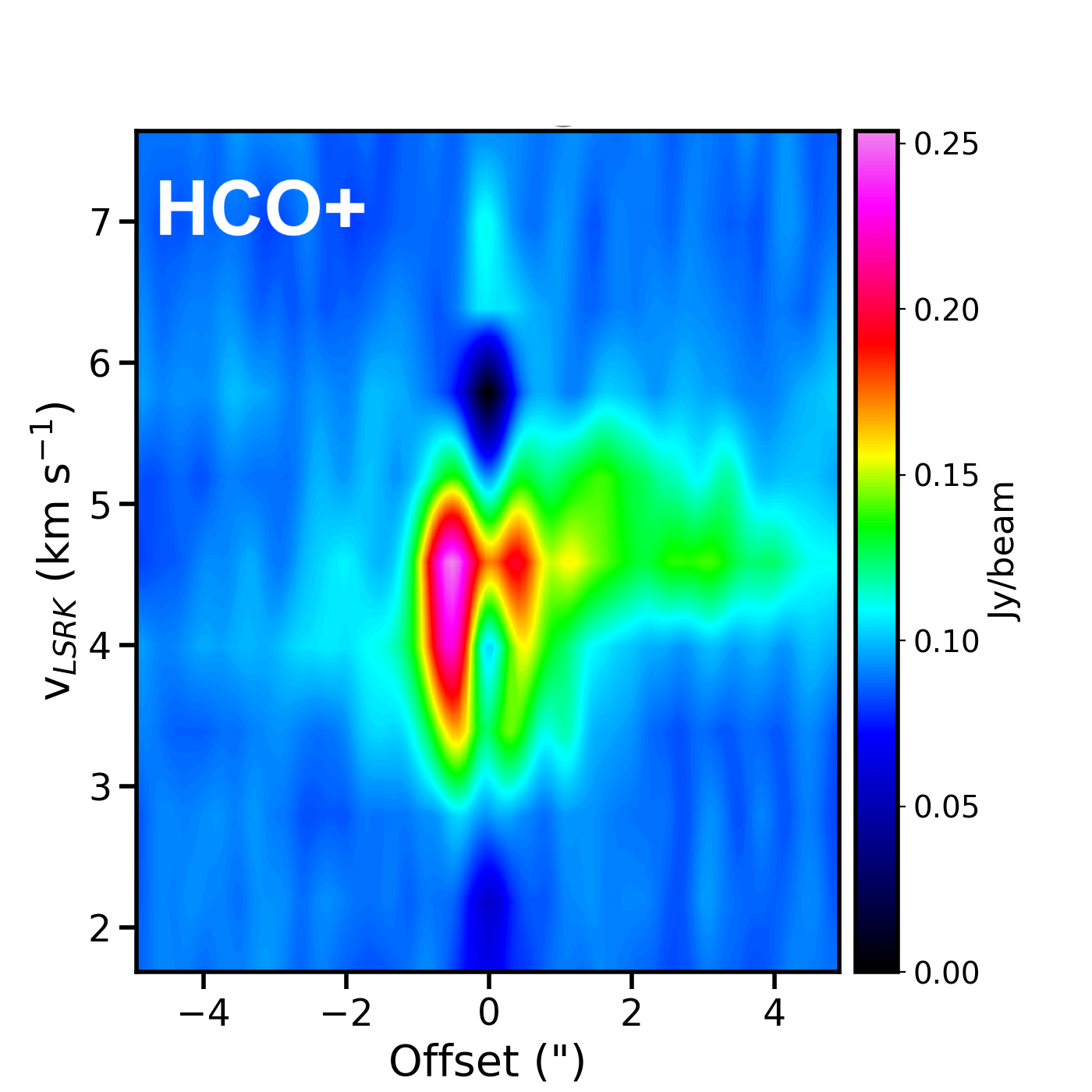}
   
    \includegraphics[width=0.33\textwidth]{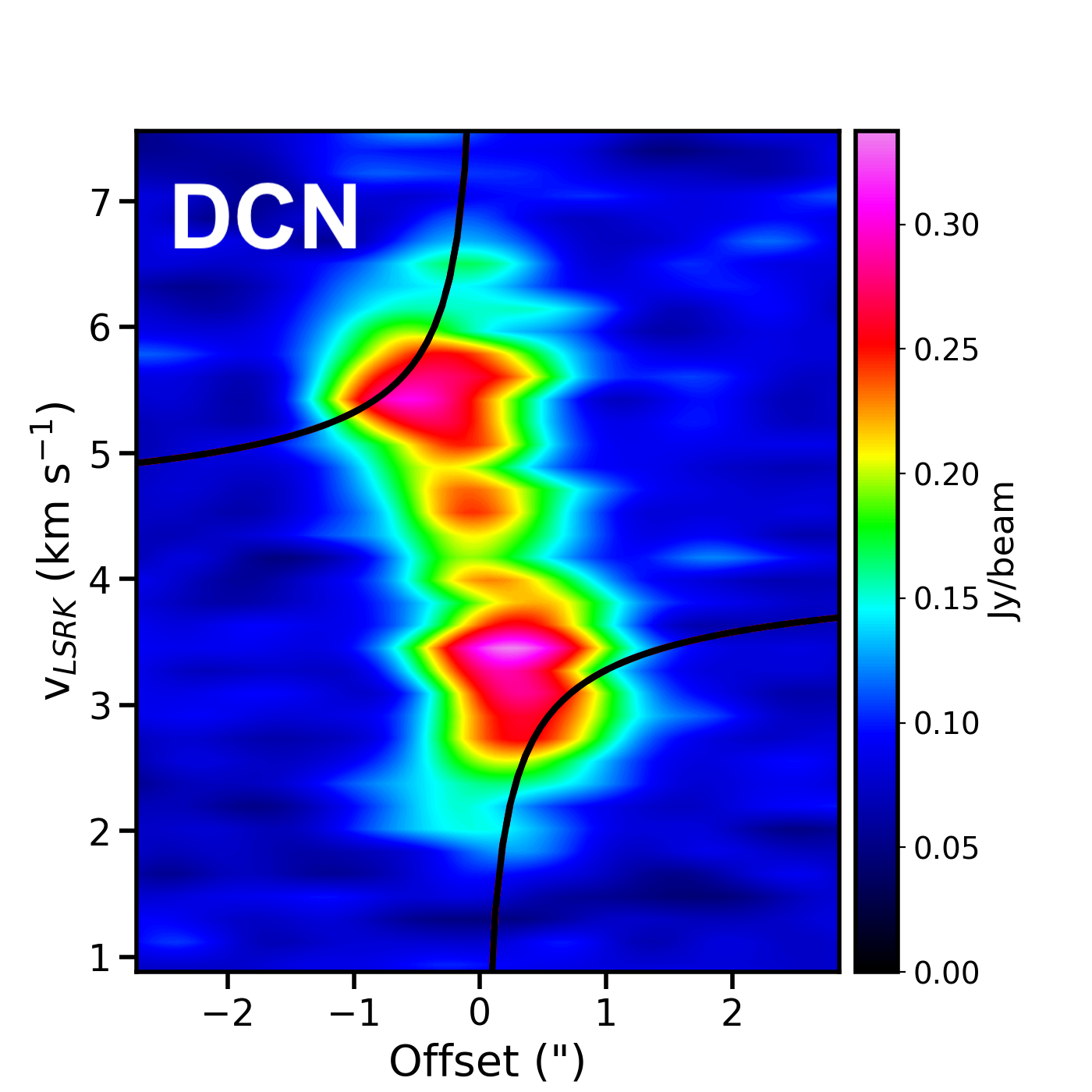}
     \includegraphics[width=0.33\textwidth]{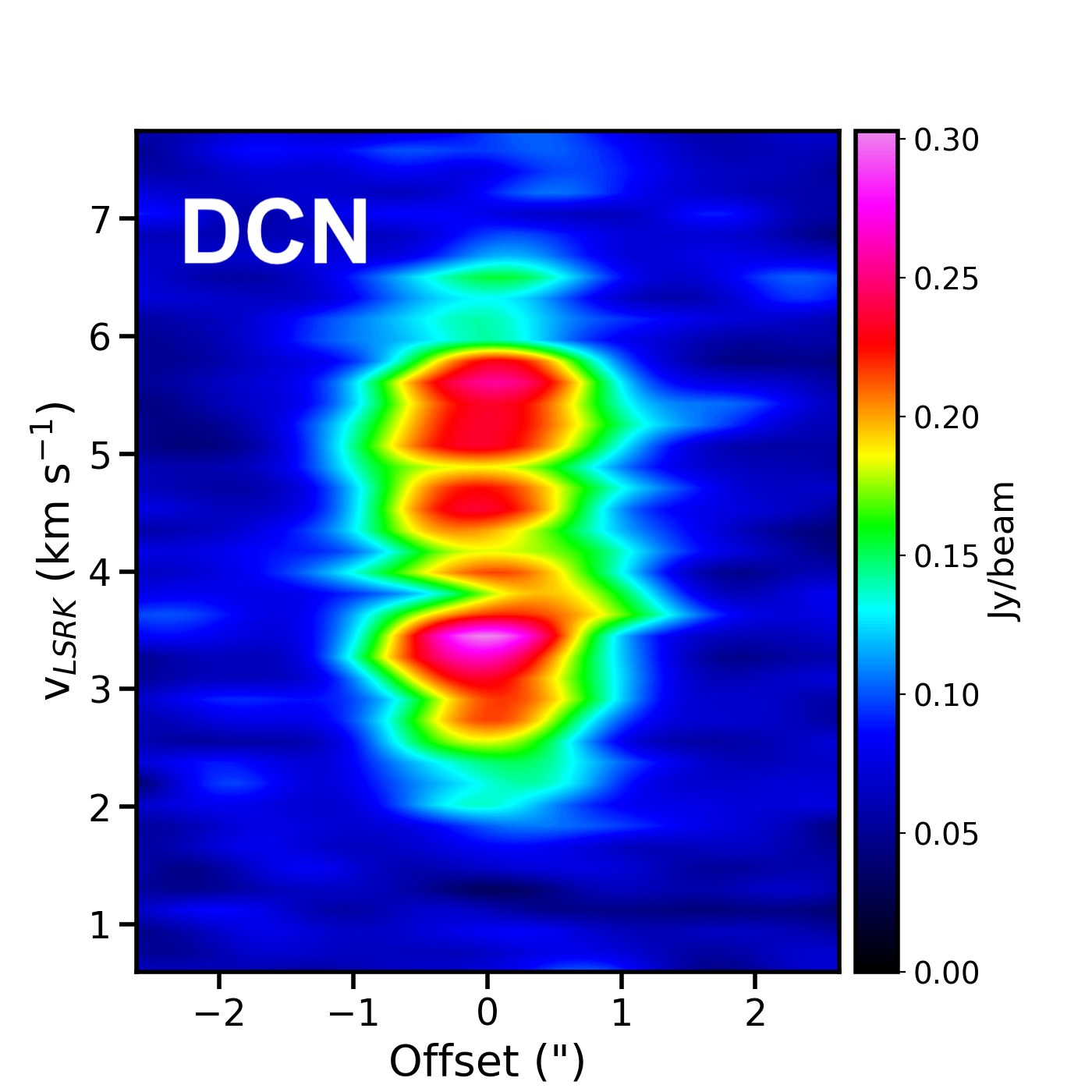}

     \caption[short]{Position-velocity diagrams extracted along the dashed lines shown in Figure \ref{Fig:mom0} for HCN, HCO$^{+}$, DCN, H$_{2}$CO, and SO. Left-Panels: PV diagrams along disc plane, i.e. P.A. 30 deg. The black curves represent the Keplerian motion around a central source of 1.3 M$_{\odot}$ \citep{Cieza2016}. Right-Panels: PV diagrams perpendicular to the disc plane, i.e. P.A. 120 deg.  These diagrams suggest infall motion perpendicular to the disc, in particular, HCN emission displays the known diamond-shaped feature, strongly indicating a gas infalling signature. }
  
\label{Fig:PV}
\end{figure*}

\begin{figure*}
    \ContinuedFloat
    \captionsetup{list=off}
    \centering
    
    \includegraphics[width=0.33\textwidth]{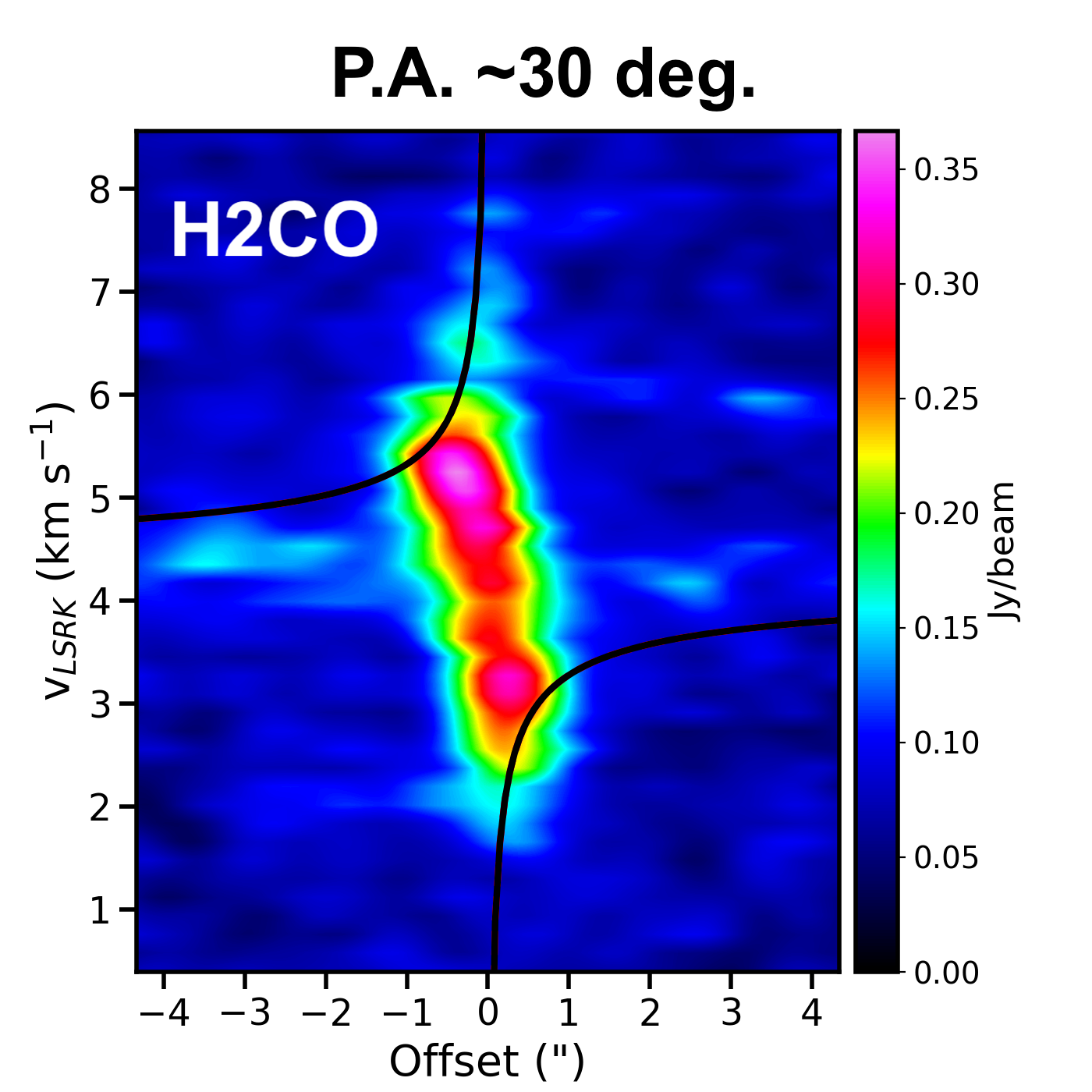}
      \includegraphics[width=0.33\textwidth]{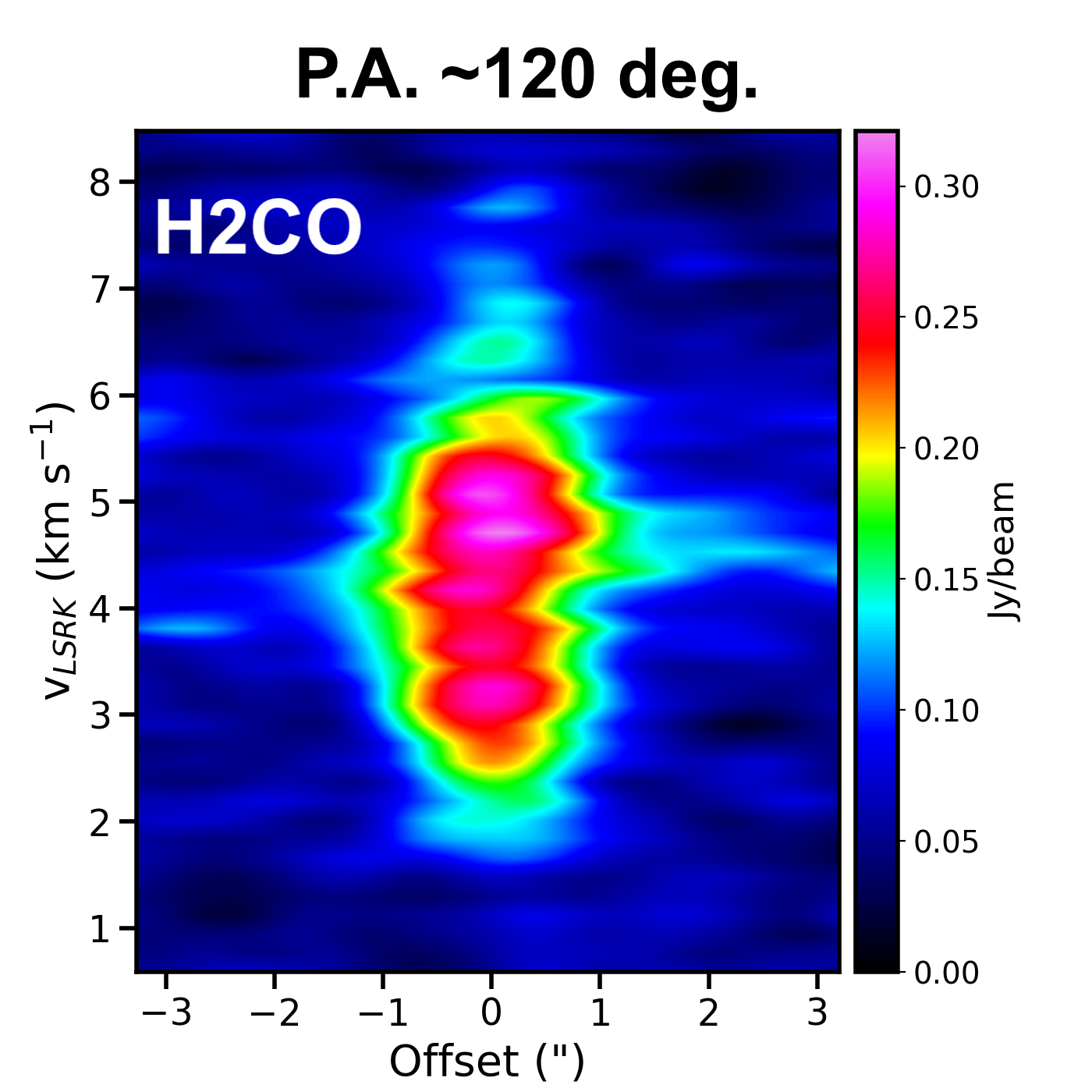} 
     
    \includegraphics[width=0.33\textwidth]{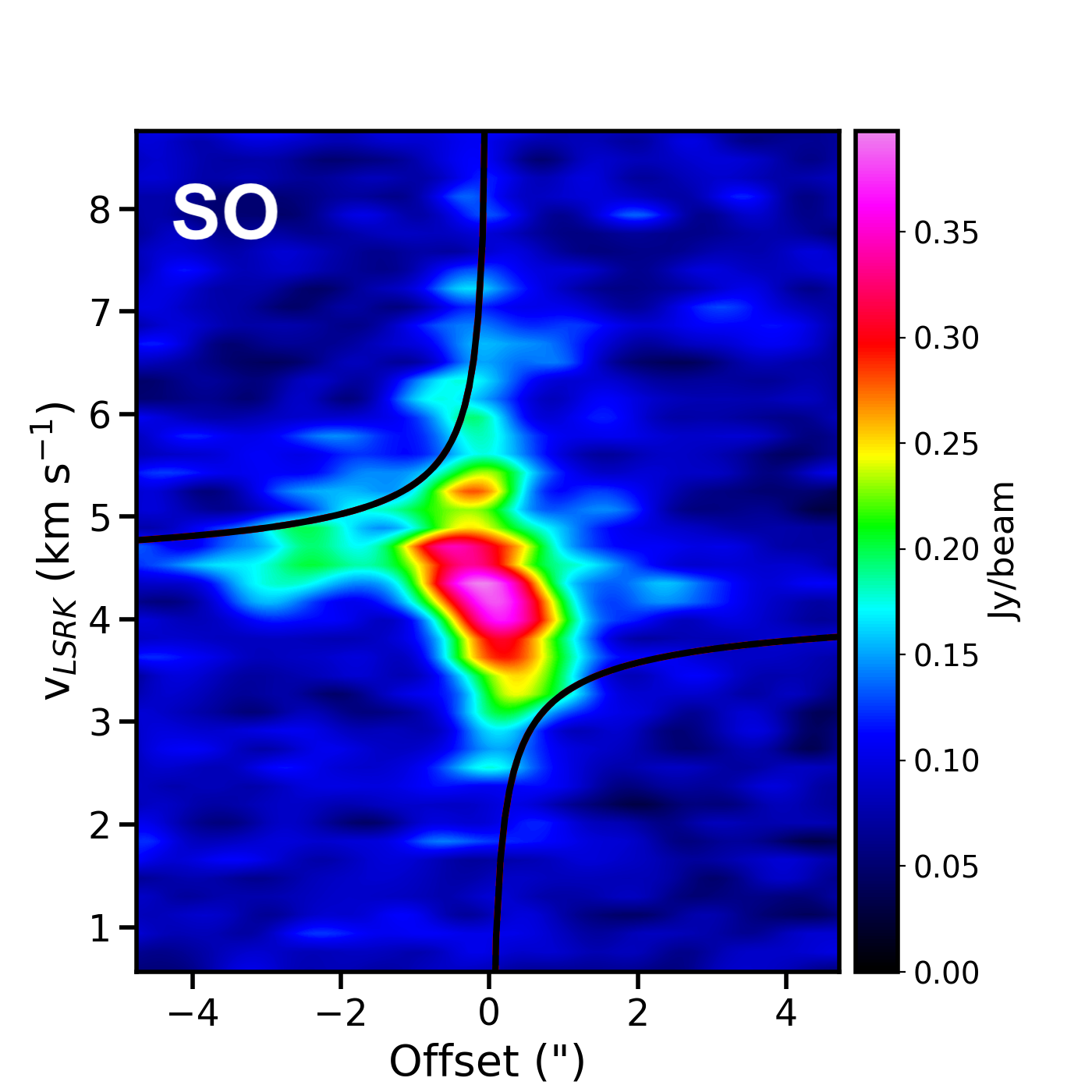}
    \includegraphics[width=0.33\textwidth]{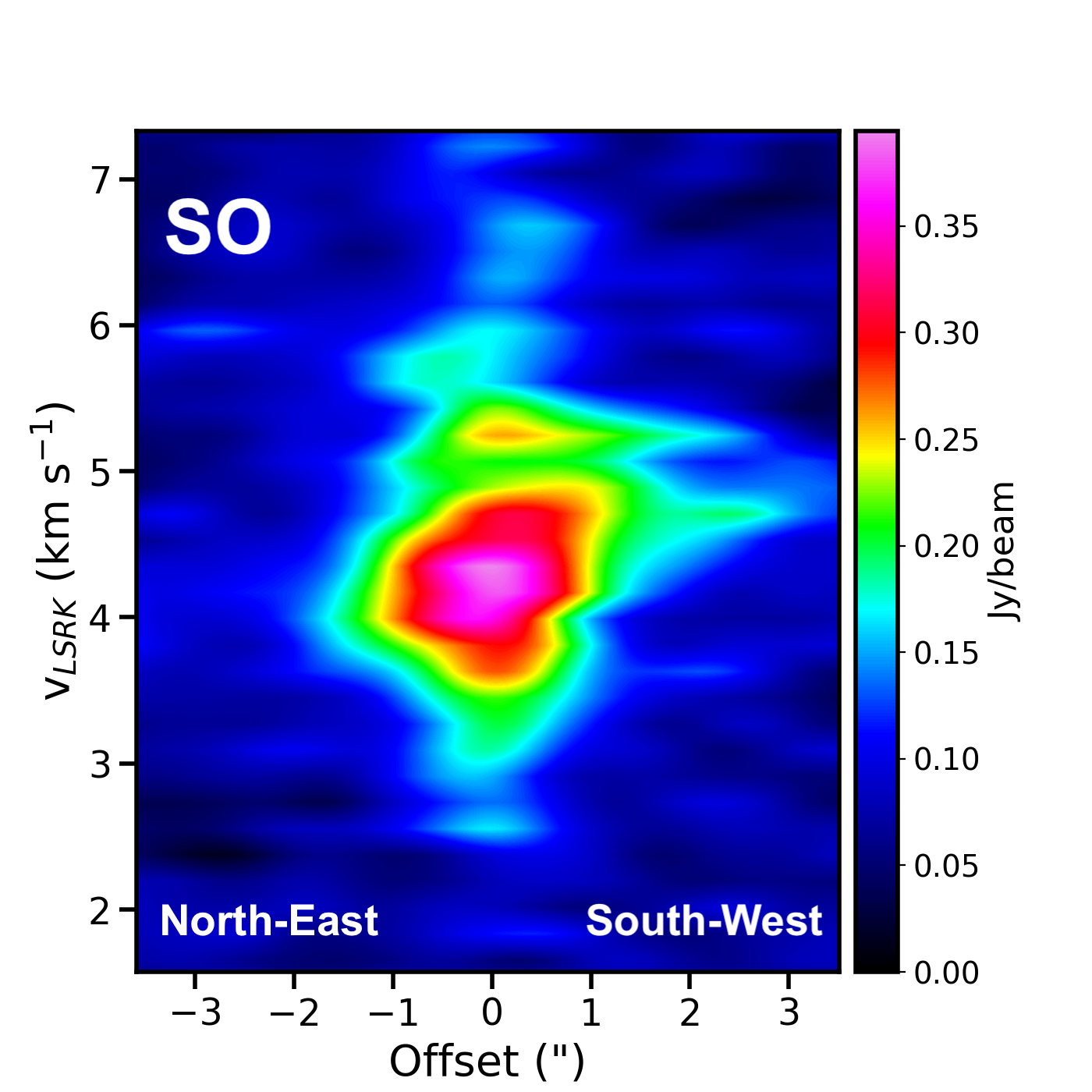} 
    \caption{Continued.}
  \end{figure*}

In order to disentangle outflowing and infalling gas from disc emission, we examine the velocity structures based on the position-velocity (PV) diagrams of the observed emission lines. To do so, we construct the PV diagrams along the horizontal direction through the centre of the disc ($\sim$ 30$^{\rm o}$), and perpendicular to it, i.e. rotational axis of $\sim$ 120$^{\rm o}$ \citep{Ruiz2017b}. The reference cuts are illustrated by magenta dashed and blue solid lines in Fig. \ref{Fig:mom1}, top-middle. Figure \ref{Fig:PV} shows PV diagrams of these emission lines for slices through the central source along and perpendicular to the outflow rotation axis\footnote{The slow outflow traced by CO emission is oriented along a P.A. of $\sim$ 120 $\rm ^{o}$ \citep{Ruiz2017b}}. We do not include a PV diagram for CH$_{3}$OH emission given the more compact nature of the disk structure traced by this molecule, thus preventing us from completely resolving any type of motion. Nevertheless, CH$_{3}$OH emission has been previously detected in V883 Ori at higher resolution data with a compact structure of radius $\sim$125 au and characterized by a Keplerian rotation motion around the central source \citep{Lee2019, vantHoff2018}.

The PV diagrams extracted along P.A.$\sim$ 30$^{\rm o}$ show blue-shifted and red-shifted components, suggestive of Keplerian rotation along the disc plane (Fig. \ref{Fig:PV}, left panels). In particular, the HCO$^{+}$, H$_{2}$CO, and SO PV diagrams show a nearly constant velocity field centered at the systemic velocity of $\sim$4.5 km s$^{-1}$, which extends up to $\sim$ 4$^{"}$ ($\sim$ 2000 au), and trace material that belongs to the surrounding envelope and material falling towards the central source. On the other hand, the PV diagrams of lines such as HCO$^{+}$ and HCN emission are indicative of infall motion along the rotational axis, i.e. P.A.$\sim$ 120$^{\rm o}$ (Fig. \ref{Fig:PV}, right panels). These emission PV diagrams present both red-shifted and blue-shifted emission at each side of the protostar position, resembling a ``diamond PV shape", which is characteristic of material infalling close to the central source \citep[e.g.][]{Tobin2012}.  The infalling motion of HCO$^{+}$ and HCN is supported by the strong red-shifted absorption features seen around 5.8 km s$^{-1}$ and across the disc, see Figures \ref{Fig:HCOChannel} and \ref{Fig:HCNChannel}.

\subsubsection{Infall motion}
\label{Sec:Infall}

\begin{figure}
\centering
    \centering
     \includegraphics[width=0.43\textwidth]{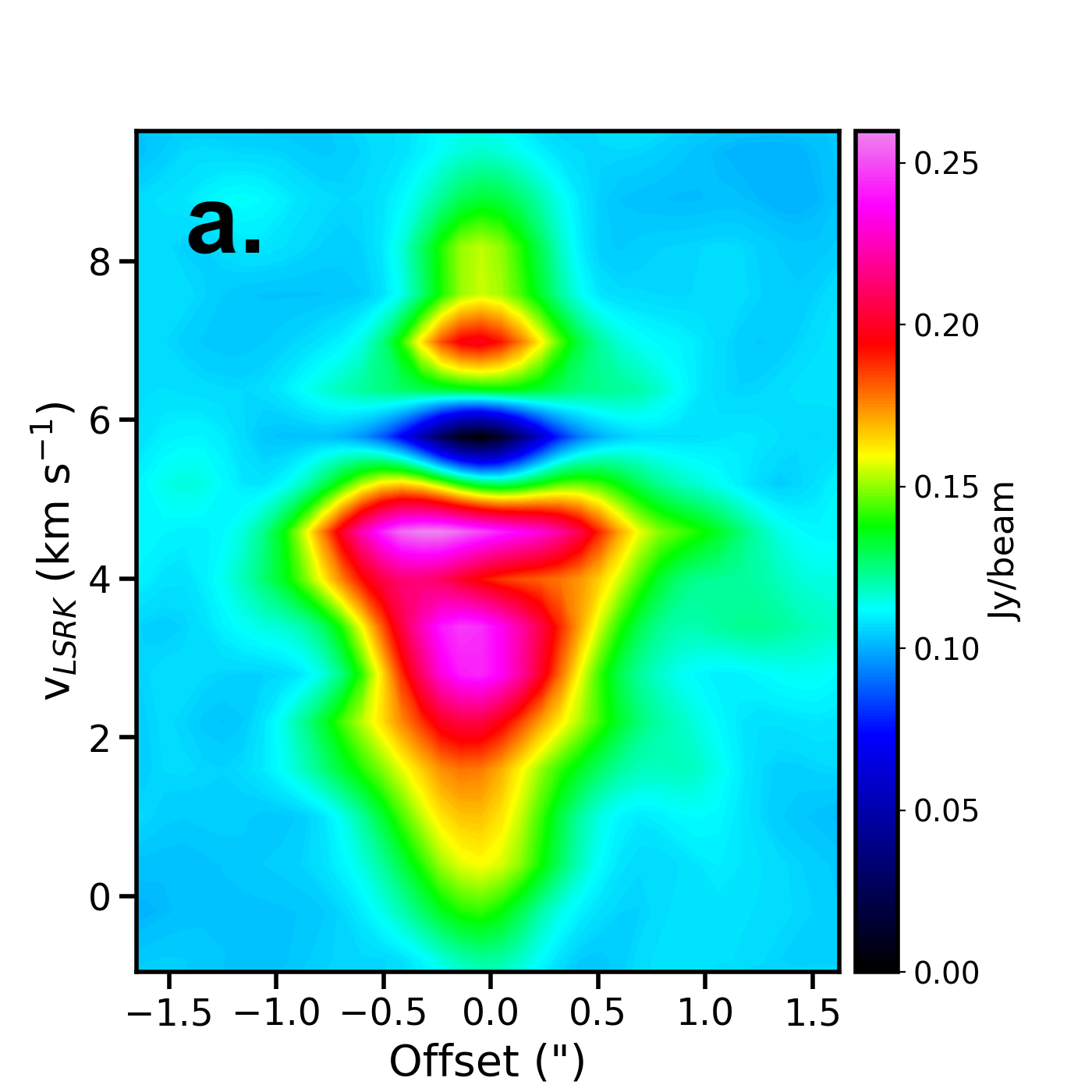}\\
    \includegraphics[width=0.43\textwidth]{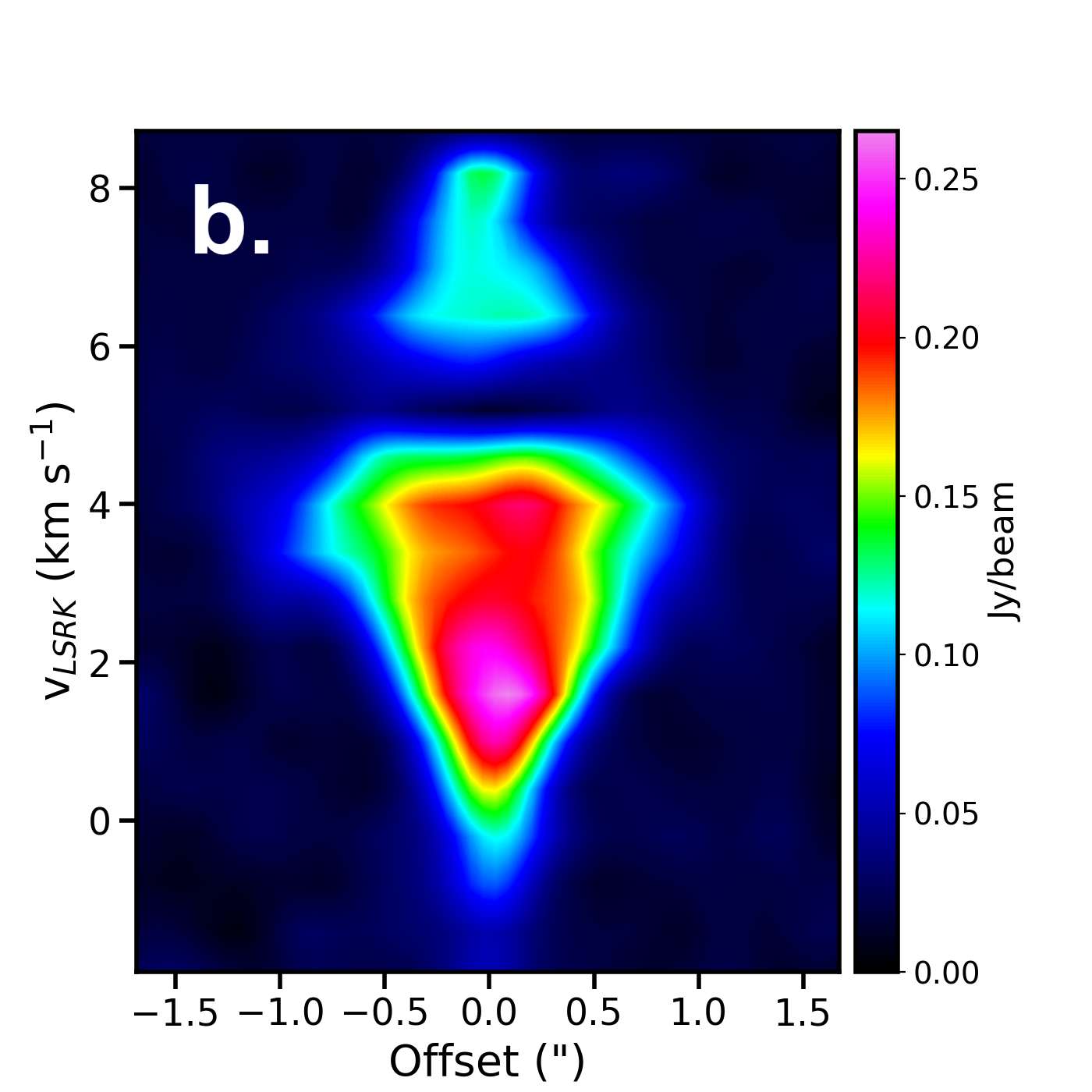}
           
%\end{subfigure}%
\caption[short]{Position-velocity diagrams of  the HCN emission line extracted along the rotational axis, i.e. P.A.$\sim$ 120$^{\rm o}$ in V883 Ori. Panel (a): PV diagram from ALMA data. Panel (b): PV diagram from synthetic HCN emission. In both panels, HCN displays the known diamond-shaped feature consistent with gas infalling towards V883 Ori.}
\label{Fig:Model}
\end{figure}

To study in more detail whether the observations show evidence of infalling motion in the innermost ($<$ 800 au) part of the V883 Ori system, we perform radiative transfer calculations via the Line Modelling Engine (LIME; Brinch $\&$ Hogerheijde 2010) code to constrain the kinematics of the infalling material along the rotational axis. To that end, we adopted the canonical ``inside-out" collapse model \citep{Shu1977} based on a molecular cloud characterized by an inner region dominated by infalling motions and an outer region in a nearly static phase. In addition, we use the kinematic and density profiles resulting from the collapse of slowly rotating material from the inner regions of a molecular cloud \citep{Ulrich1976, Cassen1981, Tereby1984}. As an initial assessment as to whether infall may be present, we compare these models with the HCN data. The non-LTE radiative transfer calculations for HCN use the collisional rate coefficients scaled to H$_{2}$ taken from the \href{https://home.strw.leidenuniv.nl/~moldata/}{LAMDA database} \citep{Dumouchel2010}. The effects of the outburst in V883 Ori is represented by a generalized radial profile of temperature as follows:

\begin{equation}
\rm T (R) = T(R_{in}) \times \left ( \frac{R}{R_{in}} \right )^{-0.5},
\end{equation}

where $\rm T(R_{in})$ $\sim$ 100 K and $\rm R_{in}$ $\sim$ 40 au. For V883 Ori, we adopted a central mass of 1.3 M$_{\odot}$, an inclination of 38 deg. and a systemic velocity 4.5 km s$^{-1}$ \citep{Cieza2016}. The initial radius of the infalling particle is assumed to be 320 au, see Table \ref{Table:RP}. In order to reproduce the red-shifted self-absorption feature together with the observed intensity levels, we have applied an HCN abundance of 1$\times$ 10$^{-8}$ and a mass infall rate of 5 $\times$ 10$^{-6}$ $\rm M_{\rm \odot}$ yr$^{-1}$ in our modeling process. To compare our model to the ALMA data, we produced synthetic line cubes with similar imaging parameters according to the data, see Sec. \ref{Sec:Observations}. Thus, the produced synthetic cubes were convolved with the observed beam size obtained for the HCN emission. Figure \ref{Fig:Model} shows PV diagrams for HCN, synthetic and observed, where the modeled velocity field and intensity levels are in good agreement with the data. The comparison in Fig. \ref{Fig:Model} hence confirms that the diamond-shaped velocity pattern traced by HCN emission in the PV diagram is consistent with the presence of material infalling toward V883 Ori in a direction perpendicular to the disk.

\section{Discussion}\label{Sec:Discussion}

\subsection{Kinematics and Structure}

Our ALMA observations reveal in the small velocity range 0 - 10 km s$^{-1}$, a dense molecular structure traced by HCN, DCN, and HCO$^{+}$ towards the central source and material likely heated by shocks detected in HCO$^{+}$, SO, and H$_{2}$CO emission at larger distances of $\sim$ 1500 au (see Figs.  \ref{Fig:mom0} and  \ref{Fig:RP}). The velocity maps and PV diagrams (Figs. \ref{Fig:mom1}, \ref{Fig:PV}) reveal that these species trace different regions of the disc-cavity system. In particular, HCN and DCN likely trace material feeding the central object and are associated with the densest molecular regions as these species tend to be abundant in regions where gas number densities range from 10$^{6}$ to 10$^{8}$ cm$^{-3}$ \citep{Willacy2007, Walsh2010}, whereas HCO$^{+}$ likely delineates an intermediate, dense, and possibly turbulent mixing layer, whose more diffuse nature indicates an interface between disc and envelope material (see also Sec. \ref{Sec:HCNHCO}). The HCO$^{+}$ emission is mostly observed at the northern side (Fig. \ref{Fig:HCOChannel}), since V883 Ori is non-uniformly embedded in the molecular cloud with a visible southern outflow carving out the surrounding material \citep{Strom1993}. These kinematic features are consistent with the previous $^{12}$CO, $^{13}$CO, and C$^{18}$O observations presented in \citet{Ruiz2017b}, where $^{12}$CO traces the southern side of the cavity, $^{13}$CO traces marginally optically thick material from the northern side of the cavity, and C$^{18}$O traces optically thin material from the disc midplane. The HCO$^{+}$ emission morphology is also consistent with that found in H$_{2}$CO and SO, which trace the northern outflow cavity wall structure, but at different scales, from the disc throughout the surrounding envelope (Figs. \ref{Fig:mom0} and \ref{Fig:mom1}).

In general, emission from molecules such as H$_{2}$CO, SO, and CH$_{3}$OH is known to be enhanced in the gas phase by shocks and has been detected in active outflow-shocked regions \citep[e.g.][]{Jorgensen2004}. Accordingly, CH$_{3}$OH is expected to be found along the cavity walls. Instead, a compact CH$_{3}$OH emission with a radius of $\sim$120 au was detected around the central source (Fig. \ref{Fig:mom0}). This indicates that the CH$_{3}$OH enhancement in V883 Ori is not directly associated with high velocity shocks but more likely results from the shear between the disc and relatively slow outflows \citep{Ruiz2017b} or any other strong mechanism that releases ices from grain mantles (e.g. X-rays), thus heating the grain surfaces and liberating the grain mantles \citep[e.g.][]{Walsh2014}, as further discussed below. This interpretation is also supported by the production of H$_{2}$CO around the central source, which likely is evaporated directly from the icy dust mantles \citep[e.g.][]{Loomis2015}, and similar to CH$_{3}$OH and HCN, only survive sputtering or desorption of grain mantles at low-shock velocities of $\sim$10 km s$^{-1}$ or less \citep{vanDishoeck1998}. The absence of velocities $>$10 km s$^{-1}$ also explains the lack of SiO emission that requires grain core or mantle destruction, a process that occurs at shock velocities in excess of$\sim$25 km s$^{-1}$ \citep{Schilke1997}. Another possible scenario, explaining an enhancement of CH$_{3}$OH and H$_{2}$CO with no SiO may be related with the shock evolution in V883 Ori, tracing a post-outburst stage within molecular depletion timescales of $\lesssim$10$^{3}$ yr \citep[e.g.][]{Mikami1992, Burkhardt2019}, where it is expected that SiO molecules are re-incorporated to the grains while some molecules such as CH$_{3}$OH and H$_{2}$CO remain in the gas-phase in presence of weak shocks.

\subsection{HCO$^{+}$ and CO Rings: Chemistry, or Optical Depth Effects?}
\label{Sec:Chemical}

At first glance, the most striking feature of the molecular line data is a ring-shaped molecular structure detected in HCO$^{+}$ with inner radius of $\sim$ 130 au (Sec. \ref{Sec:RP}), and whose inner cavity radius appears to match that of the continuum \citep[$\sim$125 au;][]{Cieza2016}. Interestingly, a similar ring-like structure has been previously observed in $^{12}$CO and $^{13}$CO emission \citep{Ruiz2017b}, where the inner depression or ``hole'' detected in these lines also has a size very similar to that estimated for the sub-millimeter continuum ($\sim$ 100 au; Figure \ref{Fig:COring}). These slight discrepancies in continuum and inner cavity radii are likely due to the different beam sizes ($\sim$0.5$^{''}$ for HCO$^{+}$ vs. $\sim$0.35$^{''}$ for the CO isotopologues). 

\subsubsection{Optical Depth Effects}
\label{Sec:Optical}

The fact that a central cavity is obtained in HCO$^{+}$, $^{12}$CO and $^{13}$CO suggests that the emission from these molecules is affected by a combination of self-absorption and/or absorption of continuum emission. This, in turn, implies that these emission lines are optically thick (as commonly found in dense YSO environments) and/or that the dust is optically thicker than the HCO$^{+}$, HCN, $^{12}$CO and $^{13}$CO emission towards the innermost region ($<$0.3$^{"}$) of the V883 Ori system. This can be inferred from the strong red-shifted absorption against the bright continuum emission detected in the PV diagrams of HCN, HCO$^{+}$, $^{12}$CO, and $^{13}$CO as well as the less deep blue-shifted absorption of the HCO$^{+}$, $^{12}$CO, and $^{13}$CO lines (see Figs. \ref{Fig:PV} and \ref{Fig:PVCO}). The strong red-shifted absorption feature can also be seen in Figs. \ref{Fig:HCOChannel} and \ref{Fig:HCNChannel}, where the HCO$^{+}$ and HCN channel maps show negative flux values mainly at the central rest line frequency, 265.886 and 267.557 GHz, respectively (i.e. $\sim$ 5.8 km s$^{-1}$). The negative values indicate that the absorption of continuum emission by foreground cold gas is responsible for this inner depression in these channel maps, given that an actual absorption by colder material in the line of sight is necessary to yield negative fluxes after the process of continuum subtraction.

In addition, it has been shown that when the line emission is optically thick, such that it significantly absorbs the continuum, this leads to overestimating the dust contribution and then underestimating the emission at the line frequency in the continuum subtraction process \citep[e.g.][]{Boehler2017, Weaver2018}. To investigate whether the ring-like feature may also be a continuum subtraction artifact, we  imaged the line data without continuum subtraction (see App. \ref{App:D}). From a visual inspection of the resulting data cubes and product maps (i.e. peak intensity maps), we can confirm that this is also an artifact of subtracting the continuum from the HCO$^{+}$ emission, indicating that the continuum is absorbing part of the line emission and that the HCO$^{+}$ emission is optically thick towards the innermost region of V883 Ori.

The combination of these two optical depth effects can then explain the observed ring-like morphology of the HCO$^{+}$ moment 0 emission map. Whereas the blue-shifted absorption can be explained by an optically thick line and disc continuum, the red-shifted absorbing layer could be either foreground infalling envelope or foreground ambient cloud material, wherein the star has a $\sim$ 1.5 km s$^{-1}$ blue-shift with respect to the ambient cloud. We note that owing to the data angular resolution of the HCO$^{+}$ emission, which allow us to probe only scales $>$ 90 au, and that disc temperatures may reach $>$ 100 K in the inner region, there may be other possible mechanisms that impact the chemistry of the disc mainly within $\sim$50 au (water-snow location) and upper surfaces at $<$ 50 au as explained below.

\subsubsection{Destruction Pathways}

Though the foregoing (optical depth) scenario appears to best explain the central cavity of $\sim$ 130 au in the HCO$^{+}$ moment 0 map, we have also considered whether the lack of HCO$^{+}$ emission might be due to chemical destruction pathways, at least to a lesser extent, in disc regions where temperatures reach out values of $>$100 K radially and vertically dependent. Such another possible scenario to explain a lack of HCO$^{+}$ emission is related to the high energy sources (e.g. X-rays, UV radiation), largely generated by the recent outburst and high accretion levels in V883 Ori, that tend to produce warm ionized layers of the disc and outflow exposed to it, influencing its chemistry. In this scenario, higher ionization rates increase the amount of H$_{3}^{+}$, which readily destroys CO to form HCO$^{+}$, primarily via the reaction:

\begin{equation}
H_{3}^{+} + CO \rightarrow HCO^{+} + H_{2}.
\end{equation}
 
To explain the observed inner depression in HCO$^{+}$, a chemical destruction pathway has been suggested requiring temperatures of $>$100 K, consistent with the outgoing outburst in V883 Ori and its effects on the surroundings \citep{Jorgensen2013, Leemker2020}. Specifically, dipolar and neutral molecules such as H$_{2}$O, which is sublimated at the inner region of V883 Ori at temperatures of $>$100 K \citep{Brown2007, Cieza2016}, can destroy HCO$^{+}$ efficiently \citep{Jorgensen2013} via the mechanism:

\begin{equation}
H_{2}O +  HCO^{+}   \rightarrow  H_{3}O^{+} + CO.
\end{equation}

If that is the case, HCO$^{+}$ may coexist with frozen water delimiting the location of the snow-line at layers higher up from the disc mid-plane, i.e. intermediate layers, and at radii $>$ 50 au, while there is a lack of HCO$^{+}$ inside the water snow-line (i.e. $<$ 50 au). However, the optically thick nature of HCO$^{+}$ in YSOs \citep[e.g.][] {Merel2021} prevents us to observationally confirm or discard this scenario. Following this approach, HCO$^{+}$ isotopologue emission can be expected to arise from denser and colder regions ($<$80 K), coexisting with frozen water as described below.

 \begin{figure}
 \centering
    \includegraphics[width=0.99\linewidth]{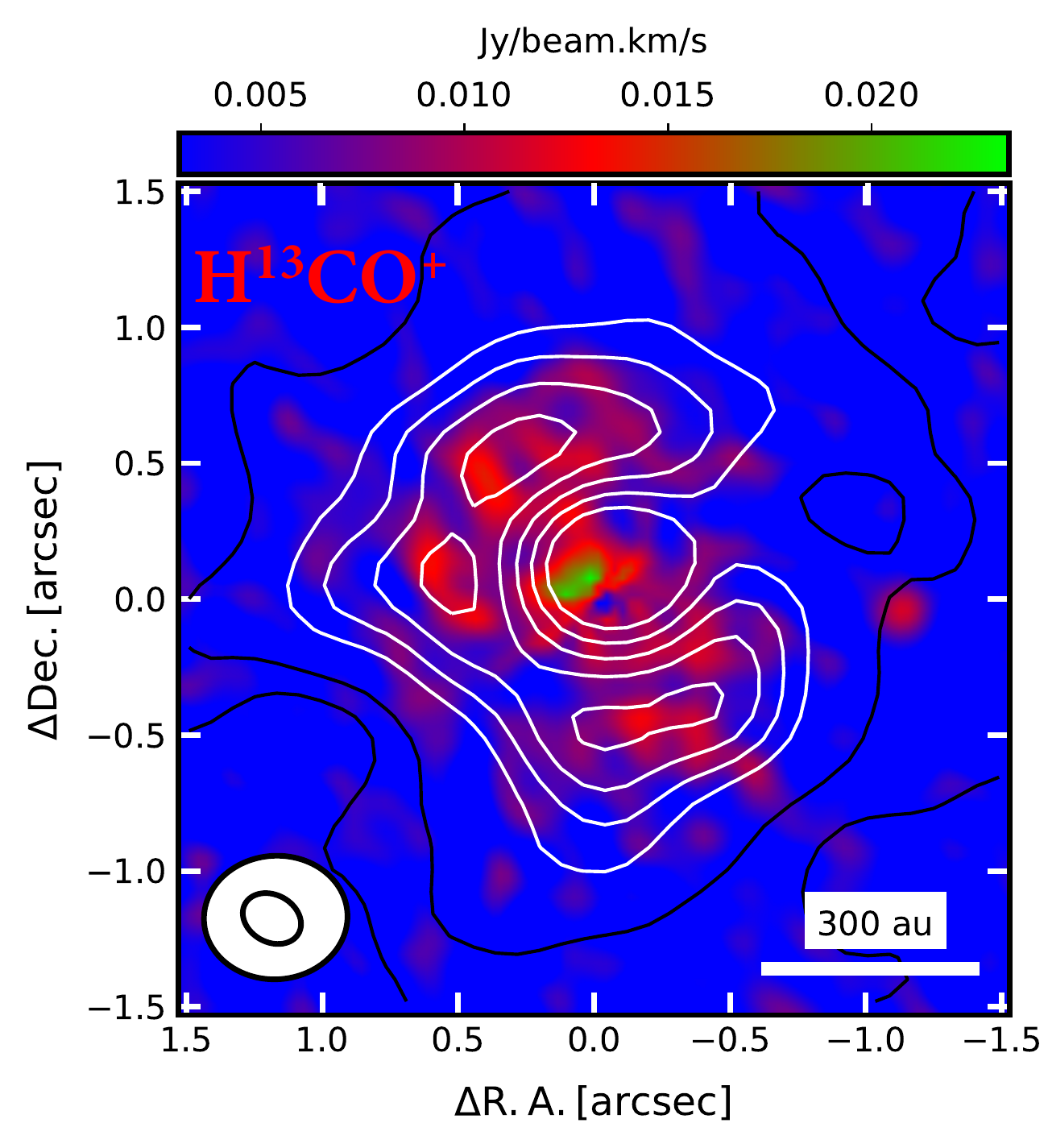}
  \caption{Moment 0 map of H$^{13}$CO$^{+}$ toward V883 Ori. White and black contours represent the ring-like structure and outflows traced by the HCO$^{+}$ emission (Fig \ref{Fig:mom0}), respectively. The corresponding beam sizes are shown at the bottom left. North is up, east is left.} 
\label{Fig:mom0_HCO13}
\end{figure}

 \begin{figure}
 \centering
     \includegraphics[width=0.99\linewidth]{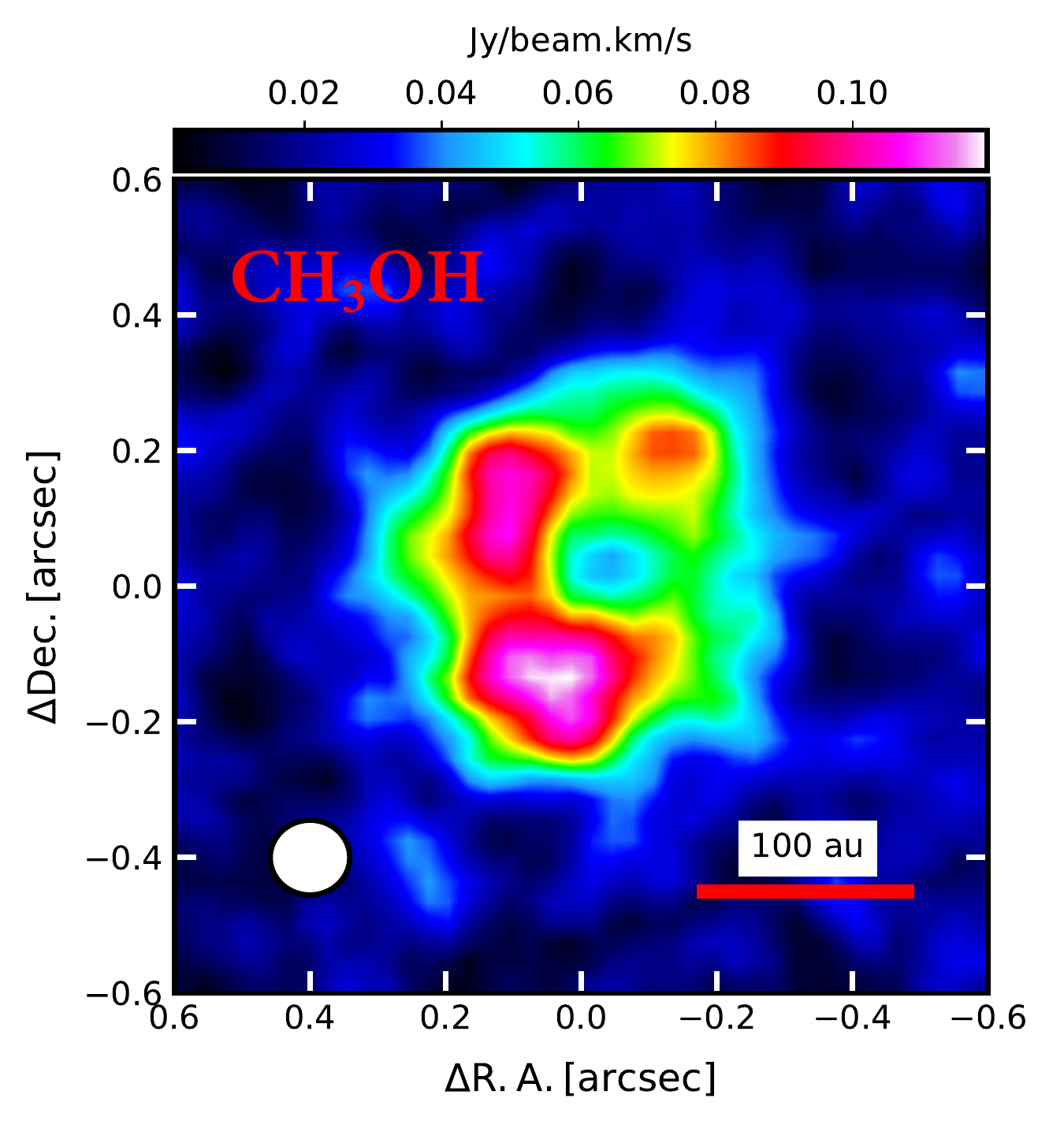}
  \caption{Moment 0 map of CH$_{3}$OH toward V883 Ori. The corresponding beam size is shown at the bottom left. North is up, east is left.} 
\label{Fig:mom0_methanol}
\end{figure}

 \begin{figure}
     \includegraphics[width=\linewidth]{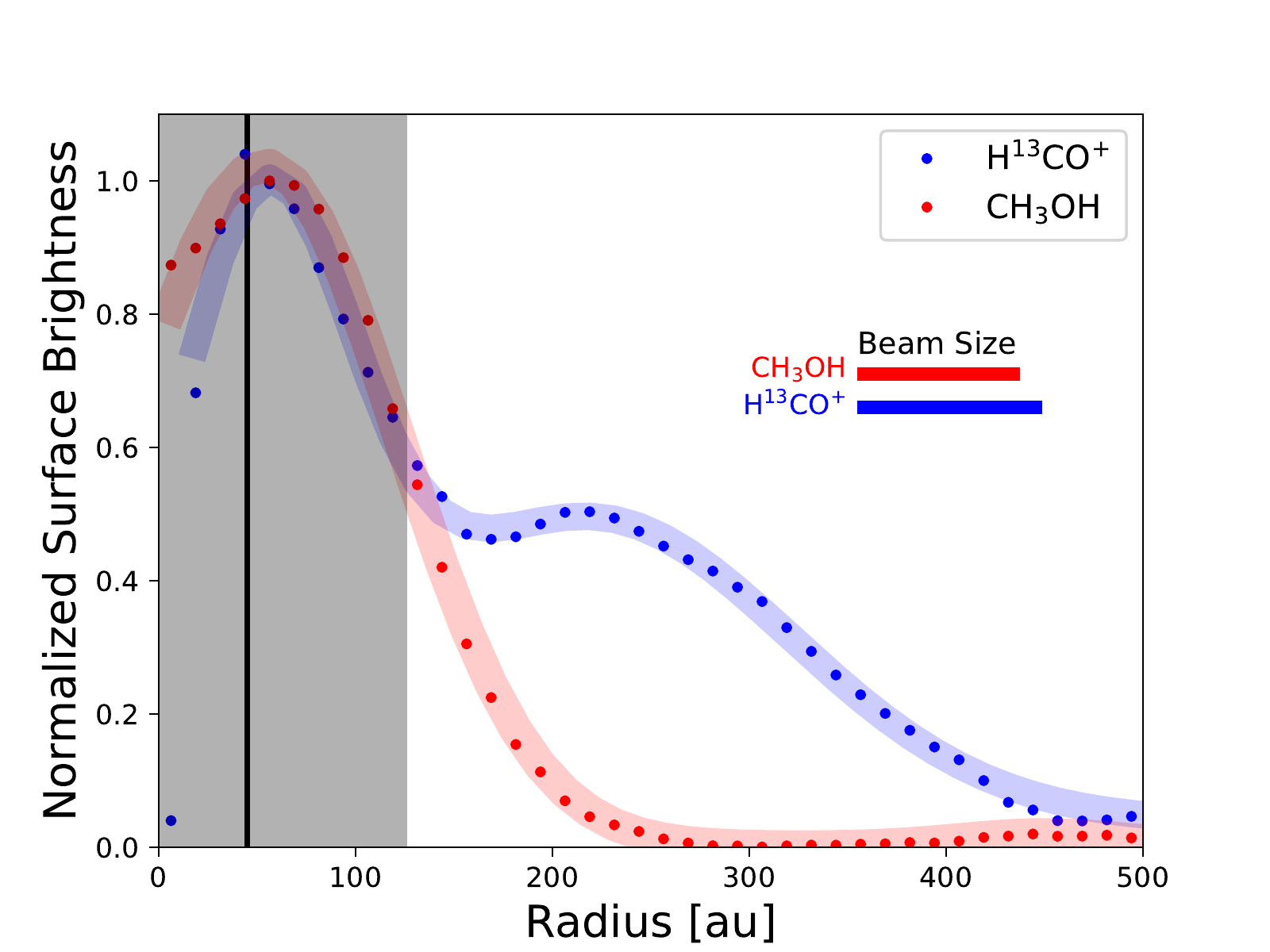}
\caption{Azimuthally averaged and de-projected radial intensity profiles from the H$^{13}$CO$^{+}$ emission after de-blending with a spatial resolution of $\sim$0.2$^{"}$ together with the CH$_{3}$OH emission convolved with a coarser beam of $\sim$ 0.2$^{"}$, see Fig. \ref{Fig:mom0_methanol}. The horizontal bars in the upper right indicate a beamwidth representative of the line. The black vertical line corresponds the snow-line location at 45 au estimated by \citet{Cieza2016}. The blue and red shaded area represents 1$\sigma$ error on the mean. The gray shaded area represents the disc region still contaminated by CH$_{3}$CHO emission making very difficult to discern the real contribution between the CH$_{3}$CHO and H$^{13}$CO$^{+}$ emission.} 
\label{Fig:RP_methanol}
\end{figure}

\begin{table}
\caption{Deconvolved radial extensions from ALMA detected lines in V883 Ori.}
\begin{tabular}{lccccccccc}
\hline Line &  FWHM & Beamwidth & Radial Extension\\
&  [au] & [au] & [au] & \\
\hline
& & ALMA cycle 4  &\\
\hline
H$^{13}$CO$^{+}$$^a$   & 515 & 83 &290 $\pm$ 20\\
CH$_{3}$OH$^a$ & 240  &80& 120 $\pm$ 4\\
\hline
\end{tabular}\\
\footnotesize{$^a$ Higher resolution ALMA data presented in \citet{Cieza2016} and \citet{Lee2019}.}
\label{Table:RP_methanol}
\end{table}

\subsection{Tracing water snow lines with HCO$^{+}$ isotopologues and CH$_{3}$OH emission}

It has been suggested that if the temperature of the dust grains rises above $\sim$ 85 K, the destruction of HCO$^{+}$ due to sublimated water is viable and CH$_{3}$OH could be also liberated from the grain surfaces through thermal desorption \citep{Brown2007, Jorgensen2013}. Hence, it is possible to trace the water snow line location determined by high temperatures in V883 Ori studying these molecules. However, as mentioned above, it is necessary to use optically thin tracers such as H$^{13}$CO$^{+}$ and HC$^{18}$O$^{+}$ lines, whose critical densities allow to probe deeper and denser regions in the system, rather than the parent chemical, HCO$^{+}$.

For our purposes, we collected archival raw data from the project 2017.1.01066.T (PI: Jeong-Eun Lee) to complement our analysis in Band 7 (275-373 GHz), which includes the H$^{13}$CO$^{+}$ J = 4-3 at 346.998 GHz transition. These observations were taken during Cycle 5 with a final synthesized beam of $\sim$0.19 arcsec $\times$ 0.17 arcsec and a spectral resolution\footnote{More observational details can be found in  \citet{Lee2019}.} of $\sim$ 0.24 km s$^{-1}$. Unfortunately, H$^{13}$CO$^{+}$ is blended with transitions from acetaldehyde (CH$_{3}$CHO) making necessary to reduce line blending effects by using the Keplerian velocity of the disc to calculate the Doppler shift of the emission in each pixel \citep[e.g.][]{Yen2016, Lee2019}. A more complete analysis of this data set was reported in \citet{Leemker2020} for more details. Using this approach, H$^{13}$CO$^{+}$ emission is detected at $>$ 5$\sigma$ but is still blended with acetaldehyde lines that are expected to peak on-source as further discussed below. Figure \ref{Fig:mom0_HCO13} shows the integrated intensity of the isotopologue H$^{13}$CO$^{+}$ J = 4-3 line towards V883 Ori. Closely matching the outer radial extension of the ring-like structure detected in HCO$^{+}$ emission, see Fig.  \ref{Fig:mom0}, H$^{13}$CO$^{+}$ is located within a radius of $\sim$ 300 au, see Table \ref{Table:RP_methanol}, and presents an inner rotating-supported Keplerian disc as revealed by its PV diagram, see Fig. \ref{Fig:PV_H13CO}. The distribution of these molecules indicates that the HCO$^{+}$ emission is indeed optically thick tracing also colder outflow material (Fig. \ref{Fig:HCOChannel}), and that its ring-like emission structure obtained in the moment-0 map is mainly a result of optical depth effects (see \ref{Sec:Optical}). The detection of the optically thin H$^{13}$CO$^{+}$ emission within a radius of 300 au also suggests a higher abundance of HCO$^{+}$ towards the disc midplane at lower temperatures (i.e. < 80 K) and higher densities, where both molecules are thought to exist due to their similar critical densities of 2.8$\times$10$^{6}$ cm$^{-3}$ for H$^{13}$CO$^{+}$ J = 4-3  and 1.4$\times$10$^{6}$ cm$^{-3}$ for HCO$^{+}$ J = 3-2 in a temperature range between 20 and 70 K \footnote{The collisional rates are adapted from the Leiden LAMDA database \citep{Schofier2005, VanderTak2020}.}.

In addition, from ALMA band-7 observations at a higher resolution (0.05$^{"}$) taken during Cycle 4 (2016.1.00728.S, PI: Lucas Cieza), \citet{Lee2019} detected a CH$_{3}$OH ring-like emission structure with an inner radius of $\sim$ 40 au (Fig.  \ref{Fig:mom0_methanol}), close to the water snow-line location at $\sim$ 42 au at the midplane \citep{Cieza2016}. These authors claimed that the dust continuum opacity is too high to detect sublimated CH$_{3}$OH in the inner region ($<$ 50 au), while around the water snow line, the continuum opacity becomes low enough allowing CH$_{3}$OH emission to be detected at larger radii. From our observations with a coarser beam ($\sim$0.5$^{"}$), CH$_{3}$OH is highly concentrated near the central object position, and spatially matches the dust continuum outer radial extension \citep[$\sim$ 125 au;][]{Cieza2016}. This  indicates that the CH$_{3}$OH emission arises in a compact, warm and dense region where the entire dust grains are heated thermally to 80-100 K, and thus the icy mantle evaporates, releasing CH$_{3}$OH into the gas phase.

In order to further compare the distribution of the detected CH$_{3}$OH and H$^{13}$CO$^{+}$ emission lines with the dust extension and water snowline location estimated previously from higher-resolution observations \citep{Cieza2016}, we additionally extracted the radial intensity profiles for the high-resolution CH$_{3}$OH emission presented in \citet{Lee2019} together with the H$^{13}$CO$^{+}$ emission line (Figs. \ref{Fig:mom0_HCO13} and \ref{Fig:mom0_methanol}). To do so, we follow a similar approach as described in Sec. \ref{Sec:RP}. For consistency, we extracted CH$_{3}$OH data cubes with the coarser resolution of the H$^{13}$CO$^{+}$ data. The resulting radial profiles are displayed in Figure \ref{Fig:RP_methanol}, and parameters are listed in Table \ref{Table:RP_methanol}. Despite the line blending in H$^{13}$CO$^{+}$ line, we have been able to estimate a radial extension of $\sim$ 290 au from the outer region of the emission, however, we are unable to resolve substructures within the very central region of the disc which is still contaminated by relatively strong CH$_{3}$CHO features. This can be seen in Fig. \ref{Fig:RP_methanol}, where the H$^{13}$CO$^{+}$ azimuthal radial profile displays an intensity break at $\sim$130 au resulting from the presence of strong CH$_{3}$CHO features that were not possible to remove in the data reduction process. This H$^{13}$CO$^{+}$ intensity break also matches up with the radial extension estimated by \citet{Cieza2016} of $\sim$125 au making more difficult its interpretation. While we do not find firm evidence of differentiation in terms of their patterns in the moment-0 map and P-V diagram, we cannot completely rule out that H$^{13}$CO$^{+}$ emission may originate more prominently from smaller radii within the snow line location ($<$ 50 au), which appears somewhat similar to the case of CH$_{3}$OH with the coarser resolution. Likewise, the H$^{13}$CO$^{+}$ emission could appear to feature an annular gap, i.e., a ring-like depression in the emission profile similar to the HCO$^{+}$ emission, but with the current data set this scenario is also inconclusive. Higher spatial and spectral resolution would be necessary to better describe the central component of the H$^{13}$CO$^{+}$ emission --a transition that is not blended with any COMs-- in V883 Ori.

Although, it is difficult to establish the exact physical origin of the CH$_{3}$OH and H$^{13}$CO$^{+}$ emission in the disc; i.e. midplane vs. surface layers, we should account that V883 Ori undergoes an outburst episode with resulting shocks that impact mostly the upper and intermediate layers of the dust grains. Thus, considering that CH$_{3}$OH may work as a good proxy for water vapor as it desorbs at similar temperatures (i.e. $\sim$ 85 K), and the location of the water snowline in V883 Ori was estimated to be at $\sim$42 au at the midplane \citep{Cieza2016}, it would mean temperatures above $\sim$100 K along intermediate and surface layers at larger radii. Accordingly, the origin of CH$_{3}$OH and H$^{13}$CO$^{+}$ may be located above and below the warm upper layer (T $\ge$ 85-100 K) in the disc, respectively.

\subsubsection{Implications of the water snow line location in the disc midplane}

The location of the water snow in the disc mid-plane is critically important for planet formation. Its position controls the efficiency in the growth of dust, planetesimal, ice giants, and the cores of gas giants \citep{Blum2008, Morbidelli2015}.  It also regulates the delivery of water to the surface of rocky planets interior to the snowline \citep{Raymond2007}. Because protoplanetary discs show a very high degree of dust settling \citep[e.g][]{Ruiz2020}, dust discs are geometrically very thin, with scales heights of only $\sim$1 au at 100 au radii \citep{Pinte2016}. Therefore, since the position of the snow line can be a strong function of the scale height, it becomes important to probe the location of the line as close to the midplane as possible, where planet formation is expected to occur.

The water snow line in protoplanetary discs around solar-type stars is expected to be located at 3-5 au in the midplane \citep{Kennedy2008}. However, in the case of FU Ori objects such as  V883 Ori, the outburst luminosity can push the snow line to much larger distances. The exact position of the water snow in the disc midplane of V883 Ori has been a subject of intense debate in recent years. Initially, \citet{Cieza2016} suggested a distance of $\sim$42 au from the star based on intensity break seen at $\sim$0.1$^{"}$ in the 1.3 mm continuum image and on the brightness temperature of these data. \citet{Schoonenberg2017} modeled the intra-band spectral index (between 218 and 230 GHz) profile of the disc and concluded that water snow line is located at $\sim$50 au.  From band-7 observations at 60 au resolution, \citet{vantHoff2018} detected CH$_{3}$OH in the V883 Ori disc up to 140 au, suggesting that the snow line of water and CH$_{3}$OH could be at distances as large as $\gtrsim$100 au. \citet{Lee2019} also detected CH$_{3}$OH at large distances from the star but explains this finding as the detection of gas-phase molecules in the surface layers of the disc. More recently, \citet{Leemker2020} argue that the ring-like shaped image of the HCO$^{+}$ line presented here, see Fig. \ref{Fig:mom0}, is caused by the destruction of HCO$^{+}$ by gas-phase water (see Eq. 2), serving as observational evidence for the water snow line being located at $\sim$ 75-120 au, but using the definition of the snowline as the mid-plane radius in where 50$\%$ of the water is in the gas phase and 50$\%$ is frozen out onto the dust grains.

As discussed in Sec. \ref{Sec:Chemical}, we argue that the dust continuum is optically thicker than the HCO$^{+}$ line at $<$0.3$^{"}$ and that HCO$^{+}$ line traces self-absorption and absorption of dust emission in the inner region of the V883 Ori system. Overall, we find that the molecular lines available reveal a complex radial and vertical structure in the V883 Ori gas disc, making it difficult to identify the location of the water snowline in the disc midplane. However, as mentioned above, the dust disc is geometrically very thin, rendering continuum data easier to interpret in terms of its thermal structure.  More recently, an analysis of the multi-frequency (44, 100, 230, 345 GHz) high-resolution (0.07$^{"}$) continuum images  of V883 Ori suggests that the temperature of the dust reaches $\sim$100 K at $<$50 au (Cieza et al., in prep), indicating that the midplane water snow line is in fact well within $\sim$50 au from the star. In particular, the discrepancy with the snowline location found by \citet{Leemker2020} could be reconciled if the current lack of HCO$^{+}$ actually traces the water vapor in the disc, but taking into account that there is a significant chemical lag following a sudden heating event \citep{Hsieh2019}. If V883 Ori has undergone a burst within the last 100$-$1000 yrs, increasing its luminosity an order of magnitude larger, it is likely that water was sublimated in the midplane out to $>$ 100 au, while destroying HCO$^{+}$ out to this radius. Currently, the luminosity of V883 Ori is lower and the dust has cooled down, which locates the 100 K radius at $\sim$50 au. Accordingly, the dust temperature traces the current water sublimation radius at the midplane, whereas HCO$^{+}$ traces water vapor at intermediate layers above the midplane where temperatures reach $>$100 K.

\subsection{Optical Depth and Physical Structure}
\label{Sec:HCNHCO}

Considering that HCN and HCO$^{+}$ are high-density tracers, it is odd that from our data a similar ring pattern is not seen in HCN. A possible explanation might be related to optical depths effects and that these molecules are tracing different regions from the surface to midplane layers, and from the disc to shocking outflows, i.e. due to variations in density and temperature. While it is expected that HCN J = 3-2 and HCO$^{+}$ J = 3-2 may be produced under similar physical conditions due to their similar dipole moments and high critical densities of 1$\times$10$^{7}$ cm$^{-3}$ for HCN J = 3-2 and 1$\times$10$^{6}$ cm$^{-3}$ for HCO$^{+}$ J = 3-2 in a temperature range between 20 and 70 K\footnote{The collisional rates are adapted from the Leiden LAMDA database \citep{Schofier2005, VanderTak2020}. For HCO$^{+}$ J=3-2 and HCN J=3-2, the collisional coefficients are from \citet{Flower1999} and \citet{Dumouchel2010}.}, different ionization rates can affect the HCO$^{+}$ abundance \citep{Krolik1983}, which is also thermalized at lower densities than HCN. As a result, HCO$^{+}$ can remain abundant and excited in diffuse gas,  such as those layers closer to the surface. This can explain the difference observed between a centralized HCN emission and a more extended HCO$^{+}$, see Fig. \ref{Fig:mom0}, whose emission bulk of a radius $\sim$ 450 au trace the denser material orbiting and feeding the central source. This is consistent with a HCN/HCO$^{+}$ ratio of $\sim$0.8 estimated within a circle region enclosed by the HCO$^{+}$ ring diameter, and tracing gas with intermediate densities between CO and HCN since HCO$^{+}$ has a critical density nearly an order of magnitude lower than HCN, i.e. densities of $\sim$10$^{6}$ cm$^{-3}$. However, higher resolution observations are needed to probe or discard the existence of a real inner hole in HCN and HCO$^{+}$ as observed in other tracers such as CH$_{3}$OH \citep{Lee2019}, since our data might have suffered significantly from beam dilution as was the case with our ALMA cycle 6 CH$_{3}$OH observation.

\subsection{Shocks vs. High-energy radiation}

An important, but yet unresolved, question concerns the impact of heating and/or ionizing sources on the reaction which leads to the production of the observed chemistry in FUor systems and its evolution. One possibility is shocks. As mentioned above, detections of CH$_{3}$OH and H$_{2}$CO tracing slow velocities -- together with a lack of SiO -- suggest weak shocks in V883 Ori, which may indicate a later stage in the shock evolution and after an outburst episode as also previously suggested by \citet{Ruiz2017b}. However, dense gas traced by HCO$^{+}$ and HCN emission indicate that the weak and slow shocks alone are insufficient to explain the measured enhancements of these molecules near the central source  \citep[e.g.][]{Tafalla2013}, specially the HCO$^{+}$ abundance variation on the 200- to 3000- au scale and observed as a sharp break in the intensity profile at $\sim$500 au. The last features suggest that an additional high-energy radiation source is also needed in order to produce these emissions concentrated towards the central star. On the one hand, the formation and fractionation of HCN are controlled by the UV radiation; the former through vibrational excitation of H$_{2}$, the latter through self-shielding of N$_{2}$. On the other hand, the HCO$^{+}$ emission can potentially be explained via X-ray driven chemistry enhancing the HCO$^{+}$ abundance in the intermediate layers close to the central source \citep{Cleeves2017}. Although V883 Ori would appear to have inefficient X-ray heating and a low ionization level based on a non-detection obtained with the XMM-Newton X-ray Observatory -- an inferred 3$\sigma$ upper limit on X-ray luminosity of log L$_{x}$ $<$ 29.7 erg s$^{-1}$ assuming an intervening absorbing column of N$_{H}$ = 6.3$\times$10$^{21}$ cm$^{-2}$ \citep{Kuhn2019} -- we must note that the non-detection does not mean that the X-rays are not present in the system, as the infalling envelope material seen along the line of sight toward the star-disk may have very high column densities which could absorb X-ray emission. That conclusion seems consistent with colors of V883 Ori. Specifically, the observed J-K color (J-K = 4.1) indicates E(J-K) $\sim$4 assuming A-type star photospheric colors, suggesting  A$_{k}$ $\sim$ 1.6 mag, A$_{v}$ $\sim$ 16 mag. This would translate to N$_{H}$ = 3.5$\times$10$^{22}$ cm$^{-2}$, leading to an estimated upper limit on intrinsic X-ray luminosity of L$_{x}$ $<$ 30.5 erg s$^{-1}$. While self-absorption by the dust continuum and cold gas appear to be the dominant effect contributing to the pronounced inner depression in HCO$^{+}$, a HCN/HCO$^{+}$ ratio of $\sim$ 0.8 indicates that, at least within a radius of $\sim$ 500 au, these tracers are being produced under similar thermal and excitation conditions, however, it remains unexplained the level of influence of shocks and/or high-energy radiation on the rich-chemistry observed in V883 Ori.

\section{Conclusions} 
\label{Sec:Conclusions}

\begin{figure*}
\includegraphics[width=0.95\textwidth]{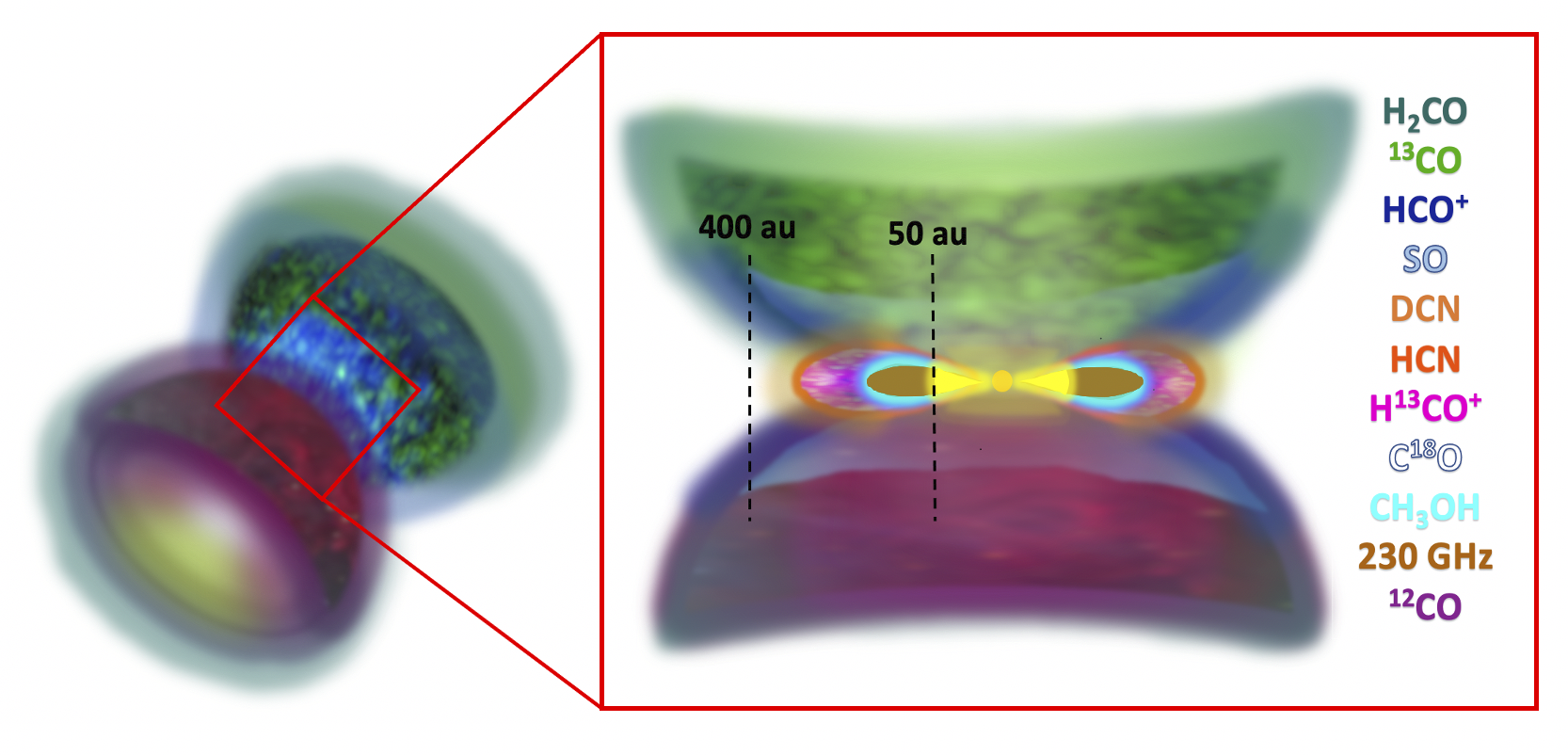}
\caption{Sketch for the observed molecules: Different chemical species trace distinct layers in the disc-outflow system of V883 Ori. CO: Probe gas opacity ($^{12}$CO tends to be optically thick, while $^{13}$CO and C$^{18}$O tend to be optically thin) with intermediate critical densities (10$^{4}$ cm$^{-3}$) in the system. HCO$^{+}$ and HCN, with critical densities of $\sim$ 10$^{6}$ cm$^{-3}$ and 10$^{7}$ cm$^{-3}$, respectively, are preferentially excited in high density environments. In addition, HCO$^{+}$ abundance can be enhanced in regions with a higher fraction of ionization. CH$_{3}$OH emission traces the breakup of the icy mantles during the dust grain destruction process, indicating the presence of slow shocks \citep{Ruiz2017b}. The bipolar slow outflow also is outlined by emission from shock tracers such as SO and H$_{2}$CO at different scales since these lines trace the disk/envelope structure around V883 Ori.}
\label{Fig:Molecules}
\end{figure*}

In this paper, we present an ALMA spectral line study of the V883 Ori system covering size scales from the disc to outflow and envelope material. To that end, we opted to combine the TP, 7-m, and 12-m arrays from ALMA, and this broad coverage allows us to detect a complex disk-outflow structure associated with V883 Ori on scales from 0.5$^{''}$ to 20$^{''}$ ($\sim$ 200 - 8000 au at 417 pc). Using HCN, HCO$^{+}$, CH$_{3}$OH, SO, DCN, and H$_{2}$CO, $^{12}$CO and $^{13}$CO emission, we have studied the kinematic and chemical structures in the circumstellar environment of V883 Ori. Fig. \ref{Fig:Molecules} presents an overview of the
results, which are summarized below:

1. V883 Ori is characterized by a very slow molecular outflow, previously detected in $^{12}$CO and $^{13}$CO. Our ALMA observations confirmed that the kinematics from more diffuse gas to the densest regions are also dominated by slow velocities in the small velocity range 0 - 10 km s$^{-1}$.

2. This FUor object presents a centralized dense structure within a radius of $\sim$ 400 au traced by HCN, HCO$^{+}$ and DCN emission together with material heated by shocks detected in CH$_{3}$OH, SO, H$_{2}$CO emission, while farther out it is a combination of disc and outflow signal traced by HCO$^{+}$, SO, and H$_{2}$CO, $^{12}$CO and $^{13}$CO emission. The extensions of the CH$_{3}$OH and dust continuum emission from the disc are similarly compact \citep{Cieza2016}. HCN rises from a slightly more compact region than DCN emission, while the bulk of the H$_{2}$CO emission matches the radial extension of HCN. In addition, slow outflows are traced by HCO$^{+}$, SO, $^{12}$CO and $^{13}$CO emission.

3. Every molecular line displays Keplerian rotation around the central stellar mass of 1.3 M$_{\odot}$ from the south-west (blue) to north-east (red) direction of the disc. In addition, high-density tracers such as HCN and HCO$^{+}$ display in-fall motion perpendicular to the disc rotation direction (P.A. $\sim$ 120 deg.).

4. HCO$^{+}$ line emission of V883 Ori extends up to $\sim$2000 au, in contrast to the more compact HCN and DCN line emission. This could indicate that HCN emission raises from deeper layers, while HCO$^{+}$ is located at intermediate and more diffuse layers in the system.

5. The various chemical tracers reveal a complex physical (radial and vertical) structure in the V883 Ori environment, with morphological differences between the tracers that are indicative of the varying dominance of shocks (e.g. SO, H$_{2}$CO) vs. radiation fields as drivers of the chemistry (e.g. HCN, HCO$^{+}$). For instance, CH$_{3}$OH and continuum emission are tightly correlated with each other. Based on their spatial distribution and high temperatures in the inner region of the V883 Ori disc, CH$_{3}$OH is most likely released from the grain mantles into the gas phase via thermal sublimation (desorption) in hot regions ($>$100 K).

6. Emission from HCO$^{+}$ reveals a pronounced inner depression or ``hole'' with a size comparable to the radial extension estimated for the 230 GHz continuum \citep[125 au; ][]{Cieza2016} and CH$_{3}$OH emission. This inner ``hole'' is likely produced by the optically thick nature of the HCO$^{+}$ and continuum emission, which can result in an artifact when over-subtracting the continuum from the HCO$^{+}$ emission over a narrow range of (mostly red-shifted) velocity channels. In this velocity range, it appears the continuum emission is absorbed by foreground, optically thick (cold, dense) HCO$^{+}$. The presence of this red-shifted absorption is hence an additional indication that HCO$^{+}$ traces gas infalling onto the V883 Ori disk.

7. Our results are consistent with the water snow currently being located at $\sim$40 to 50 au in the midplane, as derived from analyses based on the continuum. However, we can not rule out an scenario in which the observed HCO$^{+}$ ring is the result of previous outburst (100-1000 years ago) that sublimated the midplane water up to $>$100 au. In order to probe conclusively whether this scenario plays an important role in the HCO$^{+}$ destruction or optical depth effects are responsible for the observed ring-like structures in V883 Ori, future ALMA high resolution and sensitivity observations of tracers such as H$^{13}$CO$^{+}$ and HC$^{18}$O$^{+}$ are needed.

\section*{Acknowledgements}
We thank the anonymous referee for the helpful comments, and constructive remarks on this article. The National Radio Astronomy Observatory is a facility of the National Science Foundation operated under cooperative agreement by Associated Universities, Inc. This  paper  makes  use  of  the  following  ALMA  data:  ADS$/$JAO.ALMA$\#$2017.1.00015.S and ADS$/$JAO.ALMA$\#$2018.1.01131.S. ALMA  is  a  partnership of ESO (representing its member states), NSF (USA) and NINS (Japan), together with NRC (Canada), MOST and ASIAA (Taiwan), and KASI (Republic of Korea), in cooperation with the Republic of Chile. The Joint ALMA Observatory is operated by ESO, AUI/NRAO and NAOJ. J.P.W. acknowledges support from NSF grant AST-1907486. JHK's research is supported in part by NASA Exoplanets Research Program grant 80NSSC19K0292 to Rochester Institute of Technology. L.A.C  acknowledges the support from Agencia Nacional de Investigacion y Desarrollo de Chile (ANID) given by the grant FONDECYT Regular number 1211656. L.A.C acknowledges support from FONDECYT Grant 1211656 and the Millennium Nucleus  YEMS, NCN2021-080, from ANID,  Chile. M.L. acknowledges support from the Dutch Research Council (NWO) grant 618.000.001.

\section*{DATA AVAILABILITY}

The data underlying this article are available in the ALMA archive at \url{https://almascience.nrao.edu} under project codes 2018.1.01131.S and 2017.1.00015.S.

%\newpage

\bibliographystyle{mnras}
\bibliography{sample63}

\clearpage
\appendix

%\renewcommand{\thetable}{\Alph{section}\arabic{table}}
%\renewcommand{\thefigure}{\Alph{section}\arabic{figure}}
%\restartappendixnumbering

\section{Combining Arrays}
\label{App:C}
In this appendix, we present an example of the generated products after combining the ALMA 12-m, 7-m, and TP data. To do so, we have followed two sources: (1) the Casa guide\footnote{https://casaguides.nrao.edu/index.php/M100$_{-}$Band3$_{-}$Combine$_{-}$4.3} on the combination of ALMA 12-m, 7-m, and TP data with the (2) Total Power Map to Visibilities\footnote{https://github.com/tp2vis/distribute} (TP2VIS) method  \citep{Koda2019}. 

\begin{figure}
\includegraphics[width=0.4\textwidth]{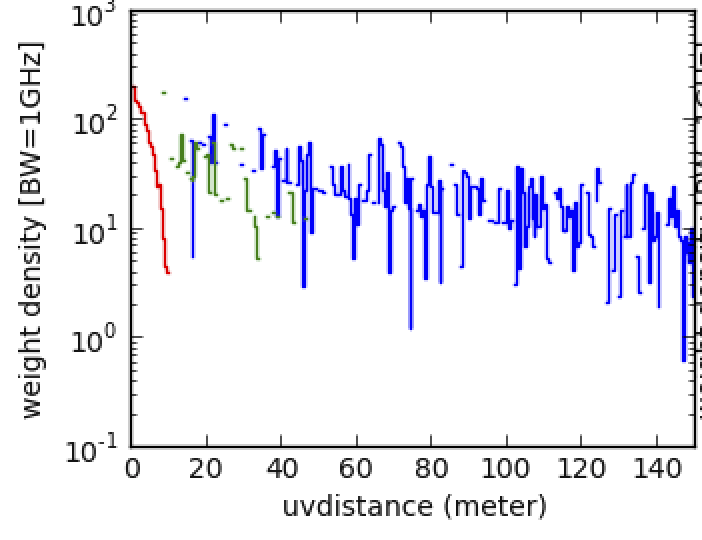}\\
\caption{Weight density as a function of \textit{uv} distance for the V883 Ori ALMA Cycle 6 observations. The red, green, and blue correspond to the
TP, 7-m, and 12-m visibilities, respectively. This plot is generated by the Tp2vispl function in order to inspect weights before concanating TP, 7-m and 12-m data.}
\label{Fig:Weight}
\end{figure}

\begin{figure*}
\includegraphics[width=0.3\textwidth]{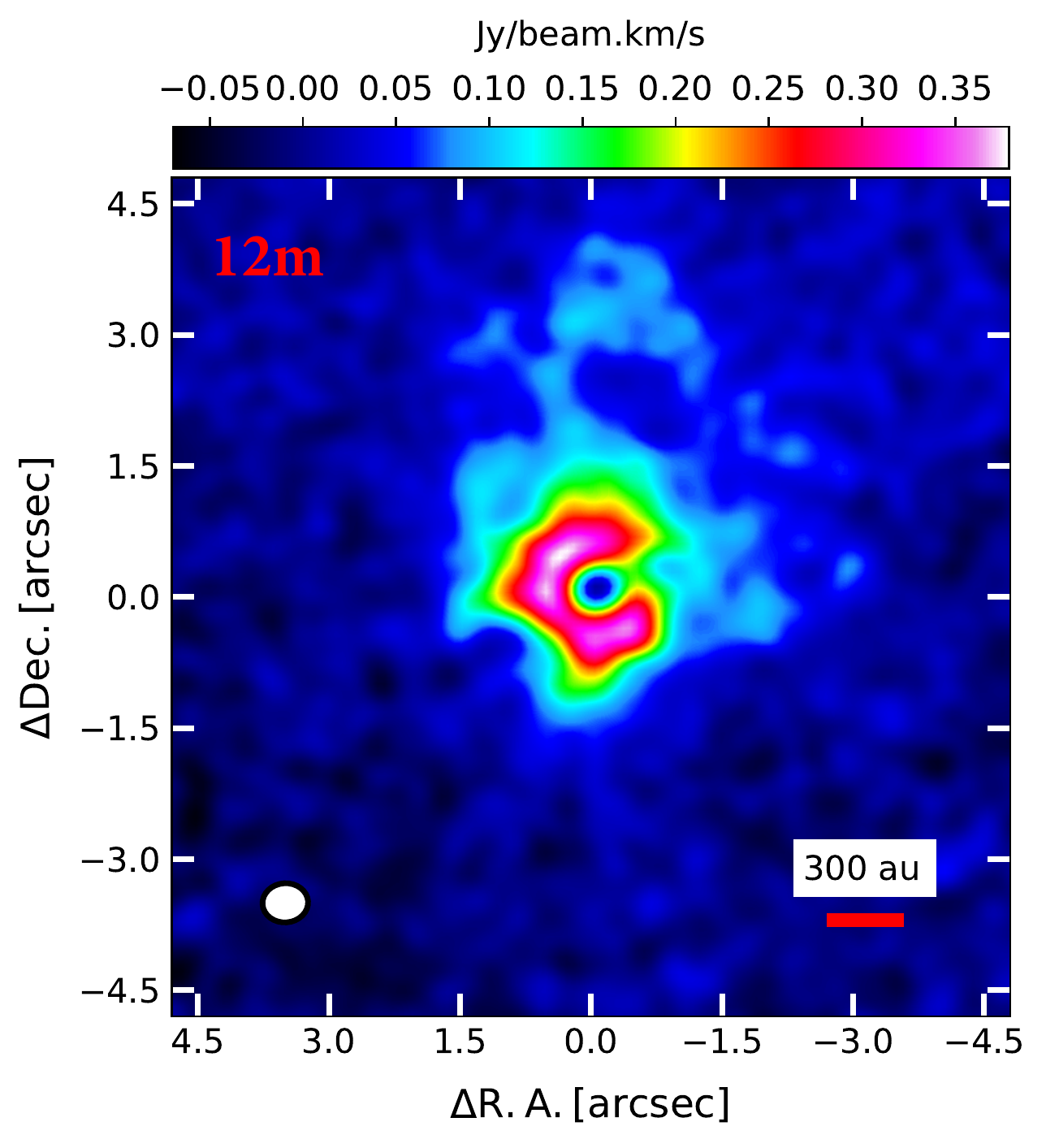}
\includegraphics[width=0.3\textwidth]{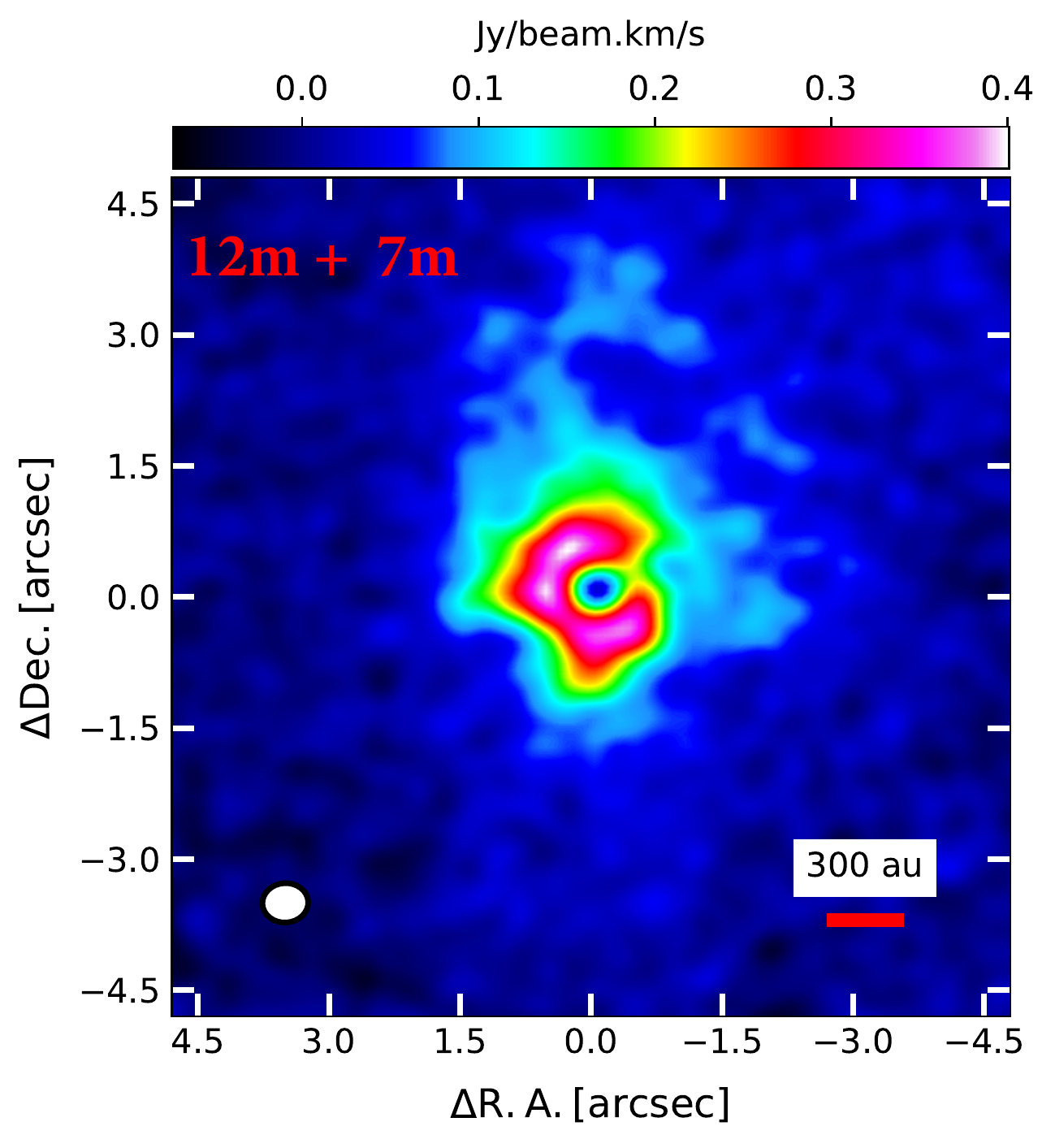}
\includegraphics[width=0.3\textwidth]{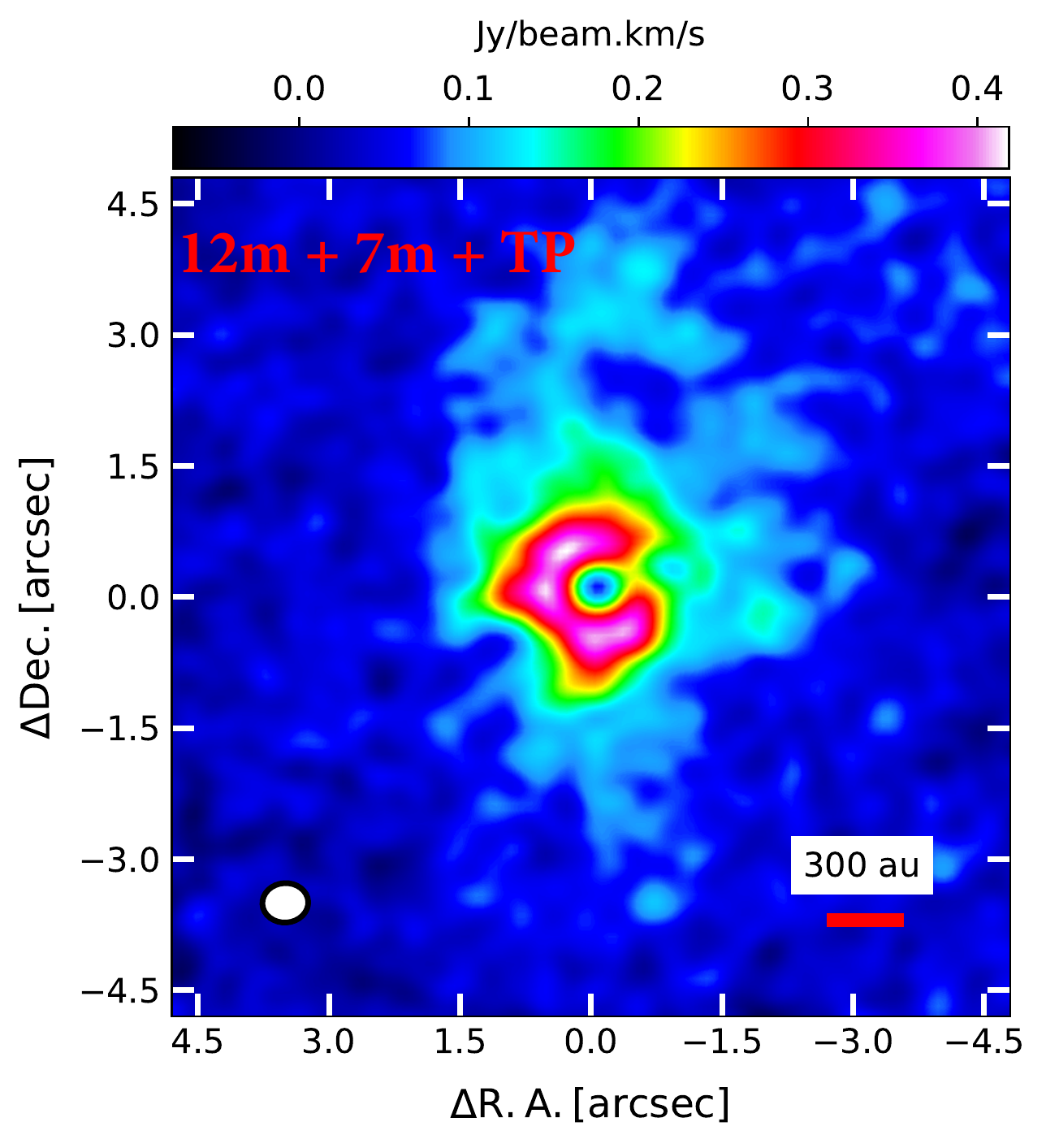}
\caption{ HCO$^{+}$ moment 0 maps of molecular emission toward V883 Ori. Left panel: HCO$^{+}$ moment 0 map using only ALMA 12-m data. Middle panel: HCO$^{+}$ moment 0 map combining 12-m and 7-m data. Right panel: HCO$^{+}$ moment 0 map with 12-m , 7-m, and TP visibilities. The line integrated emission of the HCO$^{+}$ line was performed in the range from 1.6 to 7.0 km s$^{-1}$. In each panel, the corresponding beam size is shown at the bottom left. North is up, east is left. }
\label{Fig:HCO_12m_all}
\end{figure*}

Initially, we run the Tp2vis function with three inputs: RMS, a mosaic pointing list, and the parameter \textit{nvgrp}, which controls the number of visibilities generated. For that matter, we derive the RMS noise of the TP map for each line using the IMSTAT task in CASA. We also adopt the set of the mosaic pointing coordinates of the 12-m data around which the TP visibilities are generated, and a value of 10 for  \textit{nvgrp}. Thus, the Tp2vis function sets the weights so that their sum represents the RMS noise of the original TP map. This process creates a MS of the TP data. To visualize and evaluate the relative weights among TP, 7-m, and 12-m visibilities, we use  the Tp2vispl function. Figure \ref{Fig:Weight} displays the weight density of the configurations, which appear to transition smoothly from TP to 7-m, and to 12-m. After the calibrated TP cube is converted into visibilities, and subtracting the continuum emission from the 7-m and 12-m data, we concatenate TP, 7-m, and 12-m MS with the CONCAT task to generate a single MS. We then run the TCLEAN task and obtain clean maps. For our purposes, we only present HCO$^{+}$ v=0 3-2 emission as a representative case of the line imaging process carried out in this paper combing the TP, 7-m, and 12-m data. Figure \ref{Fig:HCO_12m_all} compares the generated clean maps with Briggs value of 0.5 of the 12-m data alone, 12-m + 7-m data and combining TP${+}$7-m${+}$12-m data. To that end, we have integrated HCO$^{+}$ emission in the range from 1.6 to 7.0 km s$^{-1}$ to create the zeroth-moment maps. From these clean maps, it is evident that negative and noisier sidelobes are present in the moment-0 maps without TP data, while they are absent in the moment-0 map with TP data with a higher flux density level. Table \ref{Table:Weights} shows the missing flux of each array relative to the total flux measured by the TP array observations in the area of the TP beam size of $\sim$ 24$^{"}$ at the field center. As expected, the flux recovered by the combination of TP${+}$7-m${+}$12-m data is more evident at large scales because the \textit{uv} space is better filled with the TP data recovering the more extended emission.

\begin{table}
\caption{Missing Flux}
\begin{tabular}{lccc}
\hline
 & \multicolumn{3}{c}{\bf{Missing Flux}} \\

 \bf{Line}& 12-m & 7-m  \\
  & $\%$ & $\%$  \\
\hline
&\bf{ALMA Cycle 5}&&\\
\hline

DCN$^{a}$ v=0 J=3-2& 10 & 2 \\
H$_{2}$CO 3(2,1)-2(2,0) & 94 &85  \\
SO 3$\Sigma$ v=0 6(5)-5(4)& 72 & 55 \\

\hline
&\bf{ALMA Cycle 6}&\\
\hline

CH$_{3}$OH$^{a}$ v t=0 4(3,1)-4(2,2) & -- & -- \\
HCN v=0 J=3-2 &47 & 27 \\
HCO$^{+}$ v=0 3-2&  91 & 75\\
\hline
\hline
\end{tabular}\\
\label{Table:Weights}
\footnotesize{$^a$ No TP emission detected.}\\
\end{table}

\section{Channel maps}
\label{App:A}

\begin{figure*}
\includegraphics[width=0.95\textwidth,height=0.33\textheight]{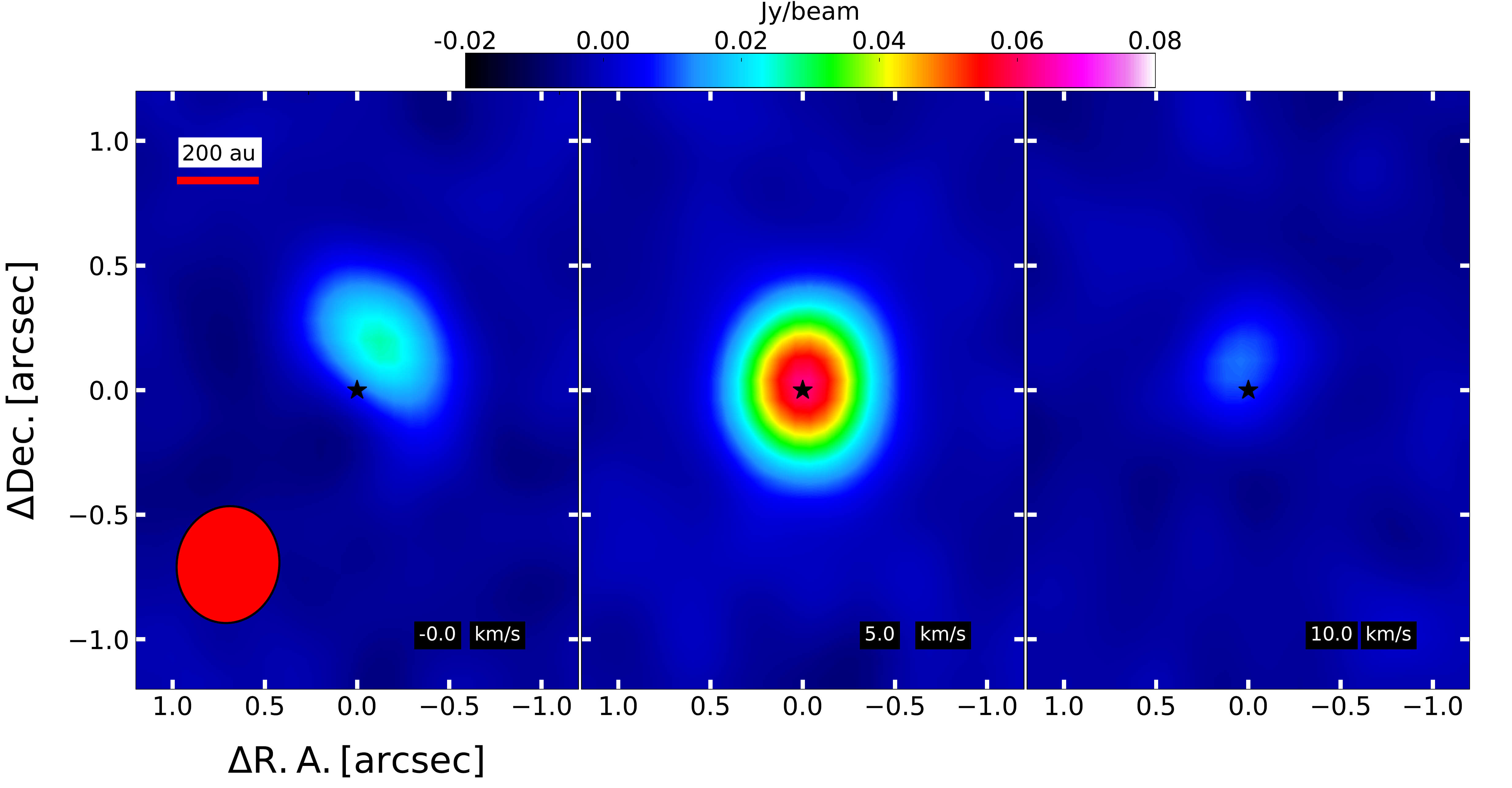}
\caption{Velocity channel maps of CH$_{3}$OH emission line towards V883 Ori. The velocity of the channel is indicated in the lower right corner of each panel.}
\label{Fig:CH3OHChannel}
\end{figure*}

\begin{figure*}
\includegraphics[width=0.95\textwidth]{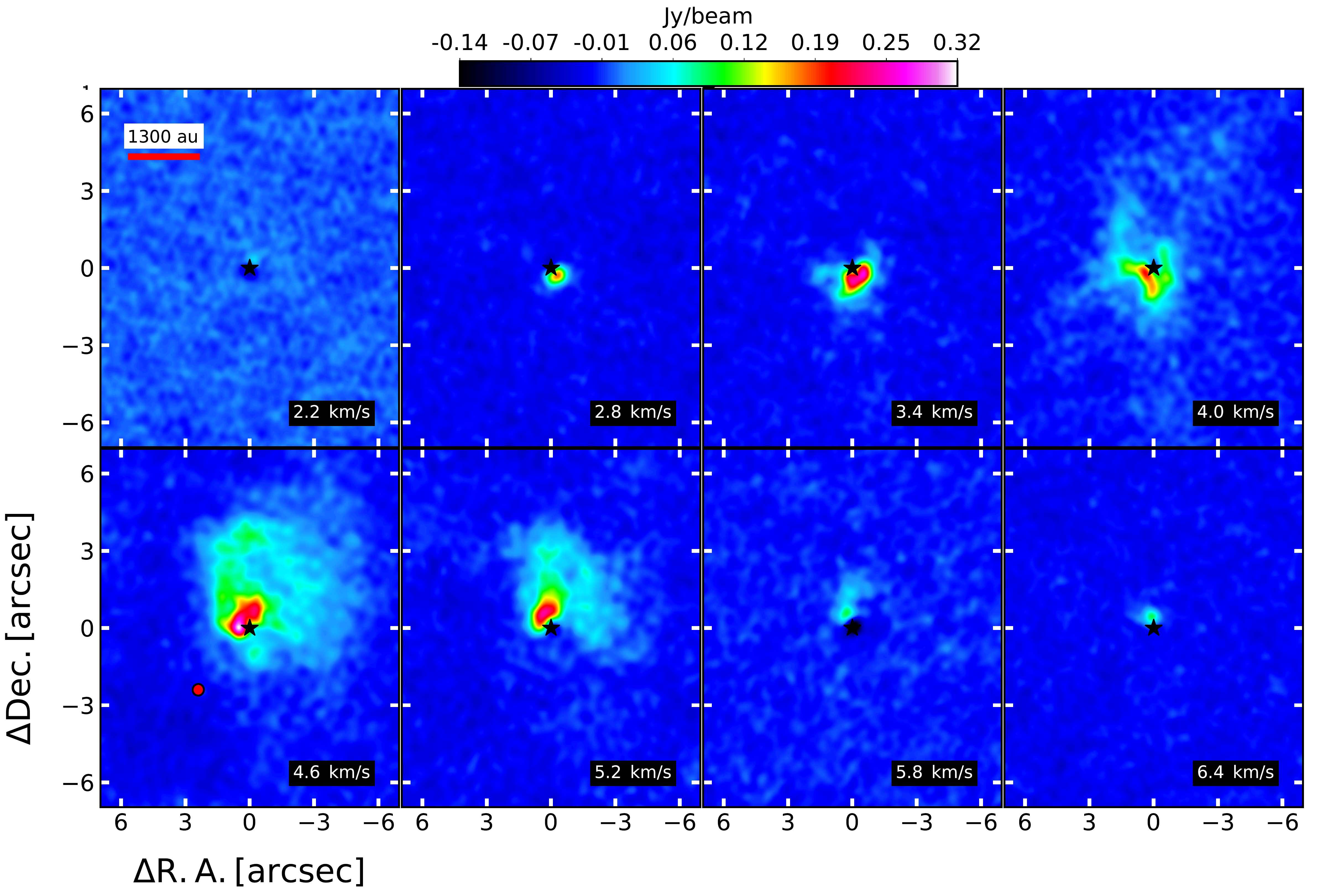}
\caption{Velocity channel maps of HCO$^{+}$ emission line towards V883 Ori. The velocity of the channel is indicated in the lower right corner of each panel. }
\label{Fig:HCOChannel}
\end{figure*}

\begin{figure*}
\includegraphics[width=0.95\textwidth]{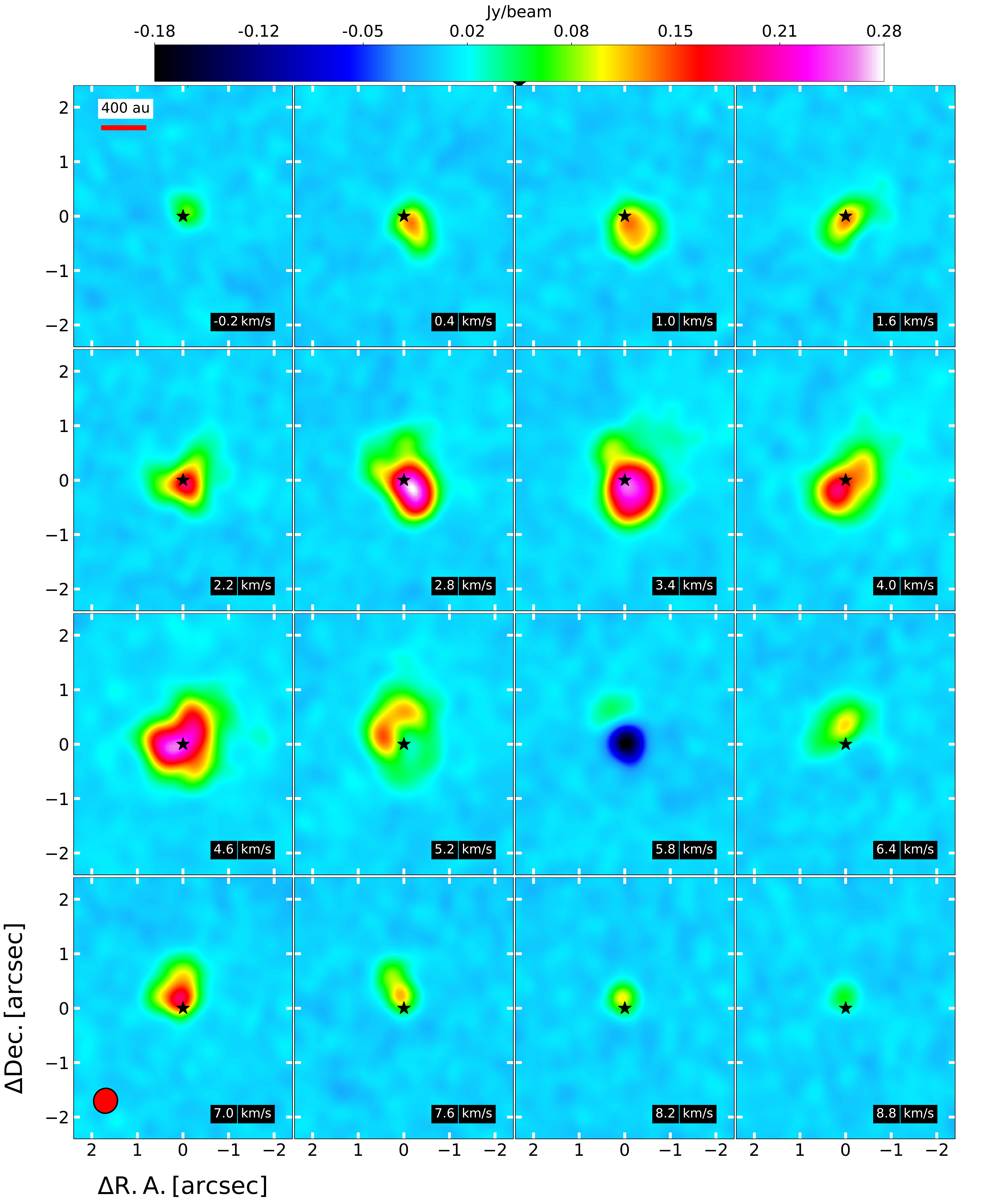}
\caption{Velocity channel maps of HCN emission line towards V883 Ori. The velocity of the channel is indicated in the lower right corner of each panel.}
\label{Fig:HCNChannel}
\end{figure*}

\begin{figure*}
\includegraphics[width=0.95\textwidth]{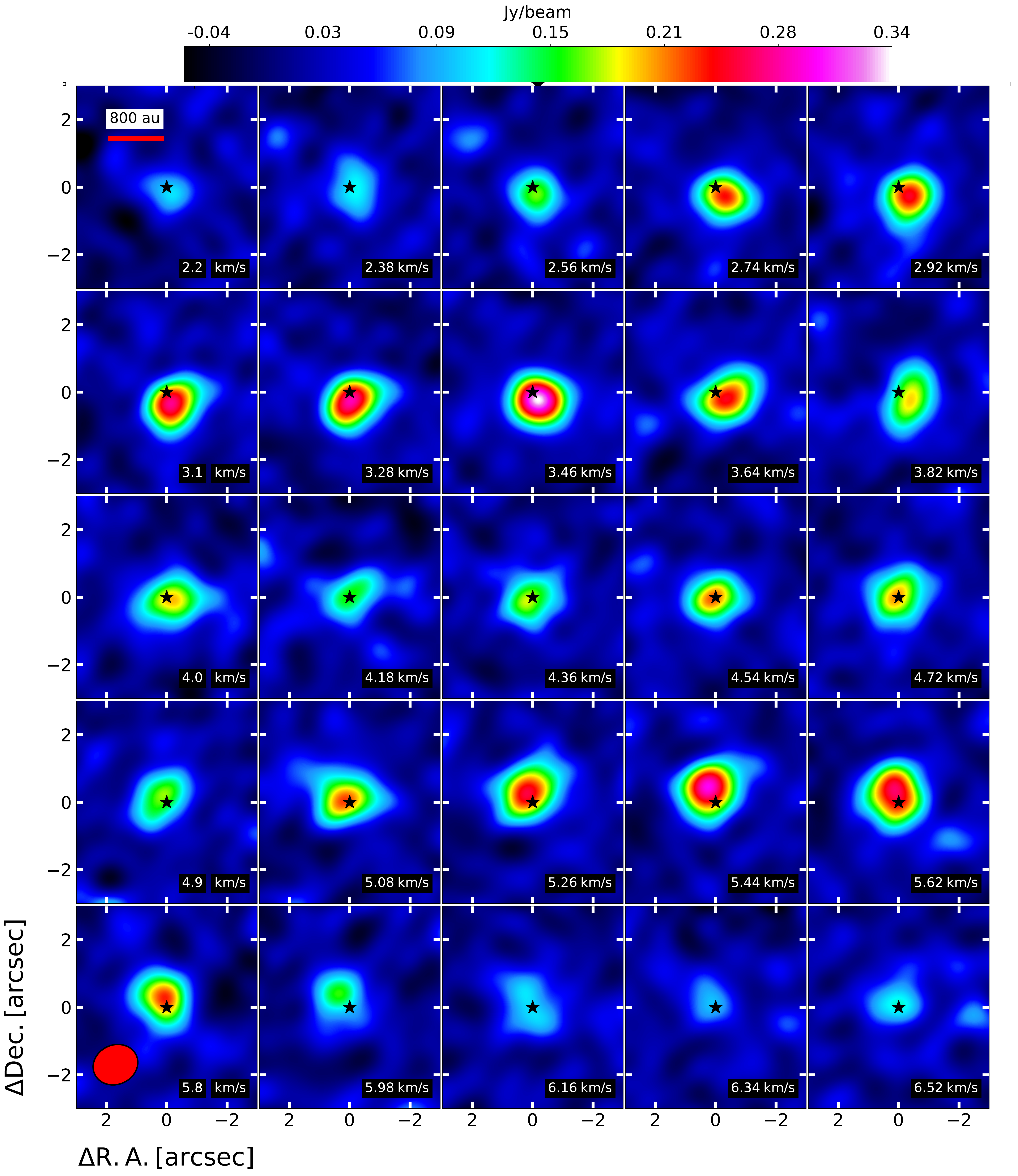}
\caption{Velocity channel maps of DCN emission line towards V883 Ori. The velocity of the channel is indicated in the lower right corner of each panel.}
\label{Fig:DCNChannel}
\end{figure*}

\begin{figure*}
\includegraphics[width=0.95\textwidth]{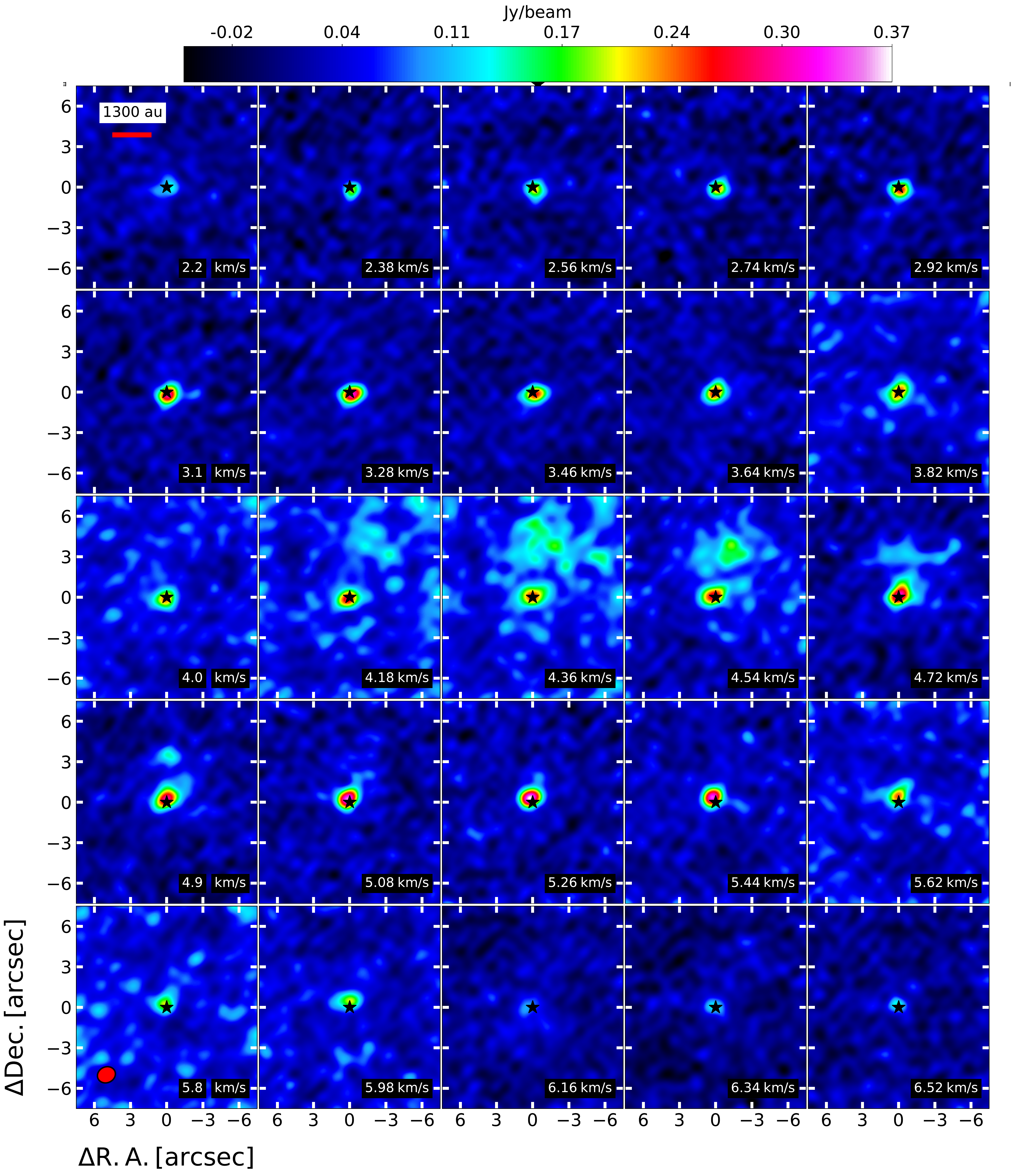}
\caption{Velocity channel maps of H$_{2}$CO emission line towards V883 Ori. The velocity of the channel is indicated in the lower right corner of each panel.}
\label{Fig:H2COChannel}
\end{figure*}

\begin{figure*}
\includegraphics[width=0.95\textwidth]{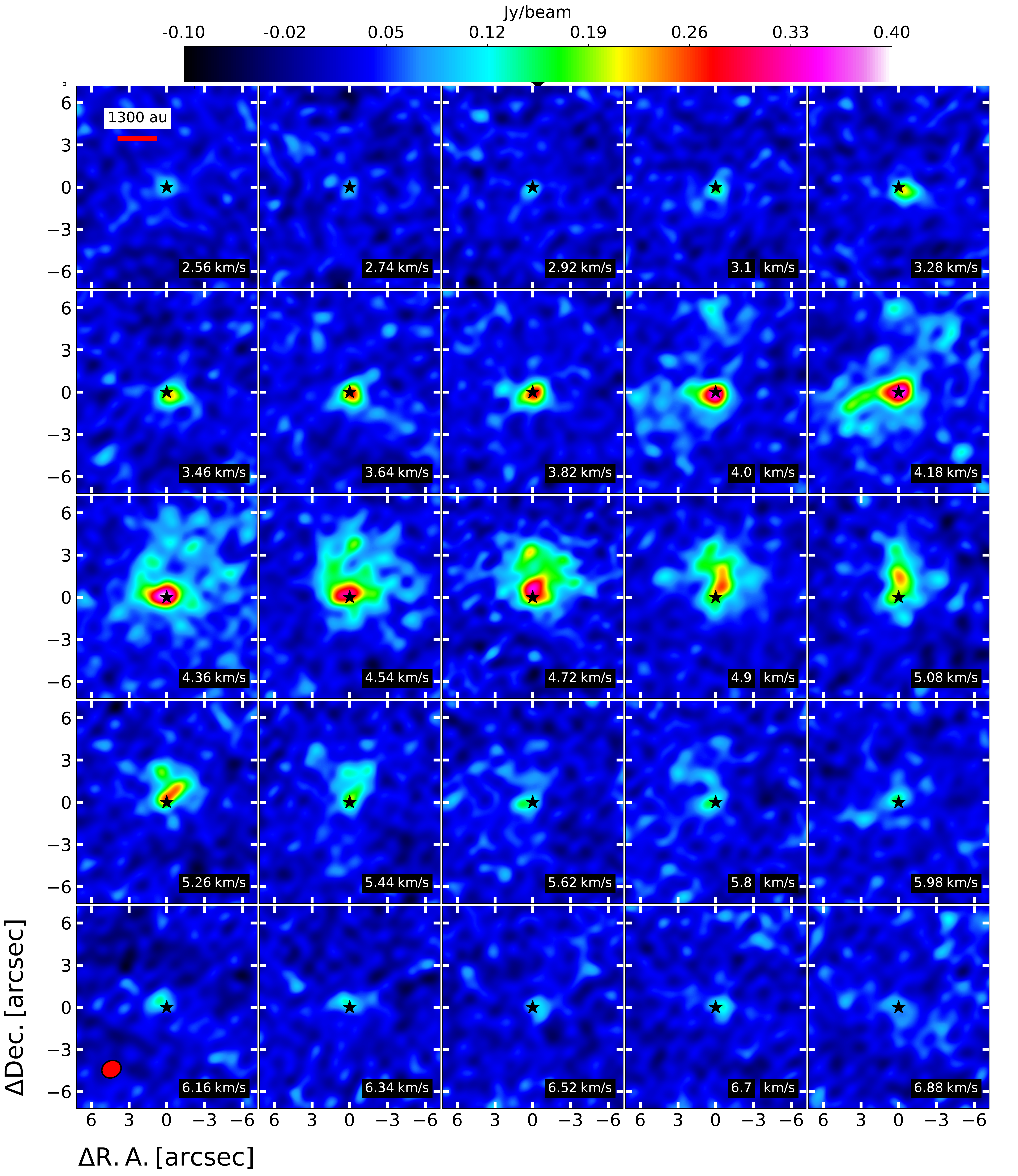}
\caption{Velocity channel maps of SO emission line towards V883 Ori. The velocity of the channel is indicated in the lower right corner of each panel.}
\label{Fig:SOChannel}
\end{figure*}

The CH$_{3}$OH, HCO$^{+}$, HCN, DCN, H$_{2}$CO, and SO channel maps are displayed in Figs.  \ref{Fig:CH3OHChannel},  \ref{Fig:HCOChannel}, \ref{Fig:HCNChannel}, \ref{Fig:DCNChannel}, \ref{Fig:H2COChannel}, and \ref{Fig:SOChannel}, respectively.

\clearpage

\clearpage

\clearpage

\section{Radial Profiles}
\label{App:B}

\begin{table*}[!h]
\caption{Deconvolved radial extensions from ALMA observations.}
\label{Table:RPCO}
\begin{tabular}{lccccccccccc}
\hline Line & Rest Freq. & Peak Flux & FWHM$^a$ & Beamwidth & Radial Extension$^b$\\
& & Jy beam$^{-1}$ km s$^{-1}$ & [au] & [au] & [au] & \\
\hline
$^{12}$CO & 230.559& 0.32 & 655 & 128 & 320 $\pm$ 20\\
$^{13}$CO & 220.398& 0.28 & 715 & 138 & 350 $\pm$ 40 \\
C$^{18}$O & 219.560& 0.22 & 620 & 137 & 310 $\pm$ 10 \\
\hline
\end{tabular}
\end{table*}

\begin{figure}
\includegraphics[width=0.5\textwidth]{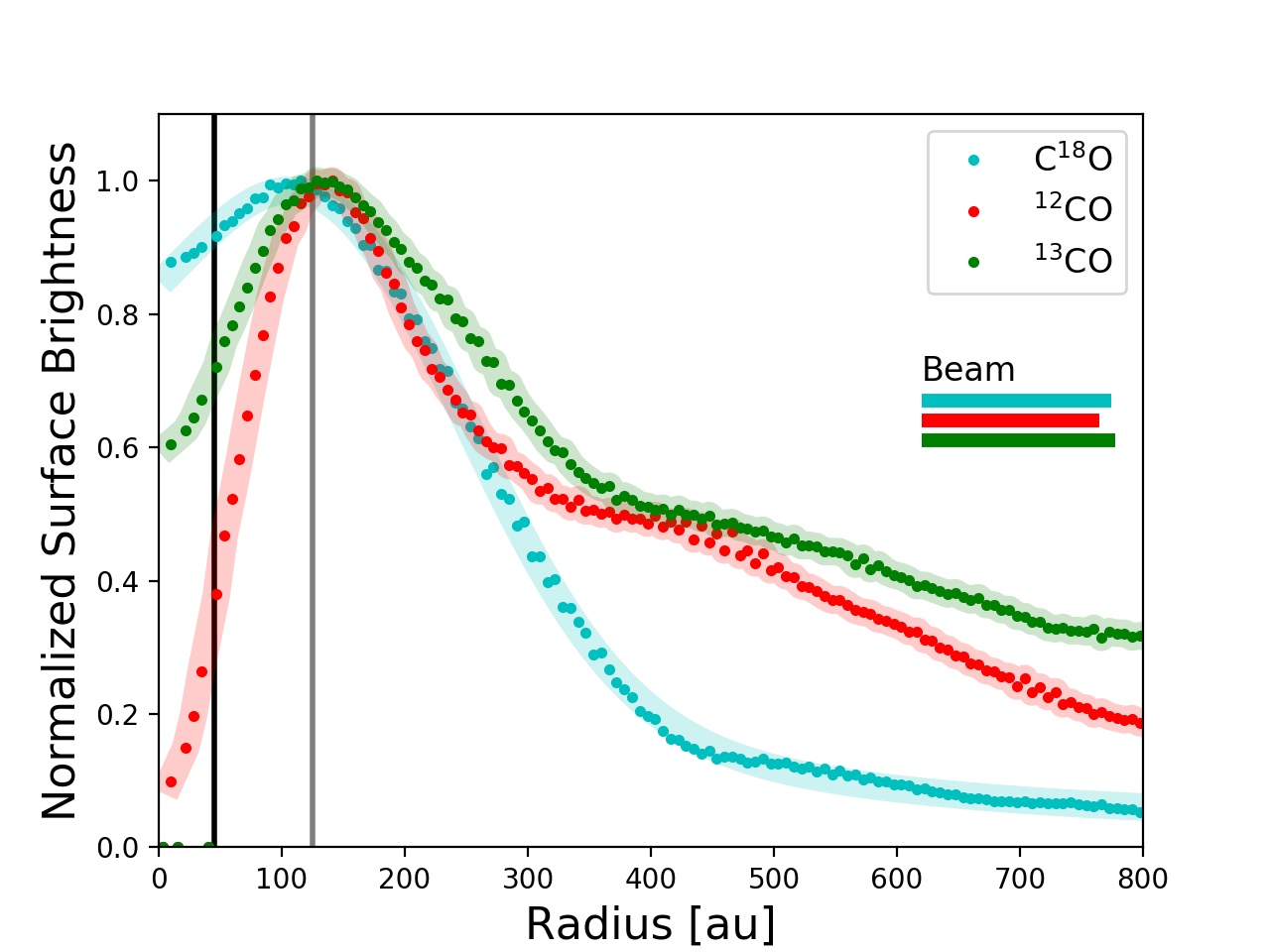}
\caption{Azimuthally averaged and de-projected radial intensity profiles for ALMA cycle 3 observations previously presented in \citet{Ruiz2017b}. The black vertical line corresponds the snow-line location at 45 au estimated by \citet{Cieza2016}; the gray vertical line ($\sim$ 125 au) indicates the radial extension of the dust and CH$_{3}$OH emission inferred by \citet{Cieza2016} and \citet{Lee2019}. }
\label{Fig:RPCO}
\end{figure}

\begin{figure}
\centering
    \centering
    \includegraphics[width=0.45\textwidth]{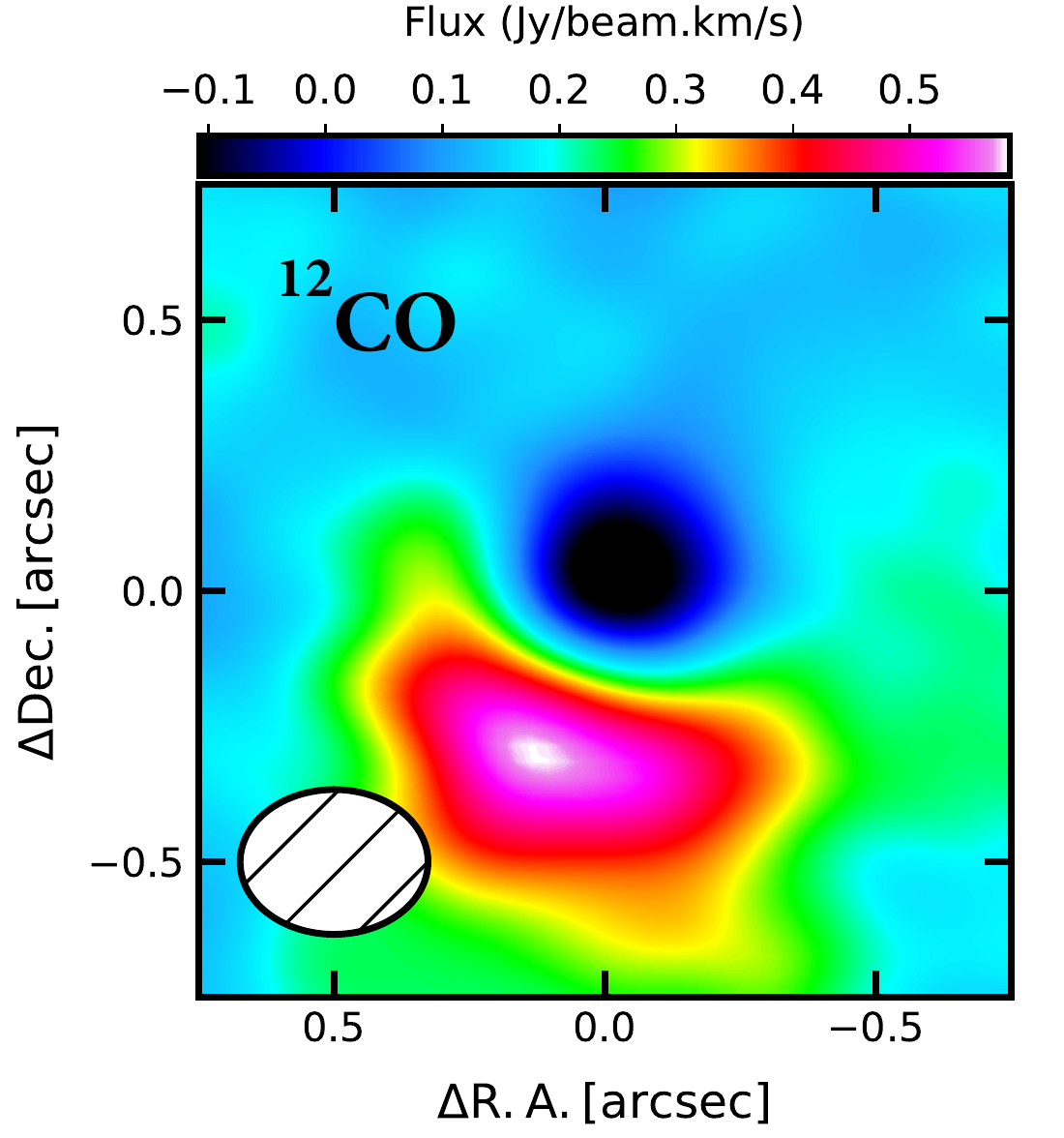}\\
    \includegraphics[width=0.45\textwidth]{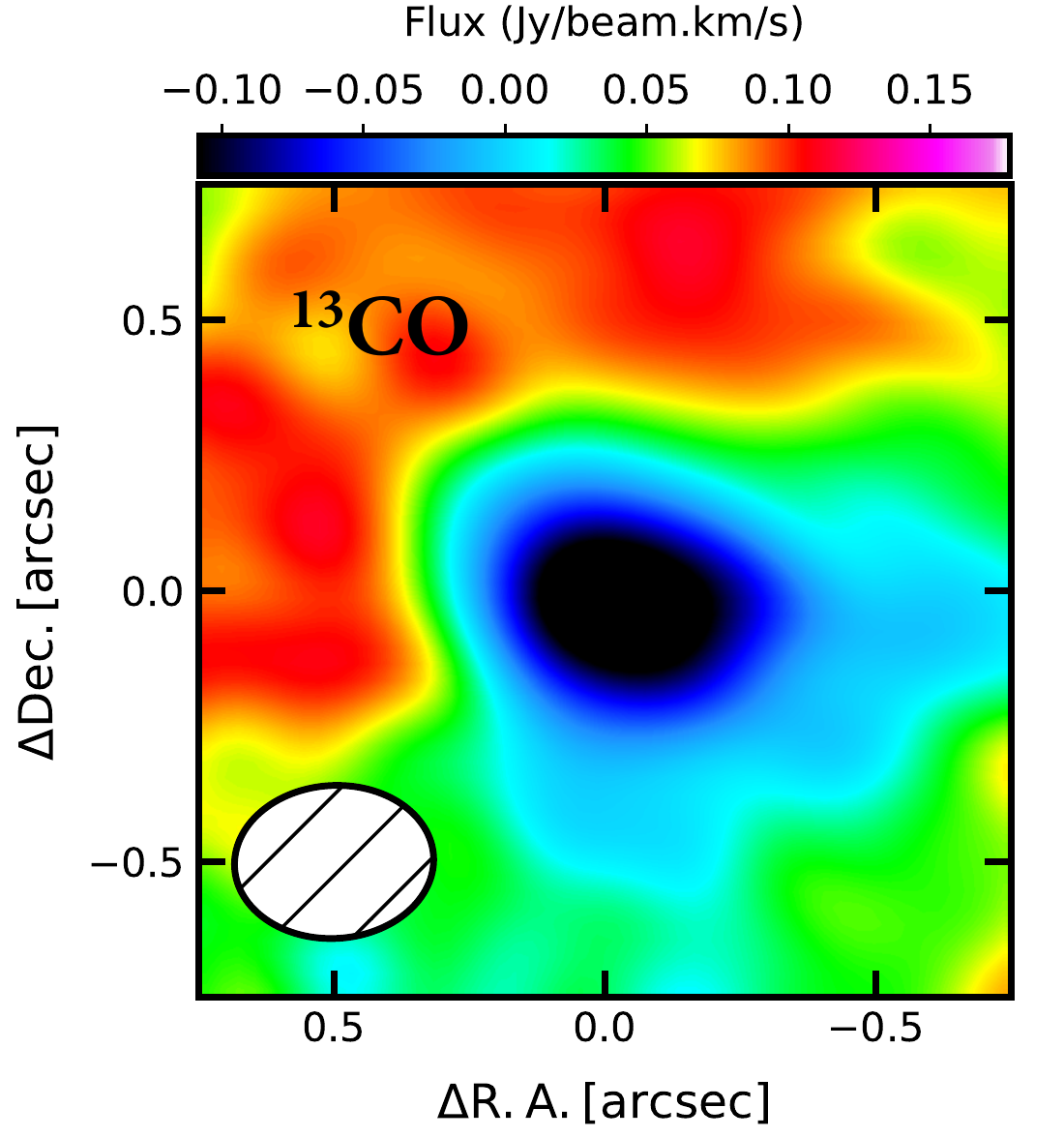}

\caption[short]{Moment 0 maps of $^{12}$CO (top) and $^{13}$CO (bottom) emission toward V883 Ori. Original images are presented in \citet{Ruiz2017b} together with reduction data details.}
\label{Fig:COring}
\end{figure}

The chemical possible pathways taken place to generate the observed emission distribution were discussed in Section \ref{Sec:Chemical}. To that end, we included the radial intensity profiles for $^{12}$CO, $^{13}$CO, and C$^{18}$O calculated from the moment zero maps presented in \citep{Ruiz2017b}. We followed the method described in Sec. \ref{Sec:Results} to characterize these molecule distributions (see Table \ref{Table:RPCO}), and using an inclination of 38 deg. for deprojecting and azimuthally averaging the radial profiles display in Figure \ref{Fig:RPCO}. We have only considered the bulk of the $^{12}$CO, $^{13}$CO, and C$^{18}$O emission at levels of $\sim$7$\sigma$. From these images, all molecule peaks are shifted further outward. At similar resolutions, these molecules peak at $\sim$ 130 au and appear depleted from the gas phase within the snow-line location and/or trace dust absorption. However, considering that the $^{12}$CO line emission is generally optically thick, and C$^{18}$O is optically thin, it is likely depletion of CO gas throughout vertical layers with the largest depletion at the disc surface, as deduced from the $^{12}$CO radial profile, see Figure \ref{Fig:RPCO}. In order to estimate the inner radial depression of $^{12}$CO, we have followed a similar approach as presented in Sec. \ref{Sec:RP} for the HCO$^{+}$ original data. Thus, using $\rm R_{r}=$135 au and a $\rm FWHM_{Beam} =$ 128, the inner depression is $\sim$ 120 au. In addition, Figure \ref{Fig:COring} displays the $^{12}$CO and $^{13}$CO moment zero maps, whose integration reveals each a pronounced inner depression or `'hole'' with a size comparable to that estimated for the 230 GHz continuum by \citet{Cieza2016}.

\begin{figure}
\centering
    \includegraphics[width=0.45\textwidth]{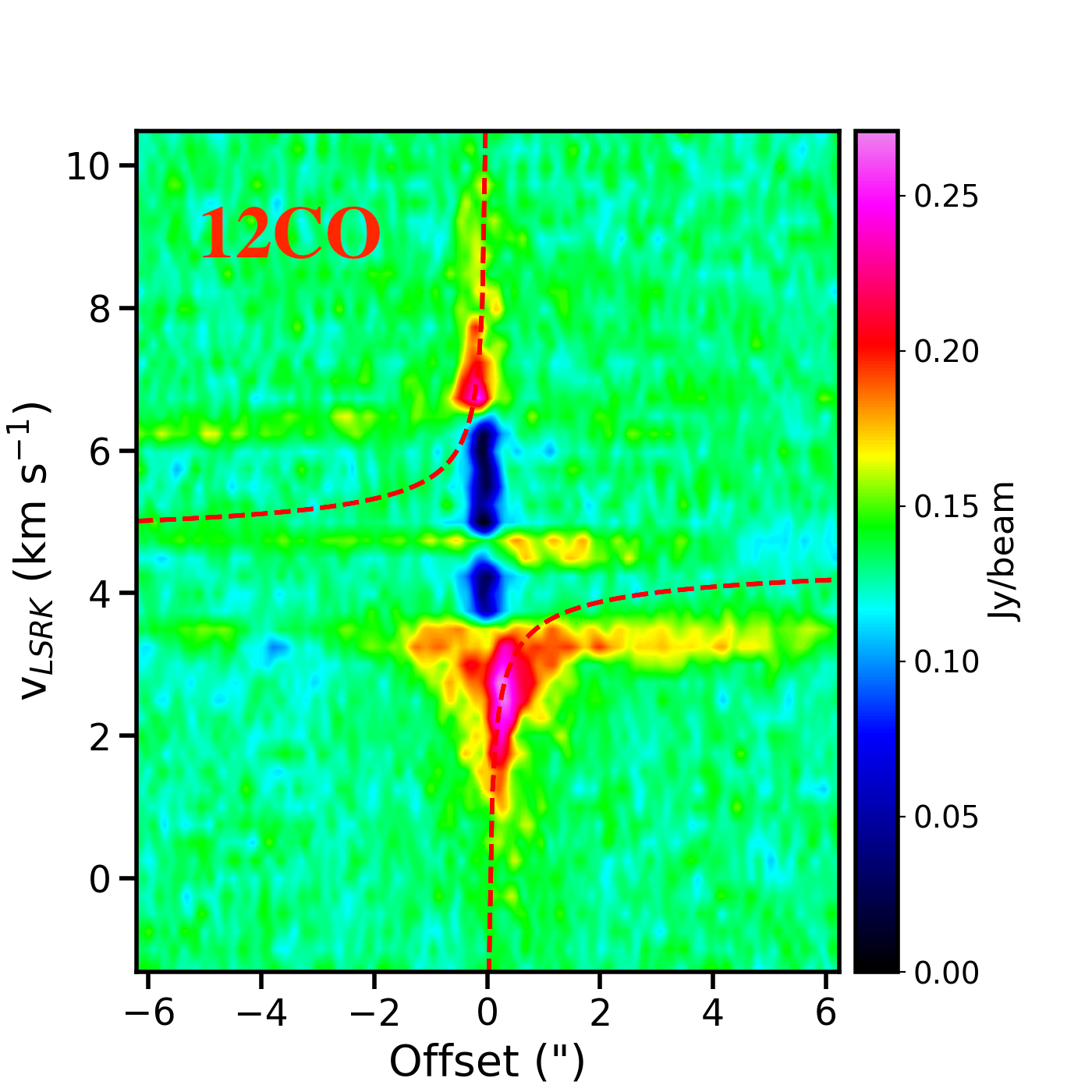}
    \includegraphics[width=0.45\textwidth]{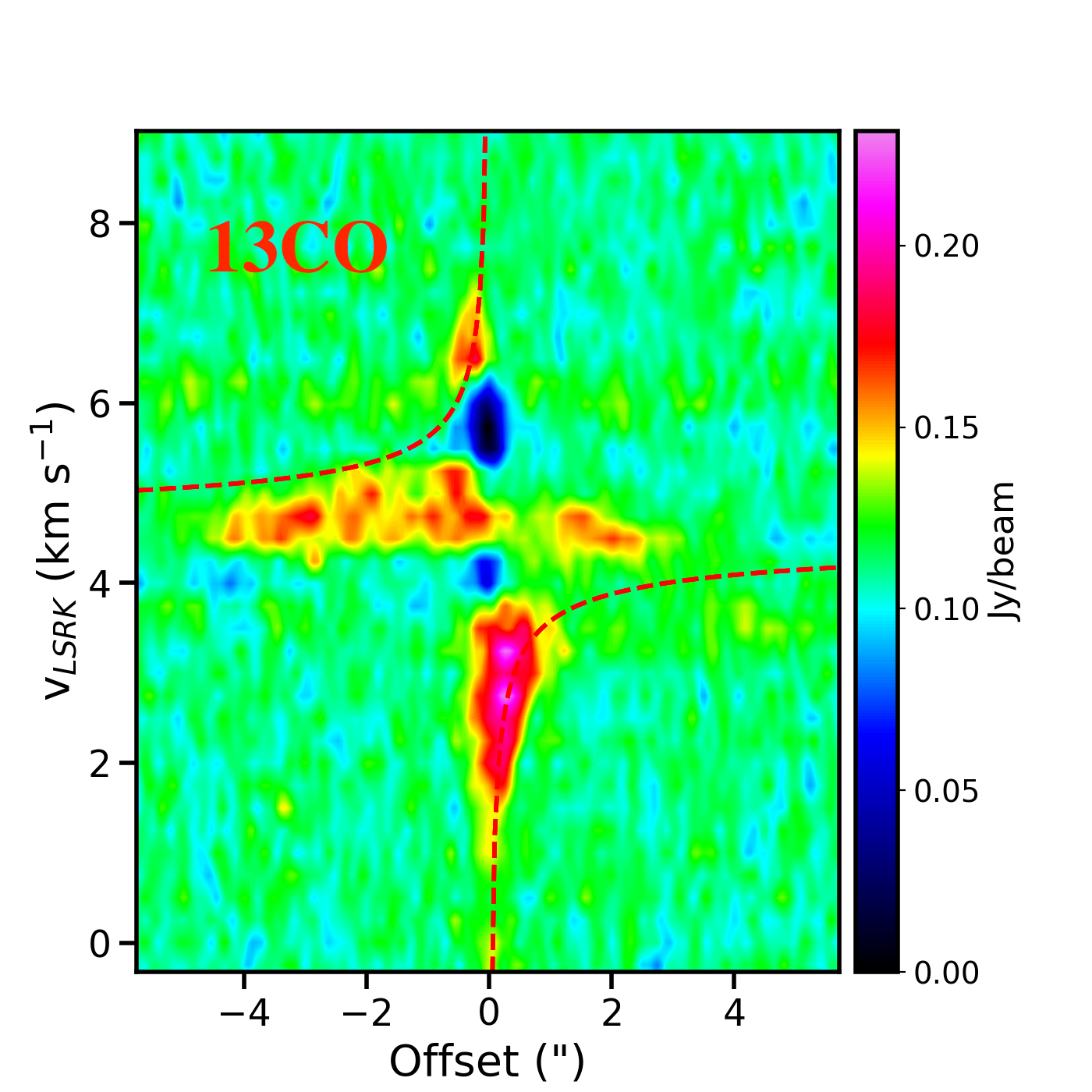}
\caption[short]{Position-velocity diagrams extracted along the disc plane, i.e. P.A. 30 deg. for $^{12}$CO and $^{13}$CO emission \citep{Ruiz2017b}. The red dashed curves represent the keplerian motion around a central source of 1.3 M$_{\odot}$ \citep{Cieza2016}.}
\label{Fig:PVCO}
\end{figure}

\begin{figure}
\centering
    \centering
     \includegraphics[width=0.45\textwidth]{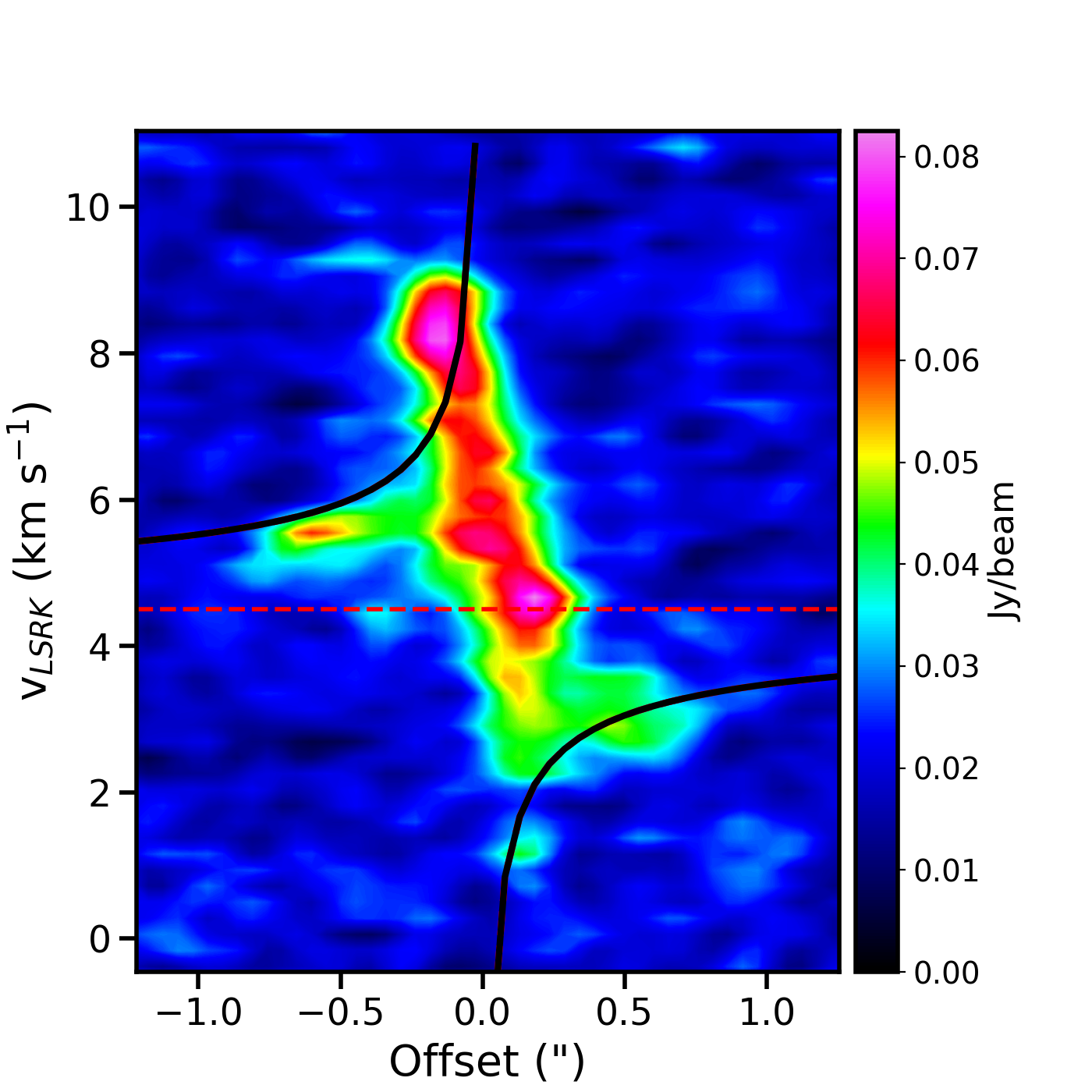}
           
\caption[short]{Position-velocity diagrams of the H$^{13}$CO$^{+}$ J = 4-3 emission line extracted along the rotating disc, i.e. P.A.$\sim$ 30$^{\rm o}$ in V883 Ori. The slice is taking from the H$^{13}$CO$^{+}$ data cube highly blended with the CH$_{3}$CHO transitions. The red dashed line indicates the systemic velocity of $\sim$4.5 km s${-1}$. The black curves show the theoretical Keplerian velocity for a 1.3 M$_{\odot}$ star. The V$_{\rm LSRK}$scale is with respect to the H$^{13}$CO$^{+}$ component detected at frequency 346.998 GHz. Gas structures within the $\sim$$\pm$0.15 $^{"}$ angular offset are dominated by a superposition of the CH$_{3}$CHO transitions and H$^{13}$CO$^{+}$ J = 4-3 emission.}
\label{Fig:PV_H13CO}
\end{figure}

In addition, the $^{12}$CO, $^{13}$CO, and H$^{13}$CO$^{+}$ emission lines also present a typical Keplerian rotation and self-absorption emission around the central source. In Figures \ref{Fig:PVCO} and \ref{Fig:PV_H13CO}, we present PV diagrams along the disc major axis with P.A. of 30 deg. overlaid with curves representing Keplerian rotation in a geometrically thin disc with an inclination of 38 deg., a central mass of 1.3 M$_{\odot}$ and systemic velocity of $\sim$ 4.5 km s$^{-1}$, assuming a distance of 417 pc.

\section{Effects of subtracting optically thick lines and continuum emission.}
\label{App:D}

\begin{figure*}
\centering
    \centering
     \includegraphics[width=0.43\textwidth]{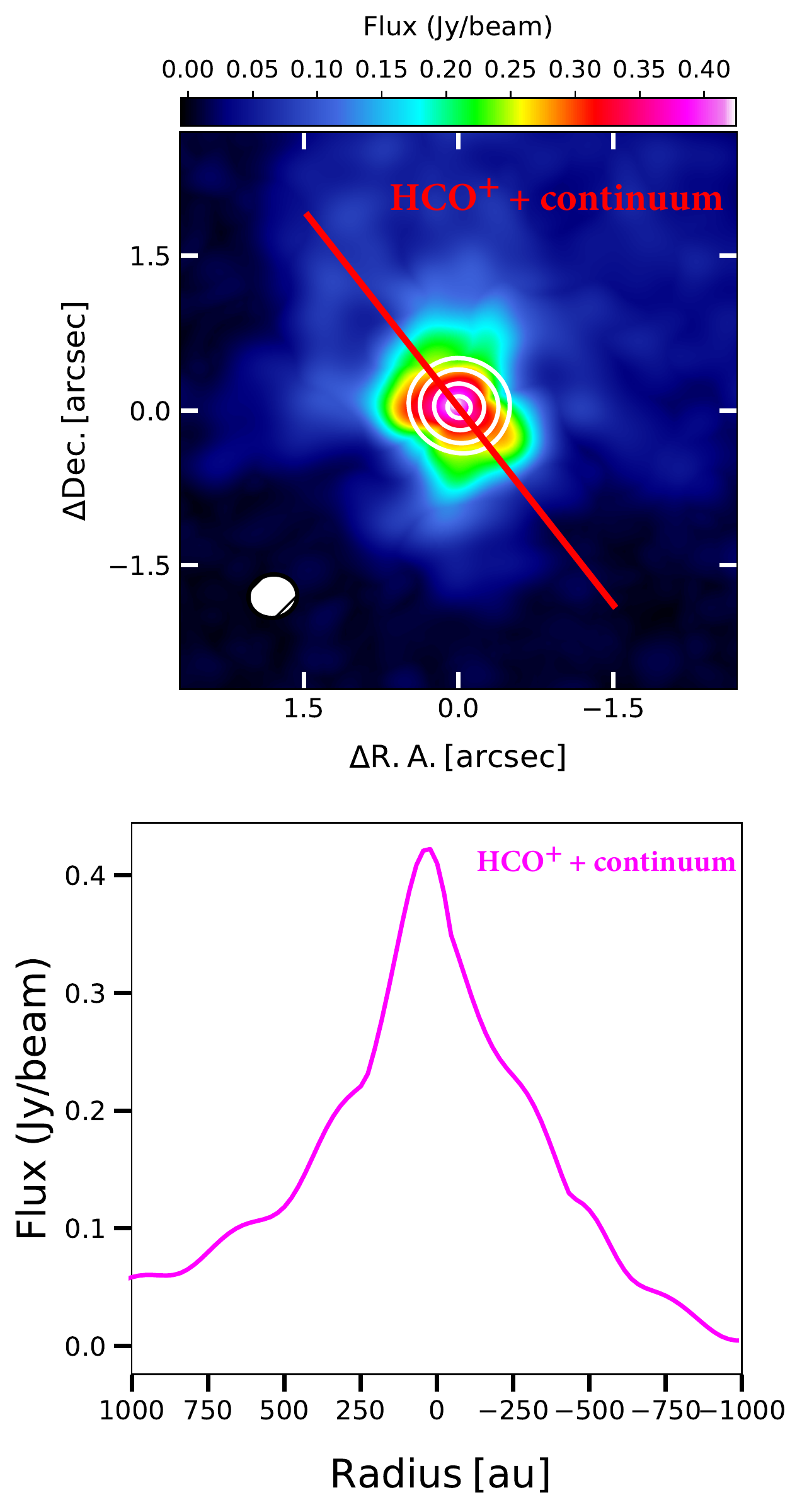}
     \includegraphics[width=0.43\textwidth]{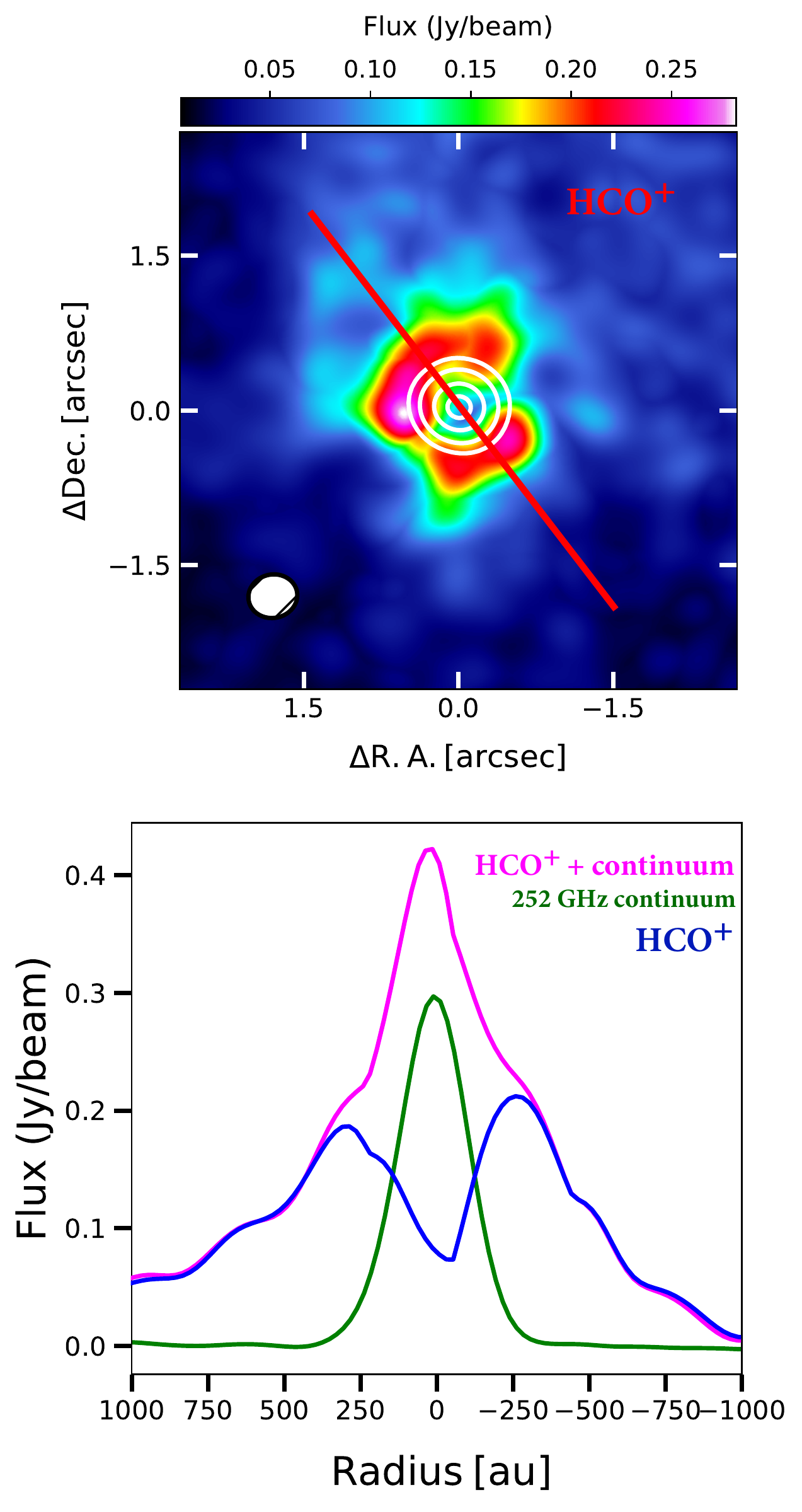}
\caption[short]{HCO$^{+}$ maximum peak intensity emission maps from data cubes obtained without (\textit{Left - Top panel}) and after (\textit{Right - Top panel}) subtracting the bright continuum. The bottom panels display the profiles extracted from the HCO$^{+}$ maximum peak intensity maps without (\textit{solid magenta line}) and after (\textit{solid blue line}) subtracting the continuum emission, and from the 252 GHz continuum emission obtained during ALMA cycle-6 (\textit{solid green line}). In the top panels, the red line represents the cut used to extract the intensity profiles. For dimensional comparison, white contours represent the dust continuum emission detected at 252 GHz at 0.1, 0.15, 0.2 and 0.25 Jy/beam intensity levels. The white ellipse in the lower left corner indicates the beam size of 0.47$^{"}$ $\times$ 0.45$^{"}$ at PA $\sim$ 77 deg.}
\label{Fig:Mom8}
\end{figure*}

\begin{figure}
\centering
    \centering
     \includegraphics[width=0.45\textwidth]{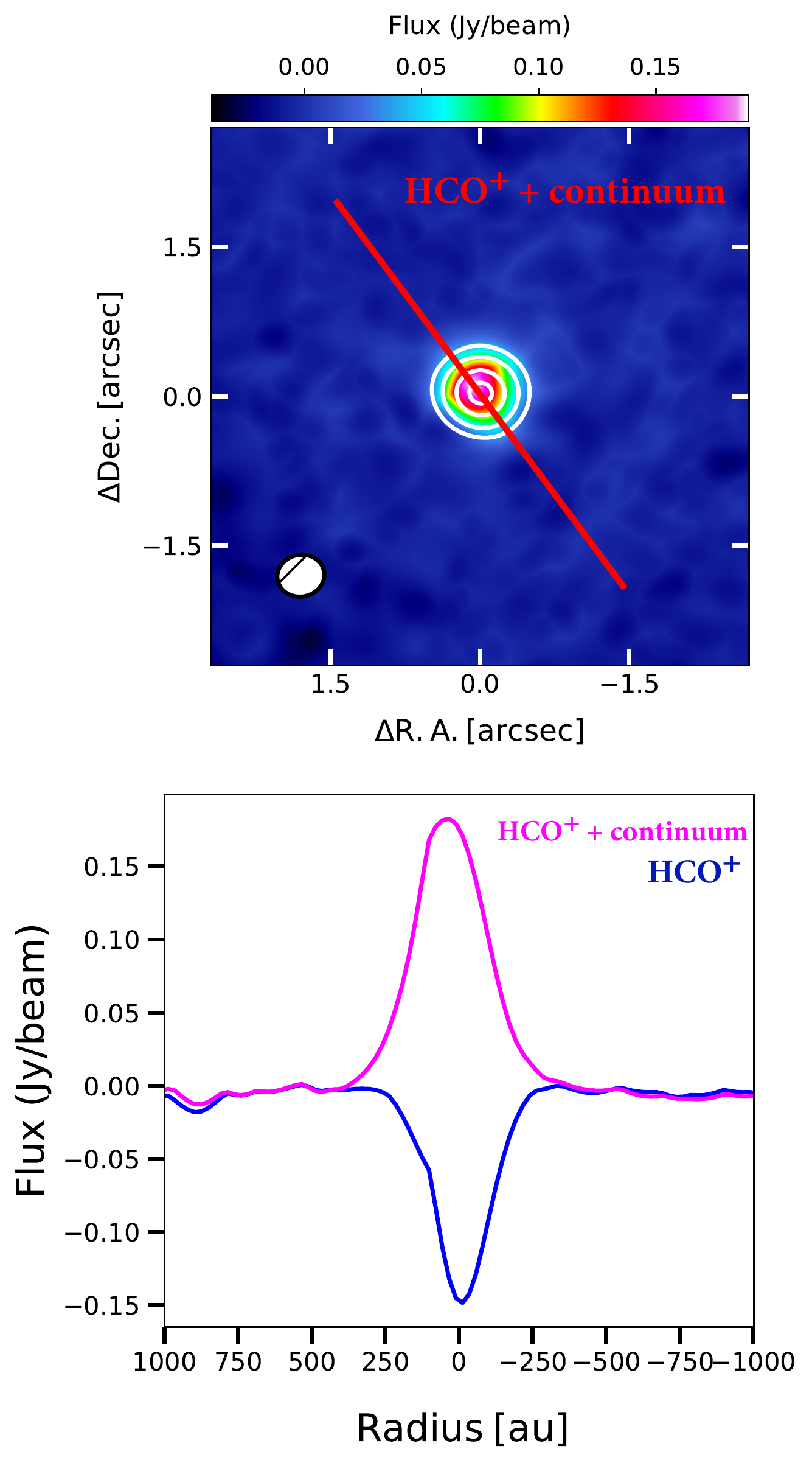}
\caption[short]{HCO$^{+}$ minimum peak intensity emission  map from the data cubesobtained without (\textit{Top panel}) subtracting the bright continuum. The bottom panel display the profiles extracted from the HCO$^{+}$ minimum peak intensity maps without (\textit{solid magenta line}) and after (\textit{solid blue line}) subtracting the continuum emission. The red line represents the cut used to extract the intensity profiles. For dimensional comparison, white contours represent the dust continuum emission detected at 252 GHz at 0.1, 0.15, 0.2 and 0.25 Jy/beam intensity levels. The white ellipse in the lower left corner indicates the beam size of 0.47$^{"}$ $\times$ 0.45$^{"}$ at PA $\sim$ 77 deg.}
\label{Fig:Mom10}
\end{figure}

In order to evaluate the amount of line emission removed by applying a flat continuum in the subtraction process of a very bright source as is the case of V883 Ori, we generated spectral line maximum and minimum peak intensity maps from HCO$^{+}$ data cubes obtained with and without subtracting the continuum emission. Figure \ref{Fig:Mom8} displays the HCO$^{+}$ peak emission calculated without subtracting the continuum emission (top -left panel) and after subtracting the continuum emission (top-right panel). Similarly, Fig. \ref{Fig:Mom10} displays the minimum intensity map without subtracting the continuum emission (top panel). In the maximum peak intensity map without subtracting the continuum, the emission significantly increases towards the center to account for the extra emission attributed to the strong continuum, while after subtracting the very bright continuum, emission is removed from the most inner disc region causing the false impression of a ``hole''. The anti-correlation between the dust and HCO$^{+}$ emission can be seen more clearly at the bottom panels of figure \ref{Fig:Mom8} showing the profiles extracted from the maximum peak intensity maps --with and without subtracting the continuum-- along the disc keplerian rotation axis ($\sim$ 30 deg; red lines drawn at the top-panels ). From these intensity profiles, it can be seen that the HCO$^{+}$ emission, shown in blue lines, drops within a radius of 125 au, which corresponds to the dust continuum radial extension \citep{Cieza2016}. This reduction is the strongest at the center of the continuum, where the brightness intensity of the continuum is expected to be the highest. That is, HCO$^{+}$ is absorbed by the optically thick continuum emission at the line frequency.

In this particular case, the minimum peak intensity maps serve to visualize the effects of applying a linear fit of the continuum based on the level measured in the line-free channel in a very bright source to subtract the dust continuum contribution. As expected for a very bright continuum emission, the minimum intensity map without subtracting the continuum emission probes the disc structure peaking at the center of the inner region. This outcome is particularly severe, as the dust in V883 Ori is much brighter due to high temperatures and densities in the inner disc. In addition, when the continuum is absorbed by foreground cold gas in the line of sight, this leads to overestimating the dust contribution and then underestimating the line emission, yielding negative flux values after the process of continuum subtraction. The profiles from the  minimum peak intensity maps shown in Fig. \ref{Fig:Mom10}b display the dust contribution (magenta line) and resulting negative flux values (blue line) after the process of continuum subtraction. Thus, the removal of line emission is maximized when both continuum and line emission are optically thick, absorbing each other, \citep{Weaver2018} as is the case for V883 Ori.

% Don't change these lines
\bsp	% typesetting comment
\label{lastpage}
\end{document}